\documentclass[a4paper,graphics,natbib,graphicx,psfig,amssymb,ccaption]{article}
\usepackage{lscape}
\usepackage{graphicx}
\newcommand{\affil}[1]{$^{\rm #1}$}
\renewcommand{\baselinestretch}{1.2}
\usepackage[authoryear]{natbib}
\bibpunct{(}{)}{;}{a}{}{,}
\date{} 

\title{\large\bf\flushleft Absolute Magnitude Calibration for Red Giants based on the Colour-Magnitude Diagrams of Galactic Clusters. III-Calibration with 2MASS}
\author{\parbox{\textwidth}{\flushleft
\vspace{-0.5cm}
{\it S. Karaali\affil{\dag, A, B}, S. Bilir\affil{A}, and E. Yaz G\"ok\c ce\affil{A}}\\
\vspace{0.4cm}
{\small \affil{A}\,Istanbul University, Faculty of Sciences, Department of Astronomy and Space Sciences, 34119, Istanbul, Turkey}\\
{\small \affil{B}\,Email: karsa@istanbul.edu.tr}}}
\begin{document}
\twocolumn[
\begin{changemargin}{.8cm}{.5cm}
\begin{minipage}{.9\textwidth}
\vspace{-1cm}
\maketitle
\small{\bf Abstract:}
We present two absolute magnitude calibrations, $M_{J}$ and $M_{K_s}$, for red 
giants with the colour magnitude diagrams of five Galactic clusters with different 
metallicities i.e. M92, M13, M71, M67, and NGC 6791. The combination of the 
absolute magnitudes of the red giant sequences with the corresponding 
metallicities provides calibration for absolute magnitude estimation for red 
giants for a given colour. The calibrations for $M_{J}$ and $M_{K_s}$ are 
defined in the colour intervals 
$1.3\leq(V-J)_{0}\leq2.8$ and $1.75 \leq (V-K_{s})_{0}\leq 3.80$ mag, 
respectively, and they cover the metallicity interval 
$-2.15 \leq \lbrack Fe/H \rbrack \leq +0.37$ dex. The absolute magnitude 
residuals obtained by the application of the procedure to another set of 
Galactic clusters lie in the intervals $-0.08<\Delta M_{J}\leq +0.34$ and 
$-0.10< \Delta M_{K_s}\leq +0.27$ mag for $M_{J}$ and $M_{K_s}$, respectively. 
The means and standard deviations of the residuals are $<\Delta M_{J}>= 0.137$ 
and $\sigma _{M_J}=0.080$, and $<\Delta M_{K_s}>=0.109$ and 
$\sigma_{M_{K_{s}}}=0.123$ mag. The derived relations are applicable to stars 
older than 4 Gyr, the age of the youngest calibrating cluster.        

\medskip
\medskip
\end{minipage}
\end{changemargin}
]
\small
\let\thefootnote\relax\footnote{\small \affil{\dag}\,Retired.}
\section{Introduction}
Systematic studies of star clusters help us to understand the Galactic structure 
and star formation processes as well as stellar evolution. By utilizing 
colour-magnitude diagrams of the stars observed in the optical/near-infrared 
(NIR) bands, it is possible to determine the underlying properties of the 
clusters such as age, metallicity and distance. Colour magnitude diagrams of 
star clusters can be used as a good distance indicators. The distance to a 
star can be evaluated by trigonometric or photometric parallaxes. 
Trigonometric parallaxes are only available for nearby stars where 
{\em Hipparcos} \citep{ESA97} is the main supplier for the data. For 
stars at large distances, the use of photometric parallaxes is unavoidable. 
In other words the study of the Galactic structure is strictly tied to 
precise determination of absolute magnitudes. 

Different methods can be used for absolute magnitude determination where 
most of them are devoted to dwarfs. The method used in the Str\"omgren's 
$uvby$-$\beta$ \citep{NS91} and in the UBV \citep{Laird88} photometry 
depends on the absolute magnitude offset from a standard main-sequence. 
In recent years the derivation of absolute magnitudes has been carried 
out by means of colour-absolute magnitude diagrams of some specific 
clusters whose metal abundances are generally adopted as the mean metal 
abundance of a Galactic population, such as thin, thick discs and halo. 
The studies of \cite{Phleps00} and \cite{Chen01} can be given as examples. 
A slightly different approach is that of \cite{Siegel02} where two relations, 
one for stars with solar-like abundances and another one for metal-poor 
stars were derived between $M_{R}$ and the colour index $R-I$, where 
$M_{R}$ is the absolute magnitude in the $R$ filter of Johnson system. 
For a star of given metallicity and colour, absolute magnitude can be 
estimated by {\em linear} interpolation of {\em two} ridgelines and 
by means of {\em linear} extrapolation beyond the metal-poor ridgeline. 

The most recent procedure  used for absolute magnitude determination consists 
of finding the most likely values of the stellar parameters, given the measured 
atmospheric ones, and the time spent by a star in each region of the H-R 
diagram. In practice, researchers select the subset of isochrones with 
$[M/H]\pm \Delta_{[M/H]}$, where $\Delta_{[M/H]}$ is the estimated error on 
the metallicity, for each set of derived $T_{eff}$, $\log g$ and $[M/H]$. 
Then a Gaussian weight is associated to each point of the selected isochrones, 
which depends on the measured atmospheric parameters and the considered errors. 
This criterion allows the algorithm to select only the points whose values are 
closed by the pipeline. For details of this procedure we cite the works of 
\cite{Breddels10} and \cite{Zwitter10}. This procedure is based on many 
parameters. Hence it provides absolute magnitudes with high accuracy. Also it 
can be applied to both dwarf and giant stars simultaneously. 

In \citet{Karaali03}, we presented a procedure for the photometric parallax 
estimation of dwarf stars which depends on the absolute magnitude offset 
from the main-sequence of the Hyades cluster. \cite{Bilir08} obtained the 
absolute magnitude calibrations of the thin disc main-sequence stars in the 
optical $M_V$ and in the near-infrared $M_{J}$ bands using the recent reduced 
{\em Hipparcos} astrometric data \citep{Leeuwen07}. \cite{Bilir09} derived a 
new luminosity colour relation based on trigonometric parallaxes for the 
thin disc main-sequence stars with Sloan Digital Sky Survey (SDSS) photometry. 
\citet{Yaz10} obtained transformation between optical and near-infrared (NIR) 
bands for red giants. \citet{Bilir12} extended this study to middle-infrared 
(MIR) bands by using Radial Velocity Experiment (RAVE) Third Data Release 
(DR3) data \citep{Siebert11}. Both works provide absolute magnitudes for 
a given photometry from another one.

\begin{table}
\setlength{\tabcolsep}{2pt}
\center
{\small
\caption{Data for five clusters. We used the data in the first line for 
each cluster for absolute magnitude calibration, whereas the ones in the 
second and third lines are for comparison purpose. $l$ and $b$ are the Galactic 
longitude and latitude of the clusters, the symbol $\mu_{0}$ indicates the 
true distance modulus of the cluster.}
    \begin{tabular}{ccccccc}
    \hline
    Cluster & $ l$ & $b$ & $E(B-V)$ & $\mu_{0}$ & $[Fe/H]$ & Ref.\\
            &  $^{(o)}$  & $^{(o)}$ & (mag)    & (mag)     & (dex)& \\
    \hline
    M92     &  68.34 & +34.86  & 0.025 & 14.72 & -2.15 & (1) \\
            &        &         & 0.020 & 14.59 & -2.31 & (2) \\
            &        &         & 0.023 & 14.55 & -2.40 & (3) \\
\hline
    M13     &  59.01 & +40.91  & 0.020 & 14.38 & -1.41 & (1) \\
            &        &         & 0.020 & 14.27 & -1.53 & (2) \\
            &        &         & 0.016 & 14.35 & -1.60 & (3) \\
\hline
    M71     &  56.75 &$-$ 4.56 & 0.280 & 12.83 & -0.78 & (4) \\
            &        &         & 0.250 & 13.03 & -0.78 & (2) \\
            &        &         & 0.220 & 13.10 & -0.80 & (3) \\
\hline
    M67     & 215.70 & +31.90  & 0.038 &  9.53 & -0.04 & (1) \\
            &        &         & 0.041 &  9.59 & -0.009& (5) \\
            &        &         & 0.050 &  9.43 & -0.09 & (6) \\
\hline
 NGC 6791   &  69.66 & +10.90  & 0.150 & 13.10 &  0.37 & (3) \\
            &        &         & 0.150 & 13.14 &  0.45 & (7) \\
            &        &         & 0.100 & 12.94 &  0.37 & (8) \\
 \hline
    \end{tabular}\\
(1) \cite{Gratton97}, (2) \cite{Harris10}, (3) \cite{Brasseur10}, 
(4) \cite{Hodder92}, (5) \cite{Sarajedini09}, (6) \cite{Hog98}, 
(7) \cite{Anthony-Twarog07}, (8) \cite{Sandage03}.
}
\end{table}

\begin{table*}
\centering
\small{
  \caption{ Comparison of the selective and total absorptions evaluated by using different extinction and colour excess ratios. The columns give: (1) the cluster, (2) adopted $E(B-V)$ colour excess, (3) $E(V-J)_p$ the colour excess evaluated by the equation $E(V-J)/E(B-V)=2.25$ and used in the paper, (4)$E(V-J)_c$ the colour excess evaluated by the equation $E(V-J)/E(B-V)=2.30$ for comparison purpose, (5) $\Delta E(V-J)$ the difference between the colour excesses in columns (3) and (4), (6) $(A_J)_p$  the total absorption evaluated by the equation $A_J/E(B-V)=0.87$ and used in the paper, (7) $(A_J)_c$ the total absorption evaluated by the equation $A_J/E(B-V)=1.70 $, and (8) $\Delta A_J$ the difference between the total absorptions in columns (6) and (7).}
\begin{tabular}{cccccccc}
\hline
       (1) &          (2) &          (3) &          (4) &          (5) &          (6) &          (7) & (8) \\
\hline
   Cluster &   $E(B-V)$ & $E(V-J)_p$ &  $E(V-J)_c$&$\Delta E(V-J)$ &$(A_J)_p$ &  $(A_J)_c$ &   $\Delta A_J$ \\
\hline
       M92 &      0.025 &      0.056 &      0.058 &      0.002 &      0.022 &      0.043 &      0.021 \\
       M13 &      0.020 &      0.045 &      0.046 &      0.001 &      0.017 &      0.034 &      0.017 \\
       M71 &      0.280 &      0.630 &      0.644 &      0.014 &      0.244 &      0.476 &      0.232 \\
       M67 &      0.038 &      0.086 &      0.087 &      0.001 &      0.033 &      0.065 &      0.032 \\
  NGC 6791 &      0.150 &      0.338 &      0.345 &      0.007 &      0.131 &      0.255 &      0.125 \\
\hline
\end{tabular}  
}
\end{table*}

In \citet[][Paper I; Paper II]{Karaali12a, Karaali12b}, we used a procedure 
for the absolute magnitude estimation of red giants by using the 
$V_{0}\times(B-V)_{0}$ and $g_{0}\times (g-r)_{0}$ apparent magnitude-colour 
diagrams of Galactic clusters with different metallicities. Here, we extend 
our procedure to Two Micron All Sky Survey \citep[2MASS;][]{Skrutskie06} 
photometry. We aim to estimate $M_{J}$ and $M_{K_s}$ absolute magnitudes for 
red giants with $J_{0} \times (V-J)_{0}$ and $K_{{s}_0} \times (V-K_{s})_{0}$ 
colour-magnitude diagrams. The outline of the paper is as follows. We present 
the data in Section 2. The procedure used for calibration is given in 
Section 3, and Section 4 is devoted to summary and discussion.

\section{Data}
We calibrated two different absolute magnitudes, $M_{J}$ and $M_{K_s}$, in 
terms of metallicity. Hence we used two different sets of data. The calibration 
of $M_{J}$ with $J_{0}$ and $(V-J)_{0}$ is given in Section 2.1, whereas the 
one for $M_{K_s}$ with $K_{{s}_0}$ and $(V-K_{s})_{0}$ is presented in Section 2.2.  

\subsection{Data for Calibration with $J_{0}$ and $(V-J)_{0}$}
Five clusters with different metallicities, i.e. M92, M13, M71, M67, and NGC 6791, 
were selected for our program (Table 1). The V magnitudes and $V-J$ colours for the clusters 
M92, M13, and M71 were taken from the tables in \cite{Brasseur10}, whereas the data 
for the clusters NGC 6791 and M67 could be provided by different procedures as 
explained in the following. We used the Fig. 1 in \citet{Brasseur10} and obtained 
a set of 30 $(M_{V}, (V-J)_{0})$ couples for the red giant branch (RGB) of the 
cluster NGC 6791. Then, we transformed the $M_V$ absolute magnitudes to $V$ 
apparent magnitudes by means of the apparent distance modulus of the cluster, 
i.e. $\mu=13.25$ mag. Finally, we de-reddened the $V$ magnitudes and combined 
them with the true colour indices $(V-J)_{0}$ and obtained the $J_{0}$ 
magnitudes. The $J_{0}$ magnitudes and $(V-J)_{0}$ colours are not available for 
the cluster M67 in the literature. Hence, we transformed the $V$, $B-V$ and $V-I$ 
data in \cite{Montgomery93} to obtain a set of ($J_{0}, (V-J)_{0}$) data for the 
RGB of M67. It turned out that 21 of the  stars in the bright stars catalogue 
in \cite{Montgomery93} were red giants. We de-reddened the $V$, $B-V$, and $V-I$ 
data of these stars and transformed them to $(V-J)_{0}$ colours by using the 
following equation of \cite{Yaz10}:

\begin{eqnarray}
(V-J)_{0}=1.080(B-V)_{0}+0.379(V-I)_{0}\nonumber \\
-0.082[Fe/H]+0.279
\end{eqnarray}                          
Then, we combined them with the $V_{0}$ magnitudes and obtained the $J_{0}$ ones. 

\begin{table*}
  \center
\setlength{\tabcolsep}{1.2pt}
\centering
  \caption{$J_0$ magnitudes and $(V-J)_0$ colours for five clusters used for 
the absolute magnitude calibration. The last two columns for the cluster M67 
indicate the $(V-I)_0$ colours and $J_0$ magnitudes of the bins used to draw 
the diagram of M67 in Fig. 1.}
\tiny{
    \begin{tabular}{ccccc|ccccc|cccccccccc}
    \hline
 $V$&$V-J$&$J$&$(V-J)_{0}$& \multicolumn{1}{c}{$J_{0}$}&$V$&$V-J$&$J$&$(V-J)_{0}$& \multicolumn{1}{c}{$J_{0}$}&$V$&$B-V$&$V-I$&$V_{0}$ & $(B-V)_{0}$ & $(V-I)_{0}$ &$(V-J)_{0}$&$J_{0}$ & $(V-J)_{0}$&$J_{0}$\\
    \hline
    \multicolumn{5}{c|}{M92}              & \multicolumn{5}{c|}{M13 (cont.)}      & \multicolumn{10}{c}{M67}                                     \\
   \hline
12.103 & 2.332 & 9.771 & 2.276 & 9.749 & 14.88 & 1.620 & 13.260 & 1.575 & 13.243 & 8.740 & 1.670 & 2.060 & 8.622 & 1.632 & 2.013 & 2.808 & 5.815 & 2.808 & 5.815\\
    12.235 & 2.262 & 9.973 & 2.206 & 9.951 & 15.036 & 1.601 & 13.435 & 1.556 & 13.418 & 8.860 & 1.590 & 1.670 & 8.742 & 1.552 & 1.623 & 2.573 & 6.169 & 2.573 & 6.169 \\
    12.362 & 2.208 & 10.154 & 2.152 & 10.132 & 15.214 & 1.569 & 13.645 & 1.524 & 13.628 & 9.370 & 1.480 & 1.500 & 9.252 & 1.442 & 1.453 & 2.390 & 6.862 & 2.390 & 6.862 \\
    12.503 & 2.147 & 10.356 & 2.091 & 10.334 & 15.408 & 1.546 & 13.862 & 1.501 & 13.845 & 9.530 & 1.380 & 1.330 & 9.412 & 1.342 & 1.283 & 2.218 & 7.194 & 2.204 & 7.373 \\
    12.655 & 2.086 & 10.569 & 2.030 & 10.547 & 15.585 & 1.527 & 14.058 & 1.482 & 14.041 & 9.690 & 1.360 & 1.330 & 9.572 & 1.322 & 1.283 & 2.196 & 7.376 & 2.049 & 8.147 \\
    12.798 & 2.046 & 10.752 & 1.990 & 10.730 & 15.747 & 1.511 & 14.236 & 1.466 & 14.219 & 9.720 & 1.370 & 1.360 & 9.602 & 1.332 & 1.313 & 2.218 & 7.384 & 1.897 & 8.915 \\
    12.933 & 1.999 & 10.934 & 1.943 & 10.912 & 16.104 & 1.478 & 14.626 & 1.433 & 14.609 & 9.840 & 1.360 & 1.300 & 9.722 & 1.322 & 1.253 & 2.185 & 7.537 & 1.794 & 9.382 \\
    13.085 & 1.964 & 11.121 & 1.908 & 11.099 & 16.433 & 1.451 & 14.982 & 1.406 & 14.965 & 10.120 & 1.300 & 1.270 & 10.002 & 1.262 & 1.223 & 2.109 & 7.894 & 1.690 & 10.337 \\
    13.236 & 1.920 & 11.316 & 1.864 & 11.294 & 16.786 & 1.424 & 15.362 & 1.379 & 15.345 & 10.300 & 1.260 & 1.230 & 10.182 & 1.222 & 1.183 & 2.050 & 8.132 & 1.585 & 11.158 \\
    13.379 & 1.889 & 11.490 & 1.833 & 11.468 & 16.952 & 1.411 & 15.541 & 1.366 & 15.524 & 10.520 & 1.230 & 1.150 & 10.402 & 1.192 & 1.103 & 1.987 & 8.415 & $-$    & $-$ \\
    13.528 & 1.857 & 11.671 & 1.801 & 11.649 & 17.112 & 1.397 & 15.715 & 1.352 & 15.698 & 10.930 & 1.150 & 1.140 & 10.812 & 1.112 & 1.093 & 1.897 & 8.915 & $-$    & $-$ \\
    13.682 & 1.818 & 11.864 & 1.762 & 11.842 & 17.271 & 1.381 & 15.890 & 1.336 & 15.873 & 11.200 & 1.080 & 1.080 & 11.082 & 1.042 & 1.033 & 1.799 & 9.283 & $-$    & $-$ \\
    13.827 & 1.783 & 12.044 & 1.727 & 12.022 & 17.420 & 1.360 & 16.060 & 1.315 & 16.043 & 11.240 & 1.100 & 1.070 & 11.122 & 1.062 & 1.023 & 1.817 & 9.305 & $-$    & $-$ \\
\cline{6-10}    13.984 & 1.761 & 12.223 & 1.705 & 12.201 & \multicolumn{5}{c|}{M71}              & 11.440 & 1.060 & 1.050 & 11.322 & 1.022 & 1.003 & 1.766 & 9.556 & $-$    & $-$\\
\cline{6-10}    14.140 & 1.727 & 12.413 & 1.671 & 12.391 & 11.997 & 4.109 & 7.888 & 3.479 & 7.644 & 12.094 & 1.007 & 1.016 & 11.976 & 0.969 & 0.969 & 1.696 & 10.280 & $-$    & $-$\\
    14.321 & 1.692 & 12.629 & 1.636 & 12.607 & 12.035 & 3.717 & 8.318 & 3.087 & 8.074 & 12.110 & 1.006 & 1.029 & 11.992 & 0.968 & 0.982 & 1.700 & 10.292 & $-$    & $-$ \\
    14.541 & 1.653 & 12.888 & 1.597 & 12.866 & 12.116 & 3.469 & 8.647 & 2.839 & 8.403 & 12.230 & 0.993 & 1.002 & 12.112 & 0.955 & 0.955 & 1.675 & 10.437 & $-$    & $-$ \\
    14.744 & 1.627 & 13.117 & 1.571 & 13.095 & 12.212 & 3.300 & 8.912 & 2.670 & 8.668 & 12.712 & 0.913 & 0.966 & 12.594 & 0.875 & 0.919 & 1.575 & 11.019 & $-$    & $-$ \\
    14.925 & 1.595 & 13.330 & 1.539 & 13.308 & 12.347 & 3.137 & 9.210 & 2.507 & 8.966 & 12.862 & 0.941 & 0.981 & 12.744 & 0.903 & 0.934 & 1.611 & 11.133 & $-$    & $-$ \\
    15.089 & 1.565 & 13.524 & 1.509 & 13.502 & 12.523 & 2.962 & 9.561 & 2.332 & 9.317 & 12.934 & 0.917 & 0.972 & 12.816 & 0.879 & 0.925 & 1.582 & 11.234 & $-$    & $-$ \\
    15.256 & 1.547 & 13.709 & 1.491 & 13.687 & 12.693 & 2.833 & 9.860 & 2.203 & 9.616 & 12.934 & 0.919 & 0.940 & 12.816 & 0.881 & 0.893 & 1.572 & 11.244 & $-$    & $-$ \\
\cline{11-20}    15.456 & 1.525 & 13.931 & 1.469 & 13.909 & 12.906 & 2.733 & 10.173 & 2.103 & 9.929 & $(V-J)_{0}$ & $M_{V}$ & $V$ & $V_{0}$ & $(V-J)_{0}$ & $J_{0}$ & $-$    & $-$    & $-$    & $-$ \\
\cline{11-16}    15.637 & 1.507 & 14.130 & 1.451 & 14.108 & 13.139 & 2.648 & 10.491 & 2.018 & 10.247 & \multicolumn{6}{c}{NGC 6791}                  & $-$    & $-$    & $-$    & $-$ \\
\cline{11-16}    15.929 & 1.479 & 14.450 & 1.423 & 14.428 & 13.305 & 2.562 & 10.743 & 1.932 & 10.499 & 2.994 & -0.175 & 13.395 & 12.930 & 2.994 & 9.936 & $-$    & $-$    & $-$    & $-$ \\
    16.118 & 1.463 & 14.655 & 1.407 & 14.633 & 13.459 & 2.516 & 10.943 & 1.886 & 10.699 & 2.935 & -0.140 & 13.430 & 12.965 & 2.935 & 10.030 & $-$    & $-$    & $-$    & $-$ \\
    16.292 & 1.448 & 14.844 & 1.392 & 14.822 & 13.646 & 2.465 & 11.181 & 1.835 & 10.937 & 2.832 & -0.078 & 13.492 & 13.027 & 2.832 & 10.195 & $-$    & $-$    & $-$    & $-$ \\
    16.453 & 1.435 & 15.018 & 1.379 & 14.996 & 13.818 & 2.422 & 11.396 & 1.792 & 11.152 & 2.742 & 0.007 & 13.577 & 13.112 & 2.742 & 10.370 & $-$    & $-$    & $-$    & $-$ \\
    16.623 & 1.421 & 15.202 & 1.365 & 15.180 & 13.993 & 2.381 & 11.612 & 1.751 & 11.368 & 2.671 & 0.085 & 13.655 & 13.190 & 2.671 & 10.519 & $-$    & $-$    & $-$    & $-$ \\
    16.801 & 1.406 & 15.395 & 1.350 & 15.373 & 14.177 & 2.342 & 11.835 & 1.712 & 11.591 & 2.592 & 0.197 & 13.767 & 13.302 & 2.592 & 10.710 & $-$    & $-$    & $-$    & $-$ \\
    16.970 & 1.391 & 15.579 & 1.335 & 15.557 & 14.356 & 2.306 & 12.050 & 1.676 & 11.806 & 2.514 & 0.295 & 13.865 & 13.400 & 2.514 & 10.886 & $-$    & $-$    & $-$    & $-$ \\
    17.130 & 1.376 & 15.754 & 1.320 & 15.732 & 14.538 & 2.274 & 12.264 & 1.644 & 12.020 & 2.443 & 0.422 & 13.992 & 13.527 & 2.443 & 11.084 & $-$    & $-$    & $-$    & $-$ \\
    17.279 & 1.368 & 15.911 & 1.312 & 15.889 & 14.736 & 2.247 & 12.489 & 1.617 & 12.245 & 2.376 & 0.541 & 14.111 & 13.646 & 2.376 & 11.270 & $-$    & $-$    & $-$    & $-$ \\
\cline{1-5}    \multicolumn{5}{c|}{M13}              & 14.986 & 2.209 & 12.777 & 1.579 & 12.533 & 2.328 & 0.647 & 14.217 & 13.752 & 2.328 & 11.424 & $-$    & $-$    & $-$    & $-$ \\
\cline{1-5}    11.895 & 2.729 & 9.166 & 2.684 & 9.1486 & 15.217 & 2.177 & 13.040 & 1.547 & 12.796 & 2.280 & 0.766 & 14.336 & 13.871 & 2.280 & 11.591 & $-$    & $-$    & $-$    & $-$ \\
    12.005 & 2.612 & 9.393 & 2.567 & 9.3756 & 15.426 & 2.152 & 13.274 & 1.522 & 13.030 & 2.221 & 0.899 & 14.469 & 14.004 & 2.221 & 11.783 & $-$    & $-$    & $-$    & $-$ \\
    12.152 & 2.474 & 9.678 & 2.429 & 9.6606 & 15.627 & 2.120 & 13.507 & 1.490 & 13.263 & 2.167 & 1.054 & 14.624 & 14.159 & 2.167 & 11.992 & $-$    & $-$    & $-$    & $-$ \\
    12.289 & 2.373 & 9.916 & 2.328 & 9.8986 & 15.854 & 2.080 & 13.774 & 1.450 & 13.530 & 2.124 & 1.180 & 14.750 & 14.285 & 2.124 & 12.161 & $-$    & $-$    & $-$    & $-$ \\
    12.487 & 2.258 & 10.229 & 2.213 & 10.212 & 16.060 & 2.070 & 13.990 & 1.440 & 13.746 & 2.088 & 1.307 & 14.877 & 14.412 & 2.088 & 12.324 & $-$    & $-$    & $-$    & $-$ \\
    12.642 & 2.182 & 10.460 & 2.137 & 10.443 & 16.268 & 2.035 & 14.233 & 1.405 & 13.989 & 2.040 & 1.461 & 15.031 & 14.566 & 2.040 & 12.526 & $-$    & $-$    & $-$    & $-$ \\
    12.787 & 2.104 & 10.683 & 2.059 & 10.666 & 16.485 & 2.021 & 14.464 & 1.391 & 14.220 & 2.010 & 1.616 & 15.186 & 14.721 & 2.010 & 12.711 & $-$    & $-$    & $-$    & $-$ \\
    12.923 & 2.053 & 10.870 & 2.008 & 10.853 & 16.688 & 1.989 & 14.699 & 1.359 & 14.455 & 1.986 & 1.735 & 15.305 & 14.840 & 1.986 & 12.854 & $-$    & $-$    & $-$    & $-$ \\
    13.051 & 1.999 & 11.052 & 1.954 & 11.035 & 16.885 & 1.977 & 14.908 & 1.347 & 14.664 & 1.963 & 1.862 & 15.432 & 14.967 & 1.963 & 13.004 & $-$    & $-$    & $-$    & $-$ \\
    13.211 & 1.944 & 11.267 & 1.899 & 11.25 & 17.084 & 1.959 & 15.125 & 1.329 & 14.881 & 1.933 & 2.002 & 15.572 & 15.107 & 1.933 & 13.174 & $-$    & $-$    & $-$    & $-$ \\
    13.36 & 1.907 & 11.453 & 1.862 & 11.436 & 17.267 & 1.931 & 15.336 & 1.301 & 15.092 & 1.909 & 2.150 & 15.720 & 15.255 & 1.909 & 13.346 & $-$    & $-$    & $-$    & $-$ \\
    13.499 & 1.875 & 11.624 & 1.830 & 11.607 & $-$    & $-$    & $-$    & $-$    & $-$    & 1.878 & 2.297 & 15.867 & 15.402 & 1.878 & 13.524 & $-$    & $-$    & $-$    & $-$ \\
    13.646 & 1.843 & 11.803 & 1.798 & 11.786 & $-$    & $-$    & $-$    & $-$    & $-$    & 1.842 & 2.445 & 16.015 & 15.550 & 1.842 & 13.708 & $-$    & $-$    & $-$    & $-$ \\
    13.800 & 1.802 & 11.998 & 1.757 & 11.981 & $-$    & $-$    & $-$    & $-$    & $-$    & 1.830 & 2.606 & 16.176 & 15.711 & 1.830 & 13.881 & $-$    & $-$    & $-$    & $-$ \\
    13.945 & 1.774 & 12.171 & 1.729 & 12.154 & $-$    & $-$    & $-$    & $-$    & $-$    & 1.800 & 2.782 & 16.352 & 15.887 & 1.800 & 14.087 & $-$    & $-$    & $-$    & $-$ \\
    14.088 & 1.748 & 12.340 & 1.703 & 12.323 & $-$    & $-$    & $-$    & $-$    & $-$    & 1.782 & 2.993 & 16.563 & 16.098 & 1.782 & 14.316 & $-$    & $-$    & $-$    & $-$ \\
    14.238 & 1.722 & 12.516 & 1.677 & 12.499 & $-$    & $-$    & $-$    & $-$    & $-$    & 1.764 & 3.161 & 16.731 & 16.266 & 1.764 & 14.502 & $-$    & $-$    & $-$    & $-$ \\
    14.393 & 1.697 & 12.696 & 1.652 & 12.679 & $-$    & $-$    & $-$    & $-$    & $-$    & 1.746 & 3.380 & 16.950 & 16.485 & 1.746 & 14.739 & $-$    & $-$    & $-$    & $-$ \\
    14.546 & 1.667 & 12.879 & 1.622 & 12.862 & $-$    & $-$    & $-$    & $-$    & $-$    & 1.728 & 3.520 & 17.090 & 16.625 & 1.728 & 14.897 & $-$    & $-$    & $-$    & $-$ \\
    14.715 & 1.643 & 13.072 & 1.598 & 13.055 & $-$    & $-$    & $-$    & $-$    & $-$    & 1.722 & 3.682 & 17.252 & 16.787 & 1.722 & 15.065 & $-$    & $-$    & $-$    & $-$ \\
    \hline

    \hline
    \end{tabular}
  \label{tab:addlabel}
}
\end{table*}

\begin{figure}
\begin{center}
\includegraphics[scale=0.35, angle=0]{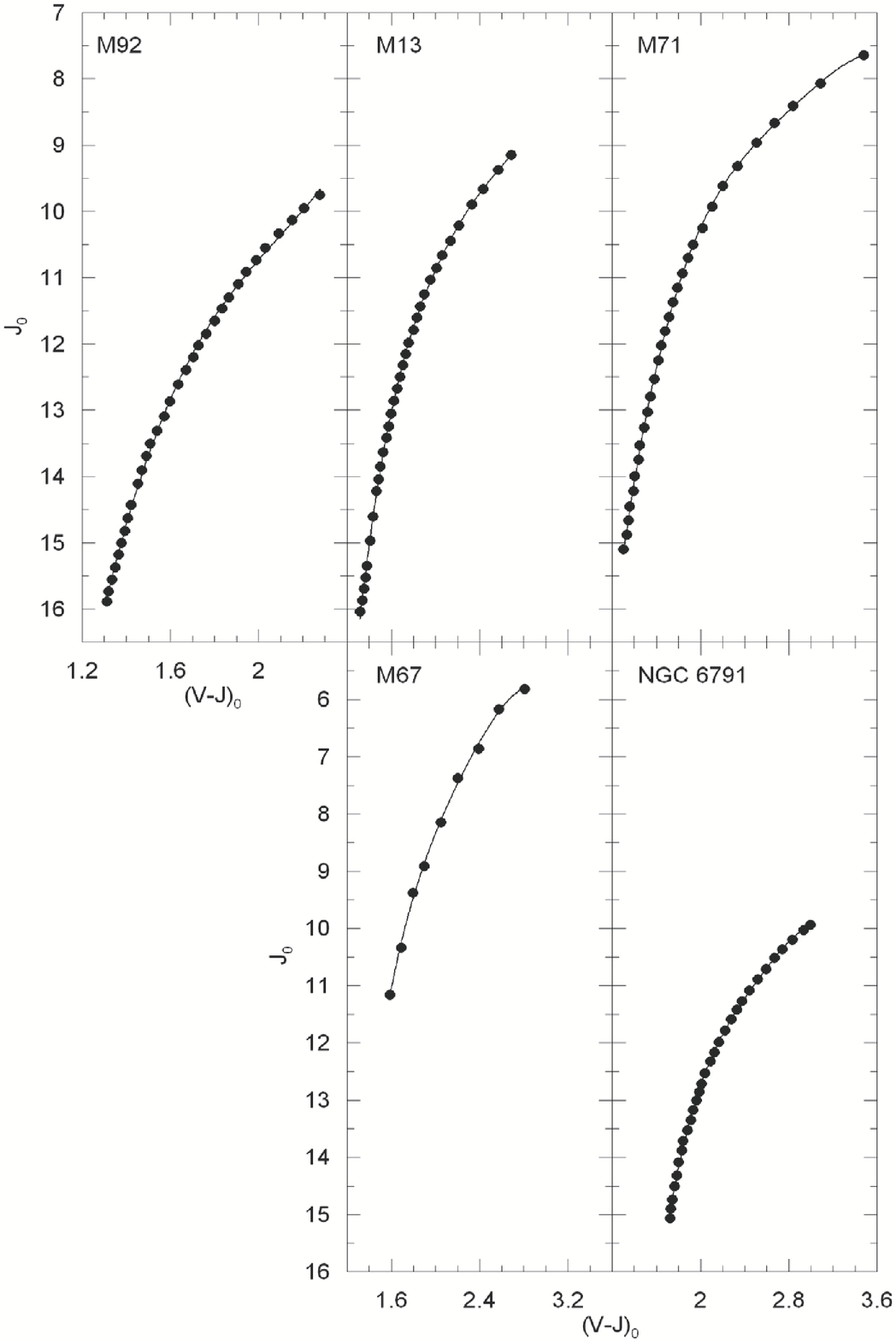} 
\caption[] {$J_{0}\times (V-J)_{0}$ colour-apparent magnitude diagrams for five 
Galactic clusters used for the absolute magnitude calibration.} 
\label{his:col}
\end{center}
\end {figure}

\begin{table}[h]
\setlength{\tabcolsep}{2.5pt}
  \center
\scriptsize{
  \caption{Numerical values of the coefficients $a_{i}$ ($i=1$, 2, 3, 4) in Eq. (2).}
    \begin{tabular}{cccccc}
    \hline
    Cluster $\rightarrow$& M92   & M13   & M71   & M67   & NGC 6791 \\
    \hline
    $(V-J)_{0}$ & $[1.30, 2.28]$ & $[1.30, 2.68]$ & $[1.30, 3.48]$ & $[1.58, 2.81]$ & $[1.72, 2.99]$ \\
    \hline
    $a_{4}$ &    8.7460 &   2.1654 &   0.5646 &   3.0424 &  3.4182 \\
    $a_{3}$ &  -67.6790 & -20.4910 &  -6.5667 & -27.8780 & -34.682 \\
    $a_{2}$ &  198.3500 &  73.7370 &  28.9320 &  97.1910 & 132.690 \\
    $a_{1}$ & -265.4200 &-121.9800 & -58.6600 &-156.3900 &-229.450 \\
    $a_{0}$ &  149.6300 &  89.1650 &  55.3480 & 106.7000 & 163.660 \\
    \hline
    \end{tabular}
  \label{tab:addlabel}
}
\end{table}

We adopted $R=A_{V}/E(B-V)=3.1$ to convert the colour excess to the extinction. 
Although different numerical values appeared in the literature for specific 
regions of our Galaxy, a single value is applicable everywhere. Then, we used 
the equations $E(V-I)/E(B-V)=1.25$ and $E(V-J)/E(B-V)=2.25$ of \cite{FM03} and 
\cite{McCall04}, respectively, to evaluate the colour excesses in $V-I$ and $V-J$ 
colours. The intrinsic colours were evaluated by the equations 
$(V-I)_{0}=(V-I)-E(V-I)$ and $(B-V)_{0}=(B-V)-E(B-V)$. 

We adopted different equations, i.e. $R=A_V/E(B-V)=4.0$ \citep{Turner12} 
and $E(V-J)/E(B-V)=2.30$ \citep{Smith87}, and evaluated the corresponding $E(V-J)$ 
selective and $A_J=1.70E(B-V)$ total absorptions for the clusters (we do not give 
the calculations here, only we remind to the reader that $E(V-J)$ is equivalent to 
$A_V-A_J$). The results are given in the Table 2. The differences between the 
$E(V-J)$ colour excesses estimated in our study and the ones in this paragraph are 
rather small. The same case holds for the $A_J$ total absorptions except the ones 
for the clusters M71 and NGC 6791, i.e. $\Delta A_J\sim0.2$ and 0.1, respectively, 
whose $E(B-V)$ colour excesses are a bit larger. We should add that the extinction 
equation of \cite{Turner12} was derived for low Galactic latitudes, i.e. Carina 
region ($b\sim0^{o}$). Whereas, the Galactic latitudes of the clusters used in our 
study are (absolutely) greater than $b=4^{o}.5$ (see Table 1). Hence, the 
extinction ratio and the colour excess ratios used in our study are preferable.

The range of the metallicity of the clusters in iron abundance is 
$-2.15 \leq [Fe/H] \leq+0.37$ dex. The $\mu_{0}$ true distance modulus, $E(B-V)$ 
colour excess, and $[Fe/H]$ iron abundance for M92, M13, M71, and M67 were taken 
from Paper I, whereas the ones for the cluster NGC 6791 are taken from 
\cite{Brasseur10}, except the metallicity which is adopted from Paper I. These 
data are given in Table 1, whereas the $J_{0}$ magnitudes and $(V-J)_{0}$ colours 
are presented in Table 3. We, then fitted the fiducial sequence of the red giants 
to a fourth degree polynomial for all clusters. The calibration of $J_{0}$ is as 
follows:  

\begin{eqnarray}
   J_{0}= \sum_{i=0}^{4}a_{i}(V-J)^{i}_0. 
\end{eqnarray}
The numerical values of the coefficients $a_{i}$ ($i=1$, 2, 3, 4) are given in 
Table 4 and the corresponding diagrams are presented in Fig. 1. The ($V-J)_{0}$ 
-interval in the second line of the table denotes the range of ($V-J)_{0}$ 
available for each cluster. 

\begin{table*}
\setlength{\tabcolsep}{3pt}
  \center
\tiny{
  \caption{$K_{s_{0}}$ magnitudes and $(V-K_s)_0$ colours for five clusters used 
for the absolute magnitude calibration. The last two columns for the cluster M67 
indicate the $(V-K_s)_0$ colours and $K_{{s}_0}$ magnitudes of the bins used to 
draw the diagram of M67 in Fig. 2.}
     \begin{tabular}{ccccc|cccccccccc}
   \hline
   $V$  & $V-K_{s}$  & $K_{s}$ & $(V-K_{s})_{0}$ & $K_{{s}_0}$ & $V$ & $V-K_{s}$  & $K_{s}$ & $(V-K_{s})_{0}$ & $K_{{s}_0}$& &  &  &  &  \\
    \hline
    \multicolumn{5}{c|}{M92}              & \multicolumn{5}{c}{M71 (cont.)}       &       &       &       &       &  \\
    \hline
    12.103 & 3.108 & 8.995  & 3.039 & 8.9855 & 12.347 & 4.164 &  8.183 & 3.391 &  8.077 & $-$   & $-$   & $-$   & $-$   & $-$ \\
    12.235 & 2.998 & 9.237  & 2.929 & 9.2275 & 12.523 & 3.970 &  8.553 & 3.197 &  8.447 & $-$   & $-$   & $-$   & $-$   & $-$ \\
    12.362 & 2.922 & 9.440  & 2.853 & 9.4305 & 12.693 & 3.800 &  8.893 & 3.027 &  8.787 & $-$   & $-$   & $-$   & $-$   & $-$ \\
    12.503 & 2.854 & 9.649  & 2.785 & 9.6395 & 12.906 & 3.619 &  9.287 & 2.846 &  9.181 & $-$   & $-$   & $-$   & $-$   & $-$ \\
    12.655 & 2.787 & 9.868  & 2.718 & 9.8585 & 13.139 & 3.481 &  9.658 & 2.708 &  9.552 & $-$   & $-$   & $-$   & $-$   & $-$ \\
    12.798 & 2.730 & 10.068 & 2.661 & 10.059 & 13.305 & 3.390 &  9.915 & 2.617 &  9.809 & $-$   & $-$   & $-$   & $-$   & $-$ \\
    12.933 & 2.675 & 10.258 & 2.606 & 10.249 & 13.459 & 3.324 & 10.135 & 2.551 & 10.029 & $-$   & $-$   & $-$   & $-$   & $-$ \\
    13.085 & 2.626 & 10.459 & 2.557 & 10.450 & 13.646 & 3.250 & 10.396 & 2.477 & 10.290 & $-$   & $-$   & $-$   & $-$   & $-$ \\
    13.236 & 2.583 & 10.653 & 2.514 & 10.644 & 13.818 & 3.187 & 10.631 & 2.414 & 10.525 & $-$   & $-$   & $-$   & $-$   & $-$ \\
    13.379 & 2.544 & 10.835 & 2.475 & 10.826 & 13.993 & 3.129 & 10.864 & 2.356 & 10.758 & $-$   & $-$   & $-$   & $-$   & $-$ \\
    13.528 & 2.505 & 11.023 & 2.436 & 11.014 & 14.177 & 3.071 & 11.106 & 2.298 & 11.000 & $-$   & $-$   & $-$   & $-$   & $-$ \\
    13.682 & 2.467 & 11.215 & 2.398 & 11.206 & 14.356 & 3.020 & 11.336 & 2.247 & 11.230 & $-$   & $-$   & $-$   & $-$   & $-$ \\
    13.827 & 2.433 & 11.394 & 2.364 & 11.385 & 14.538 & 2.974 & 11.564 & 2.201 & 11.458 & $-$   & $-$   & $-$   & $-$   & $-$ \\
    13.984 & 2.399 & 11.585 & 2.330 & 11.576 & 14.736 & 2.935 & 11.801 & 2.162 & 11.695 & $-$   & $-$   & $-$   & $-$   & $-$ \\
    14.140 & 2.357 & 11.783 & 2.288 & 11.774 & 14.986 & 2.880 & 12.106 & 2.107 & 12.000 & $-$   & $-$   & $-$   & $-$   & $-$ \\
    14.321 & 2.308 & 12.013 & 2.239 & 12.004 & 15.217 & 2.835 & 12.382 & 2.062 & 12.276 & $-$   & $-$   & $-$   & $-$   & $-$ \\
    14.541 & 2.247 & 12.294 & 2.178 & 12.285 & 15.426 & 2.798 & 12.628 & 2.025 & 12.522 & $-$   & $-$   & $-$   & $-$   & $-$ \\
    14.744 & 2.201 & 12.543 & 2.132 & 12.534 & 15.627 & 2.766 & 12.861 & 1.993 & 12.755 & $-$   & $-$   & $-$   & $-$   & $-$ \\
    14.925 & 2.162 & 12.763 & 2.093 & 12.754 & 15.854 & 2.734 & 13.120 & 1.961 & 13.014 & $-$   & $-$   & $-$   & $-$   & $-$ \\
    15.089 & 2.136 & 12.953 & 2.067 & 12.944 & 16.060 & 2.709 & 13.351 & 1.936 & 13.245 & $-$   & $-$   & $-$   & $-$   & $-$ \\
    15.256 & 2.109 & 13.147 & 2.040 & 13.138 & 16.268 & 2.687 & 13.581 & 1.914 & 13.475 & $-$   & $-$   & $-$   & $-$   & $-$ \\
    15.456 & 2.080 & 13.376 & 2.011 & 13.367 & 16.485 & 2.667 & 13.818 & 1.894 & 13.712 & $-$   & $-$   & $-$   & $-$   & $-$ \\
    15.637 & 2.054 & 13.583 & 1.985 & 13.574 & 16.688 & 2.649 & 14.039 & 1.876 & 13.933 & $-$   & $-$   & $-$   & $-$   & $-$ \\
    15.929 & 2.005 & 13.924 & 1.936 & 13.915 & 16.885 & 2.631 & 14.254 & 1.858 & 14.148 & $-$   & $-$   & $-$   & $-$   & $-$ \\
    16.118 & 1.982 & 14.136 & 1.913 & 14.127 & 17.084 & 2.606 & 14.478 & 1.833 & 14.372 & $-$   & $-$   & $-$   & $-$   & $-$ \\
    16.292 & 1.961 & 14.331 & 1.892 & 14.322 & 17.267 & 2.564 & 14.703 & 1.791 & 14.597 & $-$   & $-$   & $-$   & $-$   & $-$ \\
\cline{6-12}16.453 & 1.942 & 14.511 & 1.873 & 14.502 & $J-K_{s}$ & $K_{s}$ & $J$ & $J_{0}$ & $V_{0}$&$K_{{s}_0}$&$(V-K_{s})_{0}$ &    & \\
\cline{6-12}    16.623 & 1.922 & 14.701 & 1.853 & 14.692 & \multicolumn{8}{c}{NGC 6791}                                  & $-$   & $-$ \\
\cline{6-12}    16.801 & 1.901 & 14.900 & 1.832 & 14.891 & 1.111 & 8.693 & 9.804 & 9.717 & 12.767 & 8.636 & 4.131 & $-$  & $-$   & $-$ \\
    16.970 & 1.880 & 15.090 & 1.811 & 15.081 & 1.060 & 9.103 & 10.163 & 10.076 & 12.957 & 9.046 & 3.911 & $-$  & $-$   & $-$ \\
    17.130 & 1.857 & 15.273 & 1.788 & 15.264 & 1.016 & 9.478 & 10.494 & 10.407 & 13.143 & 9.421 & 3.722 & $-$  & $-$   & $-$ \\
    17.279 & 1.832 & 15.447 & 1.763 & 15.438 & 0.978 & 9.887 & 10.865 & 10.778 & 13.363 & 9.830 & 3.533 & $-$  & $-$   & $-$ \\
\cline{1-5}    \multicolumn{5}{c|}{M13}              & 0.930 & 10.297 & 11.227 & 11.140 & 13.591 & 10.240 & 3.351 & $-$  & $-$   & $-$ \\
\cline{1-5}    11.895 & 3.741 & 8.154 & 3.686 & 8.146 & 0.901 & 10.661 & 11.562 & 11.475 & 13.811 & 10.604 & 3.207 & $-$  & $-$   & $-$ \\
    12.005 & 3.547 & 8.458 & 3.492 & 8.450 & 0.870 & 11.060 & 11.930 & 11.843 & 14.066 & 11.003 & 3.063 & $-$  & $-$   & $-$ \\
    12.152 & 3.352 & 8.800 & 3.297 & 8.792 & 0.835 & 11.493 & 12.328 & 12.241 & 14.355 & 11.436 & 2.919 & $-$  & $-$   & $-$ \\
    12.289 & 3.223 & 9.066 & 3.168 & 9.058 & 0.800 & 11.903 & 12.703 & 12.616 & 14.640 & 11.846 & 2.794 & $-$  & $-$   & $-$ \\
    12.487 & 3.058 & 9.429 & 3.003 & 9.421 & 0.784 & 12.312 & 13.096 & 13.009 & 14.953 & 12.255 & 2.698 & $-$  & $-$   & $-$ \\
    12.642 & 2.952 & 9.690 & 2.897 & 9.682 & 0.762 & 12.617 & 13.379 & 13.292 & 15.188 & 12.560 & 2.628 & $-$  & $-$   & $-$ \\
    12.787 & 2.880 & 9.907 & 2.825 & 9.899 & 0.739 & 13.039 & 13.778 & 13.691 & 15.530 & 12.982 & 2.548 & $-$  & $-$   & $-$ \\
    12.923 & 2.818 & 10.105 & 2.763 & 10.097 & 0.721 & 13.390 & 14.111 & 14.024 & 15.827 & 13.333 & 2.494 & $-$  & $-$   & $-$ \\
    13.051 & 2.768 & 10.283 & 2.713 & 10.275 & 0.708 & 13.705 & 14.413 & 14.326 & 16.106 & 13.648 & 2.458 & $-$  & $-$   & $-$ \\
    13.211 & 2.689 & 10.522 & 2.634 & 10.514 & 0.701 & 13.999 & 14.700 & 14.613 & 16.378 & 13.942 & 2.436 & $-$  & $-$   & $-$ \\
    13.360 & 2.628 & 10.732 & 2.573 & 10.724 & 1.168 & 8.225 & 9.393 & 9.306 & $-$   & $-$   & $-$   & $-$   & $-$   & $-$ \\
    13.499 & 2.573 & 10.926 & 2.518 & 10.918 & 1.229 & 7.768 & 8.997 & 8.910 & $-$   & $-$   & $-$   & $-$   & $-$   & $-$ \\
    13.646 & 2.529 & 11.117 & 2.474 & 11.109 & 1.276 & 7.453 & 8.729 & 8.642 & $-$   & $-$   & $-$   & $-$   & $-$   & $-$ \\
    13.800 & 2.475 & 11.325 & 2.420 & 11.317 & 1.330 & 7.124 & 8.454 & 8.367 & $-$   & $-$   & $-$   & $-$   & $-$   & $-$ \\
    13.945 & 2.415 & 11.530 & 2.360 & 11.522 & 1.362 & 6.901 & 8.263 & 8.176 & $-$   & $-$   & $-$   & $-$   & $-$   & $-$ \\
\cline{6-15}14.088 & 2.379 & 11.709 & 2.324 & 11.701 &$V$ & $B-V$ &$V-I$&$V_{0}$&$(B-V)_{0}$&$(V-I)_{0}$&$(V-K_{s})_{0}$&$K_{s}$&$(V-K_{s})_{0}$&$K_{{s}_0}$ \\
\cline{6-15}14.238 & 2.342 & 11.896 & 2.287 & 11.888 & \multicolumn{10}{c}{M67}                                                      \\
\cline{6-15}14.393 & 2.298 & 12.095 & 2.243 & 12.087 & 8.740 & 1.670 & 2.060 & 8.622 & 1.632 & 2.013 & 3.796 & 4.826 & 2.135 & 10.608 \\
    14.546 & 2.249 & 12.297 & 2.194 & 12.289 & 8.860  & 1.590 & 1.670 & 8.742  & 1.552 & 1.623 & 3.538 & 5.204 & 2.292 & 9.735 \\
    14.715 & 2.215 & 12.500 & 2.160 & 12.492 & 9.370  & 1.480 & 1.500 & 9.252  & 1.442 & 1.453 & 3.291 & 5.961 & 2.447 & 8.729 \\
    14.880 & 2.173 & 12.707 & 2.118 & 12.699 & 9.530  & 1.380 & 1.330 & 9.412  & 1.342 & 1.283 & 3.062 & 6.350 & 2.570 & 8.157 \\
    15.036 & 2.145 & 12.891 & 2.090 & 12.883 & 9.690  & 1.360 & 1.330 & 9.572  & 1.322 & 1.283 & 3.026 & 6.546 & 2.740 & 7.662 \\
    15.214 & 2.104 & 13.110 & 2.049 & 13.102 & 9.720  & 1.370 & 1.360 & 9.602  & 1.332 & 1.313 & 3.053 & 6.550 & 2.817 & 7.365 \\
    15.408 & 2.072 & 13.336 & 2.017 & 13.328 & 9.840  & 1.360 & 1.300 & 9.722  & 1.322 & 1.253 & 3.017 & 6.705 & 2.901 & 7.101 \\
    15.585 & 2.050 & 13.535 & 1.995 & 13.527 & 10.120 & 1.300 & 1.270 & 10.002 & 1.262 & 1.223 & 2.901 & 7.101 & 3.039 & 6.538 \\
    15.747 & 2.031 & 13.716 & 1.976 & 13.708 & 10.300 & 1.260 & 1.230 & 10.182 & 1.222 & 1.183 & 2.817 & 7.365 & 3.291 & 5.961 \\
    16.104 & 1.974 & 14.130 & 1.919 & 14.122 & 10.520 & 1.230 & 1.150 & 10.402 & 1.192 & 1.103 & 2.740 & 7.662 & 3.538 & 5.204 \\
    16.433 & 1.945 & 14.488 & 1.890 & 14.480 & 10.930 & 1.150 & 1.140 & 10.812 & 1.112 & 1.093 & 2.594 & 8.218 & 3.796 & 4.826 \\
    16.786 & 1.907 & 14.879 & 1.852 & 14.871 & 11.200 & 1.080 & 1.080 & 11.082 & 1.042 & 1.033 & 2.451 & 8.631 & $-$   & $-$ \\
    16.952 & 1.888 & 15.064 & 1.833 & 15.056 & 11.240 & 1.100 & 1.070 & 11.122 & 1.062 & 1.023 & 2.484 & 8.638 & $-$   & $-$ \\
    17.112 & 1.868 & 15.244 & 1.813 & 15.236 & 11.440 & 1.060 & 1.050 & 11.322 & 1.022 & 1.003 & 2.406 & 8.916 & $-$   & $-$ \\
    17.271 & 1.844 & 15.427 & 1.789 & 15.419 & 12.094 & 1.007 & 1.016 & 11.976 & 0.969 & 0.969 & 2.301 & 9.675 & $-$   & $-$ \\
    17.420 & 1.814 & 15.606 & 1.759 & 15.598 & 12.110 & 1.006 & 1.029 & 11.992 & 0.968 & 0.982 & 2.303 & 9.689 & $-$   & $-$ \\
\cline{1-5}    \multicolumn{5}{c|}{M71}      & 12.230 & 0.993 & 1.002 & 12.112 & 0.955 & 0.955 & 2.272 & 9.840 & $-$   & $-$ \\
\cline{1-5}11.997 & 5.401 & 6.596 & 4.628 & 6.490 & 12.712 & 0.913 & 0.966 & 12.594 & 0.875 & 0.919 & 2.118 & 10.476 & $-$   & $-$\\
    12.035 & 4.883 & 7.152 & 4.110 & 7.046 & 12.862 & 0.941 & 0.981 & 12.744 & 0.903 & 0.934 & 2.173 & 10.571 & $-$   & $-$ \\
    12.116 & 4.552 & 7.564 & 3.779 & 7.458 & 12.934 & 0.917 & 0.972 & 12.816 & 0.879 & 0.925 & 2.127 & 10.689 & $-$   & $-$ \\
    12.212 & 4.350 & 7.862 & 3.577 & 7.756 & 12.934 & 0.919 & 0.940 & 12.816 & 0.881 & 0.893 & 2.121 & 10.695 & $-$   & $-$ \\
    \hline
    \end{tabular}%
}
  \label{tab:addlabel}%
\end{table*}%

\begin{table*}
\centering
\scriptsize{
\caption{Comparison of the selective and total absorptions evaluated by using different extinction and colour excess ratios. The columns give: (1) the cluster, (2) adopted $E(B-V)$ colour excess, (3) $E(V-K_s)_p$ the colour excess evaluated by the equation $E(V-K_s)/E(B-V)=2.72$ used in the paper, (4) $E(V-K_s)_c$ the colour excess evaluated by the equation $E(V-K_s)/E(B-V)=2.74$ for comparison purpose, (5) $\Delta E(V-K_s)$ the difference between the colour excesses in columns (3) and (4), (6) $(A_{K_{s}})_p$ the total absorption evaluated by the equation $A_{K_{s}}/E(B-V)=0.38$ used in the paper, (7)$(A_{K_{s}})_c$ the total absorption evaluated by the equation $A_{K_{s}}/E(B-V)=1.26$, and (8) $\Delta A_{K_{s}}$ the difference between the total absorptions in columns (6) and (7).}
\begin{tabular}{cccccccc}
\hline
       (1) &          (2) &          (3) &          (4) &          (5) &          (6) &          (7) & (8) \\
\hline
   Cluster &   $E(B-V)$ & $E(V-K_s)_p$ &  $E(V-K_s)_c$&$\Delta E(V-K_{s})$ &$(A_{K_{s}})_p$ &  $(A_{K_{s}})_c$ &   $\Delta A_{K_{s}}$ \\
\hline
       M92 &      0.025 &      0.068 &      0.069 &      0.001 &      0.010 &      0.032 &      0.022 \\
       M13 &      0.020 &      0.054 &      0.055 &      0.001 &      0.008 &      0.025 &      0.018 \\
       M71 &      0.280 &      0.762 &      0.767 &      0.005 &      0.106 &      0.353 &      0.246 \\
       M67 &      0.038 &      0.103 &      0.104 &      0.001 &      0.014 &      0.048 &      0.033 \\
   NGC 6791 &     0.150 &      0.408 &      0.411 &      0.003 &      0.057 &      0.189 &      0.132 \\
\hline
\end{tabular}  
}
\end{table*}

\subsection{Data for Calibration with $K_{{s}_{0}}$ and $(V-K_{s})_{0}$}
We used the data of the same clusters mentioned in Section 2.1, i.e. M92, M13, M71, M67, and NGC 6791, 
for calibration of the $M_{K_s}$ absolute magnitudes. We adopted the same colour excesses, distance 
moduli and metallicities in Table 1. The $K_{{s}_0}$ magnitudes and $(V-K_{s})_{0}$ colours for the clusters 
M92, M13, M71 were taken from the tables in \citet{Brasseur10}. Whereas, we used two different 
procedures for evaluation of the $K_{s}$ and $V-K_{s}$ data for the clusters M67 and NGC 6791, 
as explained in the following. For the cluster M67, we transformed the $V$, $B-V$ and $V-I$ data 
of \cite{Montgomery93} to $(V-K_{s})_{0}$ by the following equation in \cite{Yaz10}, and then 
we evaluated the $K_{{s}_0}$ magnitudes by the combination of $V_{0}$ magnitudes and $(V-K_{s})_{0}$ colours.  

\begin{eqnarray}
(V-K_{s})_{0} = 1.791(B-V)_{0}+0.294(V-I)_{0}\nonumber\\
-0.129[Fe/H]+0.279.
\end{eqnarray} 
The available 2MASS photometric data in \cite{Brasseur10} for the cluster NGC 6791 are the $J-K_{s}$ colours and 
$K_{s}$ magnitudes, but not the $V-K_{s}$ ones. Hence, we evaluated them by means of the $V_{0}$ and 
$J_{0}$ magnitudes for this cluster given in Table 3 in three steps. First, we plotted $V_{0}$ magnitudes 
versus $J_{0}$ magnitudes in a diagram (not given here) and obtained the following quadratic equation 
with high correlation coefficient, $R^2=0.9997$. 

\begin{eqnarray}
V_{0}=0.0457J^{2}_{0}-0.3744J_{0}+12.9000.
\end{eqnarray}
In the second step, we evaluated the $J$ magnitudes by combining $J-K_{s}$ 
and $K_{s}$, and finally we de-reddened the $J$ magnitudes and transformed 
them to $V_{0}$ magnitudes by Eq. (4). We used the equations 
$A_{J}/E(B-V)=0.87$ and $A_{K_{s}}/E(B-V)=0.38$ of \cite{SM79} for 
de-reddening of the magnitudes $J$ and $K_{s}$, respectively. As Eq. (4) 
is defined for $9.94 \leq J_{0} \leq 15.06$, we could not considered five 
bright $J_{0}$ magnitudes in Table 3. The $(V-K_{s})_{0}$ and $K_{{s}_0}$ 
data for the clusters are given in Table 5. 

As in Section 2.1, we adopted different equations, i.e. $R=A_V/E(B-V)=4.0$ 
\citep {Turner12}, $E(J-H)/E(B-V)= 0.295$, $E(H-K_s)/E(J-H)=0.49$ 
\citep{Turner11}, and evaluated the corresponding $E(V-K_s)=2.74E(B-V)$ 
selective and $A_{K_{s}}=1.26E(B-V)$ total absorptions for the clusters. The 
results are given in Table 6. The differences in $E(V-K_s)$ and 
$\Delta A_{K_{s}}$ are almost the same as in Table 2. We prefer the extinction 
ratio and the colour excess ratios used in our study due to the reason explained 
in Section 2.1.

We, fitted the $(V-K_{s})_{0}$ colours and ${K_s}_0$ magnitudes to a fourth 
degree polynomial for all clusters, except M67 for which a fifth degree 
polynomial provided higher correlation coefficient. The calibration of 
$K_{{s}_0}$ is as follows:
\begin{eqnarray}
   K_{{s}_0}=\sum_{i=0}^{5}b_{i}(V-K_{s})^{i}_0. 
\end{eqnarray}
The numerical values of the coefficients $b_{i}$ ($i=1,$ 2, 3, 4, 5) are 
given in Table 7 and the corresponding diagrams are presented in Fig. 2. 
The $(V-K_{s})_{0}$ -interval in the second line of the table indicates 
to the range of $(V-K_{s})_{0}$ available for each cluster. 

\begin{figure}
\begin{center}
\includegraphics[scale=0.45, angle=0]{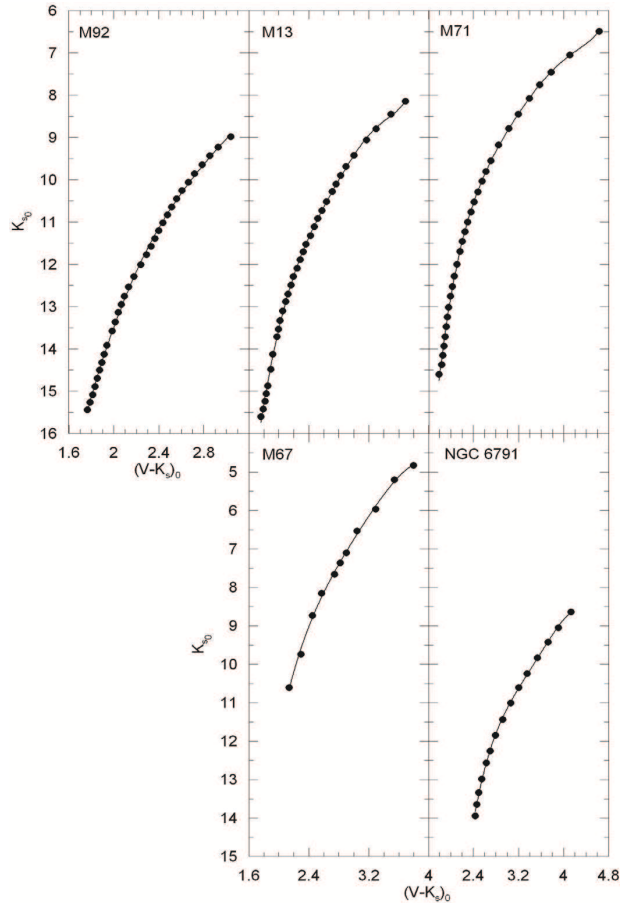} 
\caption[] {$K_{{s}_{0}} \times (V-K_{s})_{0}$ colour-apparent magnitude 
diagrams for five Galactic clusters used for the absolute magnitude 
calibration.} 
\label{his:col}
\end{center}
\end {figure}

\begin{table}
\setlength{\tabcolsep}{1.2pt}
  \center
\scriptsize{  
\caption{Numerical values of the coefficients $b_{i}$ ($i=1,$ 2, 3, 4, 5) in Eq. (5).}
    \begin{tabular}{cccccc}
    \hline
    Cluster $\rightarrow$& M92   & M13   & M71$^{*}$   & M67   & NGC 6791\\
    \hline
$(V-K_{s})_{0}$ $\rightarrow$& $[1.76, 3.04]$ & $[1.76, 3.69]$ & $[1.79, 3.10]$ & $[2.14, 3.80]$ & $[2.44, 4.13]$ \\
    \hline
    $b_{5}$ &  $-$     & $-$      & $-$    & 1.5121 & $-$\\
    $b_{4}$ &   2.2478 &   0.9441 &  0.3976 & -21.290 &   1.2082 \\
    $b_{3}$ & -22.7640 & -11.2930 & -5.6464 & 117.700 & -16.8610 \\
    $b_{2}$ &  87.9930 &  51.2230 & 30.0540 &-317.680 &  88.1870 \\
    $b_{1}$ &-157.9500 &-106.9700 &-72.7860 & 412.050 &-207.2100 \\
    $b_{0}$ & 123.5300 &  97.8320 & 76.9680 &-191.240 & 196.4700 \\
    \hline
    \end{tabular}\\
(*)The $(V-K_{s})_{0}$ domain of the cluster M71 is in fact $[1.79, 4.63]$. However, we restricted it with an upper limit of 3.10 mag due to some uncertainties claimed by the authors.  
  \label{tab:addlabel}
}
\end{table}

\begin{table}
\setlength{\tabcolsep}{3pt}
  \center
\scriptsize{
\caption{$M_J$ absolute magnitudes estimated for a set of $(V-J)_0$ colours for five clusters used in the calibration.}
    \begin{tabular}{cccrcc}
    \hline
    Cluster$\rightarrow$& M92   & M13   & \multicolumn{1}{c}{M71} & M67   & NGC 6791 \\
    \hline
    $(V-J)_{0}$ & \multicolumn{5}{c}{$M_{J}$} \\
    \hline
    1.30  &  1.364 &  1.992 & \multicolumn{1}{c}{2.341}  & $-$    & $-$ \\
    1.35  &  0.620 &  1.275 & \multicolumn{1}{c}{1.774}  & $-$    & $-$ \\
    1.40  & -0.025 &  0.629 & \multicolumn{1}{c}{1.251}  & $-$    & $-$ \\
    1.45  & -0.584 &  0.049 & \multicolumn{1}{c}{0.767}  & $-$    & $-$ \\
    1.50  & -1.073 & -0.472 & \multicolumn{1}{c}{0.321}  & $-$    & $-$ \\
    1.55  & -1.501 & -0.938 & \multicolumn{1}{c}{-0.091} & $-$    & $-$ \\
    1.60  & -1.881 & -1.356 & \multicolumn{1}{c}{-0.469} & 1.505  & $-$ \\
    1.65  & -2.222 & -1.731 & \multicolumn{1}{c}{-0.817} & 1.048  & $-$ \\
    1.70  & -2.532 & -2.068 & \multicolumn{1}{c}{-1.137} & 0.635  & $-$ \\
    1.75  & -2.817 & -2.370 & \multicolumn{1}{c}{-1.431} & 0.261  & 1.571 \\
    1.80  & -3.084 & -2.643 & \multicolumn{1}{c}{-1.700} & -0.080 & 1.083 \\
    1.85  & -3.336 & -2.890 & \multicolumn{1}{c}{-1.948} & -0.391 & 0.655 \\
    1.90  & -3.576 & -3.114 & \multicolumn{1}{c}{-2.175} & -0.678 & 0.278 \\
    1.95  & -3.806 & -3.320 & \multicolumn{1}{c}{-2.383} & -0.944 & -0.053 \\
    2.00  & -4.026 & -3.509 & \multicolumn{1}{c}{-2.574} & -1.192 & -0.345 \\
    2.05  & -4.235 & -3.684 & \multicolumn{1}{c}{-2.750} & -1.425 & -0.604 \\
    2.10  & -4.431 & -3.847 & \multicolumn{1}{c}{-2.912} & -1.646 & -0.835 \\
    2.15  & -4.609 & -4.001 & \multicolumn{1}{c}{-3.061} & -1.857 & -1.042 \\
    2.20  & -4.766 & -4.146 & \multicolumn{1}{c}{-3.199} & -2.058 & -1.231 \\
    2.25  & -4.894 & -4.285 & \multicolumn{1}{c}{-3.327} & -2.253 & -1.404 \\
    2.30  & $-$    & -4.417 & \multicolumn{1}{c}{-3.447} & -2.439 & -1.566 \\
    2.35  & $-$    & -4.545 & \multicolumn{1}{c}{-3.559} & -2.619 & -1.717 \\
    2.40  & $-$    & -4.667 & \multicolumn{1}{c}{-3.664} & -2.792 & -1.862 \\
    2.45  & $-$    & -4.783 & \multicolumn{1}{c}{-3.763} & -2.956 & -2.001 \\
    2.50  & $-$    & -4.895 & \multicolumn{1}{c}{-3.857} & -3.111 & -2.135 \\
    2.55  & $-$    & -5.000 & \multicolumn{1}{c}{-3.947} & -3.255 & -2.266 \\
    2.60  & $-$    & -5.097 & \multicolumn{1}{c}{-4.033} & -3.386 & -2.393 \\
    2.65  & $-$    & -5.186 & \multicolumn{1}{c}{-4.116} & -3.501 & -2.516 \\
    2.70  & $-$    & $-$    & \multicolumn{1}{c}{-4.197} & -3.598 & -2.634 \\
    2.75  & $-$    & $-$    & \multicolumn{1}{c}{-4.275} & -3.672 & -2.745 \\
    2.80  & $-$    & $-$    & \multicolumn{1}{c}{-4.352} & -3.719 & -2.848 \\
    2.85  & $-$    & $-$    & \multicolumn{1}{c}{-4.427} & $-$    & -2.940 \\
    2.90  & $-$    & $-$    & \multicolumn{1}{c}{-4.500} & $-$    & -3.019 \\
    2.95  & $-$    & $-$    & \multicolumn{1}{c}{-4.572} & $-$    & -3.080 \\
    3.00  & $-$    & $-$    & -4.642 			 & $-$    & -3.120 \\
    3.05  & $-$    & $-$    & -4.711 			 & $-$    & $-$ \\
    3.10  & $-$    & $-$    & -4.778 			 & $-$    & $-$ \\
    3.15  & $-$    & $-$    & -4.843 			 & $-$    & $-$ \\
    3.20  & $-$    & $-$    & -4.905 			 & $-$    & $-$ \\
    3.25  & $-$    & $-$    & -4.965 			 & $-$    & $-$ \\
    3.30  & $-$    & $-$    & -5.021 			 & $-$    & $-$ \\
    3.35  & $-$    & $-$    & -5.073  			 & $-$    & $-$ \\
    3.40  & $-$    & $-$    & -5.120 			 & $-$    & $-$ \\
    3.45  & $-$    & $-$    & -5.162 			 & $-$    & $-$ \\
    \hline
    \end{tabular}
  \label{tab:addlabel}
}
\end{table}

\section{The Procedure}
\subsection{Absolute Magnitude as a Function of Metallicity}
We adopted the procedure in Paper II which consists of calibration of an absolute magnitude as a function of metallicity. We calibrated the $M_{J}$ and $M_{K_s}$, absolute magnitudes in terms of metallicity for a given $(V-J)_{0}$ and $ (V-K_{s})_{0}$ colour, respectively.

\begin{table}
\setlength{\tabcolsep}{3pt}
  \center
\scriptsize{
  \caption{$M_J$ absolute magnitudes and $[Fe/H]$ metallicities for seven $(V-J)_0$ -intervals.}
    \begin{tabular}{ccc}
    \hline
    $(V-J)_{0}$ & $[Fe/H]$ (dex)& $M_{J}$ \\
    \hline
    1.30  & -2.15 & 1.364 \\
          & -1.41 & 1.992 \\
          & -0.78 & 2.341 \\
    \hline
    1.60  & -2.15 & -1.881 \\
          & -1.41 & -1.356 \\
          & -0.78 & -0.469 \\
          & -0.04 &  1.505 \\
    \hline
    1.90  & -2.15 & -3.576 \\
          & -1.41 & -3.114 \\
          & -0.78 & -2.175 \\
          & -0.04 & -0.678 \\
          &  0.37 &  0.278 \\
    \hline
    2.10  & -2.15 & -4.431 \\
          & -1.41 & -3.847 \\
          & -0.78 & -2.912 \\
          & -0.04 & -1.646 \\
          &  0.37 & -0.835 \\
    \hline
    2.25  & -2.15 & -4.894 \\
          & -1.41 & -4.285 \\
          & -0.78 & -3.327 \\
          & -0.04 & -2.253 \\
          &  0.37 & -1.404 \\
    \hline
    2.50  & -1.41 & -4.895 \\
          & -0.04 & -3.111 \\
          &  0.37 & -2.135 \\
    \hline
    2.65  & -1.41 & -5.186 \\
          & -0.04 & -3.501 \\
          &  0.37 & -2.516 \\
    \hline
    \end{tabular}
  \label{tab:addlabel}
}
\end{table}

\subsubsection{Calibration of $M_{J}$ in terms of Metallicity}
We estimated the $M_{J}$ absolute magnitudes for the $(V-J)_{0}$ colours given in Table 8 for the clusters M92, M13, M71, M67, and NGC 6791 by combining the $J_{0}$ apparent magnitudes evaluated by using Eq. (2) and the true distance modulus ($\mu_{0}$) of the cluster in question, i.e.

\begin{eqnarray}
M_{J}=J_{0}-\mu_{0}.     
\end{eqnarray}
Then, we plotted the absolute magnitudes versus $(V-J)_{0}$ colours. Fig. 3 shows that the absolute magnitude is metallicity dependent. It increases (algebraically) with increasing metallicity and decreasing colour. 

\begin{figure}
\begin{center}
\includegraphics[scale=0.35, angle=0]{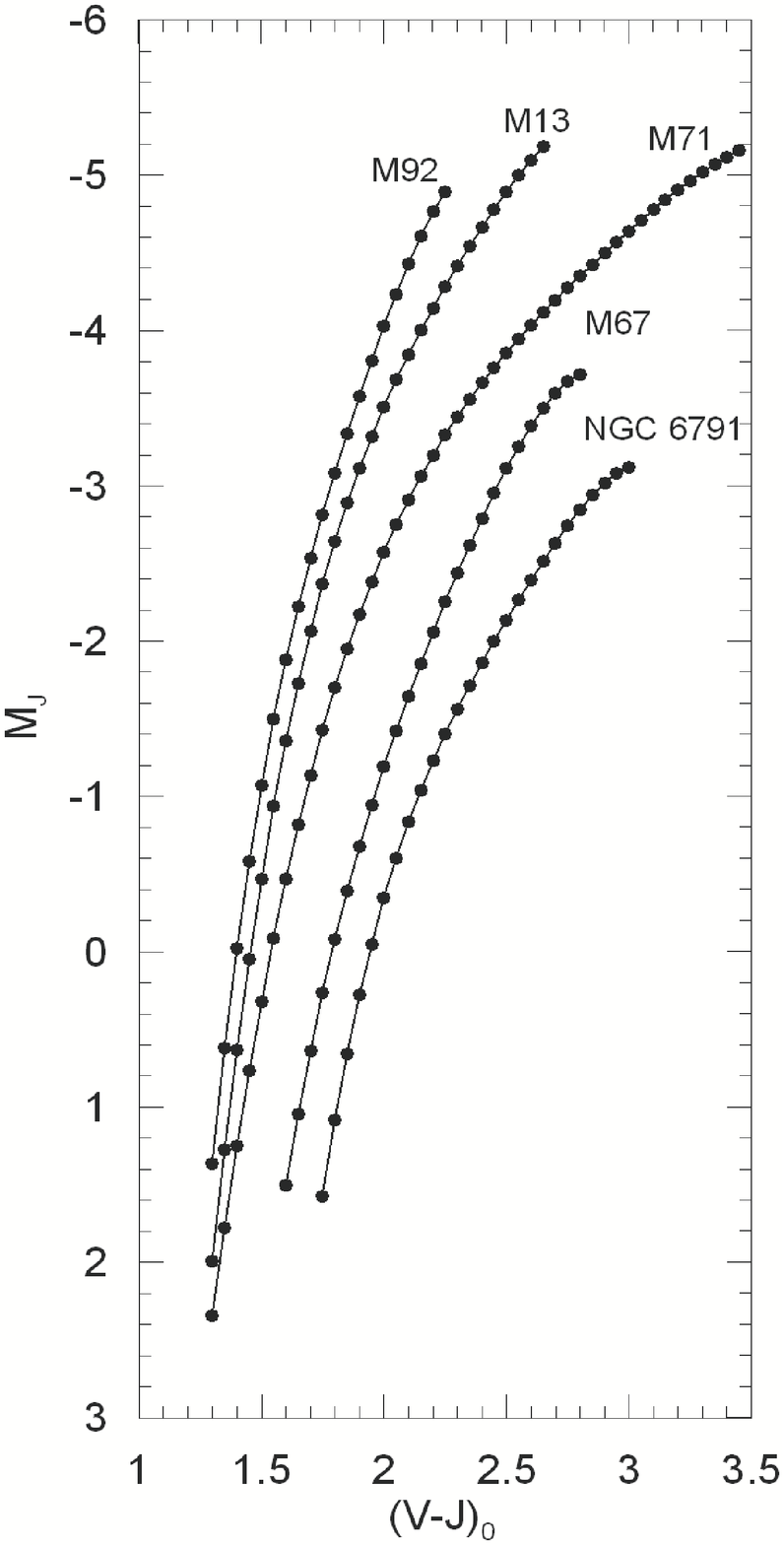} 
\caption[] {$M_{J}\times(V-J)_{0}$ colour-absolute magnitude diagrams for five clusters used for the absolute magnitude calibration.} 
\label{his:col}
\end{center}
\end {figure}

\begin{figure}
\begin{center}
\includegraphics[scale=0.35, angle=0]{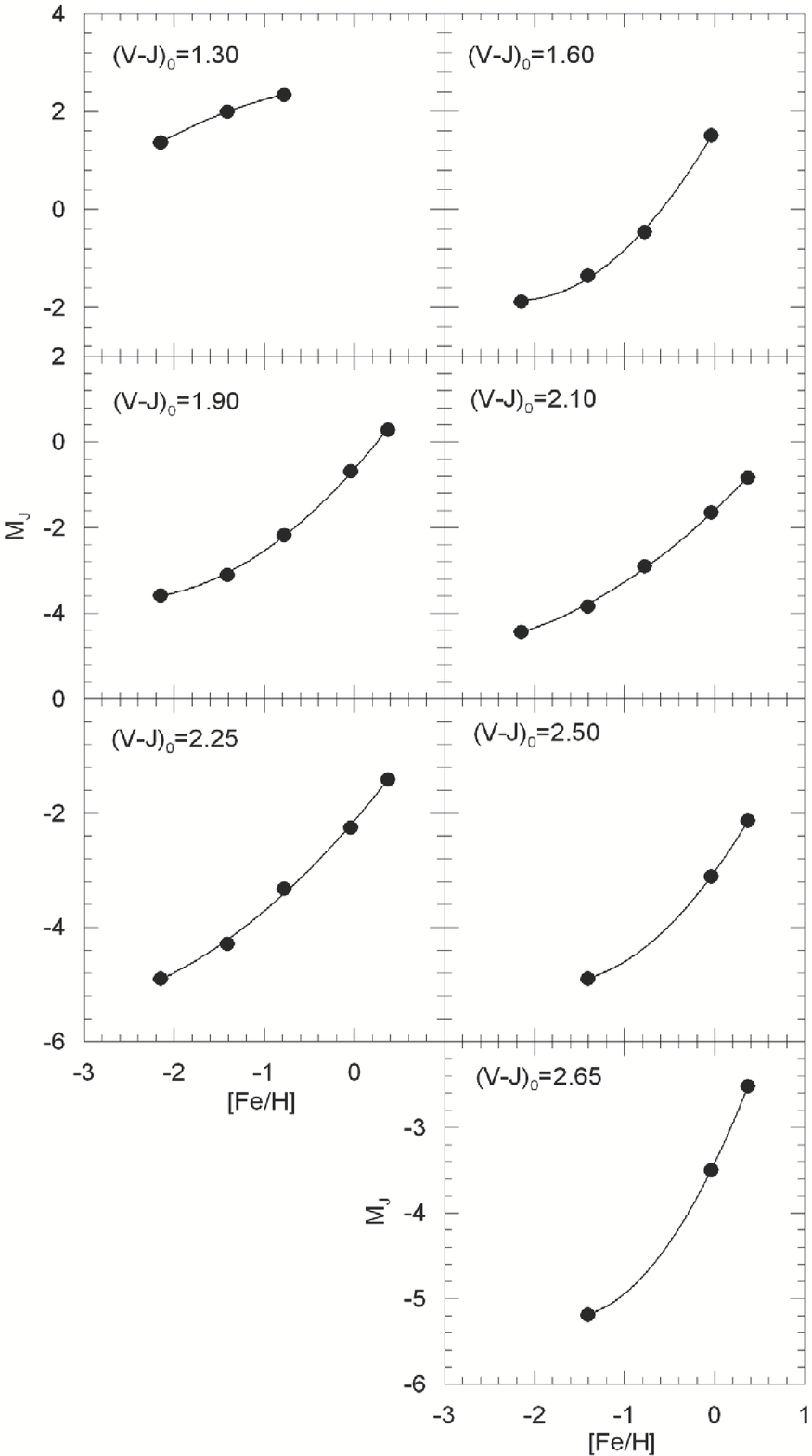} 
\caption[] {Calibration of the absolute magnitude $M_{J}$ as a function of metallicity $[Fe/H]$ for seven colour-indices.} 
\label{his:col}
\end{center}
\end {figure}

\begin{table*}
\setlength{\tabcolsep}{.8pt}
\renewcommand{\baselinestretch}{0.5}  
 \center
\tiny{
\caption{$M_J$ absolute magnitudes estimated for five Galactic clusters and the numerical values of $c_i$ ($i=$ 0, 1, 2) coefficients in Eq. (7). Eleventh and twenty second-th columns give the range of the metallicity $[Fe/H]$ (dex) for the star whose absolute magnitude would be estimated. $R^2$ is the square of the correlation coefficient.}
   \begin{tabular}{ccccccccccc|ccccccccccc}
   \hline
    Cluster $\rightarrow$& M92   & M13   & M71   & M67   & NGC 6791  &       &       &       &       & \multicolumn{1}{r}{} & Cluster $\rightarrow$& M92   & M13   & M71   & M67   & NGC 6791  &       &       &       &       &  \\
    \hline
     $(V-J)_{0}$ & \multicolumn{5}{c}{$M_{J}$} & $c_{0}$    & $c_{1}$    & $c_{2}$     & $R^{2}$  & $[Fe/H]$-int.&  $(V-J)_{0}$& \multicolumn{5}{c}{$M_{J}$}             & $c_{0}$    & $c_{1}$    & $c_{2}$     & $R^{2}$   & $[Fe/H]$-int.\\
    \hline

    1.30  & 1.364 & 1.992 & 2.341 & $-$    & $-$    & 2.5339 & 0.0791 & -0.2162 & 1     & [-2.15, -0.78] & 2.06  & -4.275 & -3.717 & -2.783 & -1.470 & -0.652 & -1.4238 & 2.0438 & 0.3285 & 0.9991 & [-2.15,  0.37] \\
    1.31  & 1.207 & 1.843 & 2.224 & $-$    & $-$    & 2.4912 & 0.1978 & -0.1859 & 1     & [-2.15, -0.78] & 2.07  & -4.315 & -3.750 & -2.816 & -1.515 & -0.699 & -1.4661 & 2.0284 & 0.3220 & 0.9991 & [-2.15,  0.37] \\
    1.32  & 1.054 & 1.696 & 2.109 & $-$    & $-$    & 2.4485 & 0.3140 & -0.1557 & 1     & [-2.15, -0.78] & 2.08  & -4.354 & -3.783 & -2.849 & -1.559 & -0.745 & -1.5077 & 2.0134 & 0.3156 & 0.9992 & [-2.15,  0.37] \\
    1.33  & 0.905 & 1.553 & 1.996 & $-$    & $-$    & 2.4057 & 0.4277 & -0.1257 & 1     & [-2.15, -0.78] & 2.09  & -4.393 & -3.815 & -2.880 & -1.603 & -0.790 & -1.5486 & 1.9990 & 0.3095 & 0.9992 & [-2.15,  0.37] \\
    1.34  & 0.760 & 1.412 & 1.884 & $-$    & $-$    & 2.5413 & 0.8220 & $-$    & 0.9979 & [-2.15, -0.78] & 2.10  & -4.431 & -3.847 & -2.912 & -1.646 & -0.835 & -1.5888 & 1.9851 & 0.3037 & 0.9992 & [-2.15,  0.37] \\
    1.35  & 0.620 & 1.275 & 1.774 & $-$    & $-$    & 2.4436 & 0.8438 & $-$    & 0.9990 & [-2.15, -0.78] & 2.11  & -4.468 & -3.879 & -2.943 & -1.689 & -0.878 & -1.6284 & 1.9717 & 0.2981 & 0.9992 & [-2.15,  0.37] \\
    1.36  & 0.484 & 1.140 & 1.666 & $-$    & $-$    & 2.3464 & 0.8639 & $-$    & 0.9997 & [-2.15, -0.78] & 2.12  & -4.504 & -3.910 & -2.973 & -1.731 & -0.920 & -1.6675 & 1.9587 & 0.2927 & 0.9992 & [-2.15,  0.37] \\
    1.37  & 0.351 & 1.008 & 1.560 & $-$    & $-$    & 2.2497 & 0.8825 & $-$    & 1     & [-2.15, -0.78] & 2.13  & -4.540 & -3.940 & -3.003 & -1.773 & -0.962 & -1.7059 & 1.9463 & 0.2876 & 0.9992 & [-2.15,  0.37] \\
    1.38  & 0.222 & 0.879 & 1.455 & $-$    & $-$    & 2.1535 & 0.8995 & $-$    & 0.9999 & [-2.15, -0.78] & 2.14  & -4.575 & -3.971 & -3.032 & -1.815 & -1.002 & -1.7439 & 1.9343 & 0.2828 & 0.9992 & [-2.15,  0.37] \\
    1.39  & 0.097 & 0.752 & 1.352 & $-$    & $-$    & 2.0579 & 0.9152 & $-$    & 0.9996 & [-2.15, -0.78] & 2.15  & -4.609 & -4.001 & -3.061 & -1.857 & -1.042 & -1.7813 & 1.9228 & 0.2782 & 0.9992 & [-2.15,  0.37] \\
    1.40  & -0.025 & 0.629 & 1.251 & $-$    & $-$    & 1.9629 & 0.9294 & $-$    & 0.9990 & [-2.15, -0.78] & 2.16  & -4.642 & -4.030 & -3.090 & -1.898 & -1.081 & -1.8182 & 1.9118 & 0.2740 & 0.9992 & [-2.15,  0.37] \\
    1.41  & -0.143 & 0.508 & 1.151 & $-$    & $-$    & 1.8686 & 0.9424 & $-$    & 0.9982 & [-2.15, -0.78] & 2.17  & -4.675 & -4.060 & -3.118 & -1.938 & -1.120 & -1.8547 & 1.9012 & 0.2700 & 0.9991 & [-2.15,  0.37] \\
    1.42  & -0.258 & 0.389 & 1.053 & $-$    & $-$    & 2.0168 & 1.3377 & 0.1301 & 1     & [-2.15, -0.78] & 2.18  & -4.706 & -4.089 & -3.145 & -1.979 & -1.158 & -1.8907 & 1.8911 & 0.2664 & 0.9991 & [-2.15,  0.37] \\
    1.43  & -0.370 & 0.273 & 0.956 & $-$    & $-$    & 1.9728 & 1.4259 & 0.1564 & 1     & [-2.15, -0.78] & 2.19  & -4.736 & -4.118 & -3.172 & -2.019 & -1.195 & -1.9263 & 1.8815 & 0.2632 & 0.9990 & [-2.15,  0.37] \\
    1.44  & -0.478 & 0.160 & 0.861 & $-$    & $-$    & 1.9286 & 1.5114 & 0.1823 & 1     & [-2.15, -0.78] & 2.20  & -4.766 & -4.146 & -3.199 & -2.058 & -1.231 & -1.9615 & 1.8724 & 0.2603 & 0.9990 & [-2.15,  0.37] \\
    1.45  & -0.584 & 0.049 & 0.767 & $-$    & $-$    & 1.8843 & 1.5943 & 0.2075 & 1     & [-2.15, -0.78] & 2.21  & -4.794 & -4.175 & -3.226 & -2.098 & -1.267 & -1.9963 & 1.8638 & 0.2578 & 0.9989 & [-2.15,  0.37] \\
    1.46  & -0.687 & -0.060 & 0.675 & $-$    & $-$    & 1.8397 & 1.6746 & 0.2322 & 1     & [-2.15, -0.78] & 2.22  & -4.821 & -4.203 & -3.252 & -2.137 & -1.302 & -2.0308 & 1.8857 & 0.2557 & 0.9987 & [-2.15,  0.37] \\
    1.47  & -0.787 & -0.166 & 0.584 & $-$    & $-$    & 1.7949 & 1.7523 & 0.2564 & 1     & [-2.15, -0.78] & 2.23  & -4.846 & -4.230 & -3.277 & -2.176 & -1.337 & -2.0649 & 1.8480 & 0.2540 & 0.9986 & [-2.15,  0.37] \\
    1.48  & -0.885 & -0.270 & 0.495 & $-$    & $-$    & 1.7498 & 1.8273 & 0.2799 & 1     & [-2.15, -0.78] & 2.24  & -4.871 & -4.258 & -3.303 & -2.214 & -1.371 & -2.0987 & 1.8409 & 0.2527 & 0.9985 & [-2.15,  0.37] \\
    1.49  & -0.980 & -0.372 & 0.407 & $-$    & $-$    & 1.7046 & 1.8997 & 0.3028 & 1     & [-2.15, -0.78] & 2.25  & -4.894 & -4.285 & -3.327 & -2.253 & -1.404 & -2.1322 & 1.8343 & 0.2519 & 0.9983 & [-2.15,  0.37] \\
    1.50  & -1.073 & -0.472 & 0.321 & $-$    & $-$    & 1.6590 & 1.9694 & 0.3251 & 1     & [-2.15, -0.78] & 2.26  & -4.915 & -4.312 & -3.352 & -2.290 & -1.437 & -2.1654 & 1.8282 & 0.2516 & 0.9981 & [-2.15,  0.37] \\
    1.51  & -1.163 & -0.569 & 0.236 & $-$    & $-$    & 1.6133 & 2.0366 & 0.3467 & 1     & [-2.15, -0.78] & 2.27  & -4.935 & -4.339 & -3.376 & -2.328 & -1.470 & -2.1984 & 1.8226 & 0.2517 & 0.9979 & [-2.15,  0.37] \\
    1.52  & -1.251 & -0.664 & 0.152 & $-$    & $-$    & 1.5673 & 2.1011 & 0.3677 & 1     & [-2.15, -0.78] & 2.28  & -4.954 & -4.365 & -3.400 & -2.365 & -1.502 & -2.2311 & 1.8176 & 0.2524 & 0.9976 & [-2.15,  0.37] \\
    1.53  & -1.336 & -0.758 & 0.070 & $-$    & $-$    & 1.5210 & 2.1630 & 0.3879 & 1     & [-2.15, -0.78] & 2.29  & $-$    & -4.391 & -3.424 & -2.403 & -1.534 & -2.2417 & 1.7305 & 0.1585 & 0.9965 & [-1.41,  0.37] \\
    1.54  & -1.420 & -0.849 & -0.011 & $-$    & $-$    & 1.4745 & 2.2223 & 0.4075 & 1     & [-2.15, -0.78] & 2.30  & $-$    & -4.417 & -3.447 & -2.439 & -1.566 & -2.2734 & 1.7244 & 0.1577 & 0.9960 & [-1.41,  0.37] \\
    1.55  & -1.501 & -0.938 & -0.091 & $-$    & $-$    & 1.4277 & 2.2791 & 0.4264 & 1     & [-2.15, -0.78] & 2.31  & $-$    & -4.443 & $-$    & -2.476 & -1.597 & -2.3960 & 2.0134 & 0.3982 & 1     & [-1.41,  0.37] \\
    1.56  & -1.581 & -1.025 & -0.169 & $-$    & $-$    & 1.3806 & 2.3332 & 0.4445 & 1     & [-2.15, -0.78] & 2.32  & $-$    & -4.469 & $-$    & -2.512 & -1.627 & -2.4319 & 2.0230 & 0.4102 & 1     & [-1.41,  0.37] \\
    1.57  & -1.659 & -1.111 & -0.246 & $-$    & $-$    & 1.3333 & 2.3849 & 0.4620 & 1     & [-2.15, -0.78] & 2.33  & $-$    & -4.494 & $-$    & -2.548 & -1.658 & -2.4675 & 2.0328 & 0.4222 & 1     & [-1.41,  0.37] \\
    1.58  & -1.735 & -1.194 & -0.321 & 1.702 & $-$    & 1.8084 & 3.2112 & 0.7317 & 0.9984 & [-2.15, -0.04] & 2.34  & $-$    & -4.520 & $-$    & -2.584 & -1.688 & -2.5029 & 2.0426 & 0.4342 & 1     & [-1.41,  0.37] \\
    1.59  & -1.809 & -1.276 & -0.396 & 1.603 & $-$    & 1.7099 & 3.1824 & 0.7231 & 0.9987 & [-2.15, -0.04] & 2.35  & $-$    & -4.545 & $-$    & -2.619 & -1.717 & -2.5379 & 2.0525 & 0.4463 & 1     & [-1.41,  0.37] \\
    1.60  & -1.881 & -1.356 & -0.469 & 1.505 & $-$    & 1.6134 & 3.1544 & 0.7148 & 0.9989 & [-2.15, -0.04] & 2.36  & $-$    & -4.569 & $-$    & -2.654 & -1.747 & -2.5726 & 2.0624 & 0.4583 & 1     & [-1.41,  0.37] \\
    1.61  & -1.952 & -1.435 & -0.541 & 1.410 & $-$    & 1.5188 & 3.1272 & 0.7069 & 0.9991 & [-2.15, -0.04] & 2.37  & $-$    & -4.594 & $-$    & -2.689 & -1.776 & -2.6070 & 2.0723 & 0.4702 & 1     & [-1.41,  0.37] \\
    1.62  & -2.022 & -1.511 & -0.612 & 1.317 & $-$    & 1.4262 & 3.1008 & 0.6992 & 0.9993 & [-2.15, -0.04] & 2.38  & $-$    & -4.618 & $-$    & -2.724 & -1.805 & -2.6412 & 2.0821 & 0.4821 & 1     & [-1.41,  0.37] \\
    1.63  & -2.090 & -1.586 & -0.682 & 1.225 & $-$    & 1.3354 & 3.0752 & 0.6918 & 0.9995 & [-2.15, -0.04] & 2.39  & $-$    & -4.643 & $-$    & -2.758 & -1.833 & -2.6750 & 2.0917 & 0.4938 & 1     & [-1.41,  0.37] \\
    1.64  & -2.157 & -1.659 & -0.750 & 1.136 & $-$    & 1.2464 & 3.0502 & 0.6846 & 0.9996 & [-2.15, -0.04] & 2.40  & $-$    & -4.667 & $-$    & -2.792 & -1.862 & -2.7085 & 2.1013 & 0.5053 & 1     & [-1.41,  0.37] \\
    1.65  & -2.222 & -1.731 & -0.817 & 1.048 & $-$    & 1.1592 & 3.0258 & 0.6777 & 0.9997 & [-2.15, -0.04] & 2.41  & $-$    & -4.690 & $-$    & -2.825 & -1.890 & -2.7417 & 2.1106 & 0.5167 & 1     & [-1.41,  0.37] \\
    1.66  & -2.286 & -1.801 & -0.883 & 0.962 & $-$    & 1.0737 & 3.0019 & 0.6709 & 0.9998 & [-2.15, -0.04] & 2.42  & $-$    & -4.714 & $-$    & -2.859 & -1.918 & -2.7746 & 2.1197 & 0.5278 & 1     & [-1.41,  0.37] \\
    1.67  & -2.349 & -1.870 & -0.948 & 0.878 & $-$    & 0.9899 & 2.9786 & 0.6642 & 0.9999 & [-2.15, -0.04] & 2.43  & $-$    & -4.737 & $-$    & -2.891 & -1.946 & -2.8072 & 2.1286 & 0.5388 & 1     & [-1.41,  0.37] \\
    1.68  & -2.411 & -1.937 & -1.012 & 0.795 & $-$    & 0.9078 & 2.9558 & 0.6577 & 0.9999 & [-2.15, -0.04] & 2.44  & $-$    & -4.761 & $-$    & -2.924 & -1.973 & -2.8394 & 2.1372 & 0.5494 & 1     & [-1.41,  0.37] \\
    1.69  & -2.472 & -2.003 & -1.075 & 0.714 & $-$    & 0.8272 & 2.9333 & 0.6513 & 1     & [-2.15, -0.04] & 2.45  & $-$    & -4.783 & $-$    & -2.956 & -2.001 & -2.8713 & 2.1454 & 0.5597 & 1     & [-1.41,  0.37] \\
    1.70  & -2.532 & -2.068 & -1.137 & 0.635 & $-$    & 0.7482 & 2.9113 & 0.6449 & 1     & [-2.15, -0.04] & 2.46  & $-$    & -4.806 & $-$    & -2.988 & -2.028 & -2.9028 & 2.1533 & 0.5697 & 1     & [-1.41,  0.37] \\
    1.71  & -2.591 & -2.131 & -1.198 & 0.557 & $-$    & 0.6706 & 2.8896 & 0.6387 & 1     & [-2.15, -0.04] & 2.47  & $-$    & -4.829 & $-$    & -3.019 & -2.055 & -2.9339 & 2.1608 & 0.5794 & 1     & [-1.41,  0.37] \\
    1.72  & -2.649 & -2.193 & -1.258 & 0.481 & 1.895 & 0.6564 & 3.0018 & 0.6839 & 0.9996 & [-2.15,  0.37] & 2.48  & $-$    & -4.851 & $-$    & -3.050 & -2.082 & -2.9647 & 2.1678 & 0.5887 & 1     & [-1.41,  0.37] \\
    1.73  & -2.706 & -2.253 & -1.316 & 0.406 & 1.784 & 0.5706 & 2.9565 & 0.6684 & 0.9997 & [-2.15,  0.37] & 2.49  & $-$    & -4.873 & $-$    & -3.081 & -2.109 & -2.9950 & 2.1743 & 0.5975 & 1     & [-1.41,  0.37] \\
    1.74  & -2.762 & -2.312 & -1.374 & 0.333 & 1.676 & 0.4867 & 2.9127 & 0.6533 & 0.9998 & [-2.15,  0.37] & 2.50  & $-$    & -4.895 & $-$    & -3.111 & -2.135 & -3.0250 & 2.1804 & 0.6059 & 1     & [-1.41,  0.37] \\
    1.75  & -2.817 & -2.370 & -1.431 & 0.261 & 1.571 & 0.4047 & 2.8703 & 0.6387 & 0.9999 & [-2.15,  0.37] & 2.51  & $-$    & -4.916 & $-$    & -3.141 & -2.162 & -3.0546 & 2.1859 & 0.6139 & 1     & [-1.41,  0.37] \\
    1.76  & -2.872 & -2.427 & -1.487 & 0.190 & 1.468 & 0.3245 & 2.8291 & 0.6245 & 0.9999 & [-2.15,  0.37] & 2.52  & $-$    & -4.937 & $-$    & -3.170 & -2.188 & -3.0837 & 2.1908 & 0.6213 & 1     & [-1.41,  0.37] \\
    1.77  & -2.926 & -2.483 & -1.541 & 0.121 & 1.368 & 0.2461 & 2.7893 & 0.6107 & 1     & [-2.15,  0.37] & 2.53  & $-$    & -4.958 & $-$    & -3.199 & -2.214 & -3.1123 & 2.1951 & 0.6282 & 1     & [-1.41,  0.37] \\
    1.78  & -2.979 & -2.537 & -1.595 & 0.053 & 1.271 & 0.1695 & 2.7507 & 0.5973 & 1     & [-2.15,  0.37] & 2.54  & $-$    & -4.979 & $-$    & -3.227 & -2.240 & -3.1405 & 2.1987 & 0.6346 & 1     & [-1.41,  0.37] \\
    1.79  & -3.032 & -2.591 & -1.648 & -0.014 & 1.176 & 0.0945 & 2.7133 & 0.5842 & 0.9999 & [-2.15,  0.37] & 2.55  & $-$    & -5.000 & $-$    & -3.255 & -2.266 & -3.1683 & 2.2017 & 0.6404 & 1     & [-1.41,  0.37] \\
    1.80  & -3.084 & -2.643 & -1.700 & -0.080 & 1.083 & 0.0213 & 2.6771 & 0.5715 & 0.9999 & [-2.15,  0.37] & 2.56  & $-$    & -5.020 & $-$    & -3.283 & -2.292 & -3.1955 & 2.2039 & 0.6455 & 1     & [-1.41,  0.37] \\
    1.81  & -3.135 & -2.694 & -1.752 & -0.144 & 0.993 & -0.0504 & 2.6420 & 0.5591 & 0.9999 & [-2.15,  0.37] & 2.57  & $-$    & -5.040 & $-$    & -3.309 & -2.317 & -3.2222 & 2.2054 & 0.6500 & 1     & [-1.41,  0.37] \\
    1.82  & -3.186 & -2.745 & -1.802 & -0.208 & 0.905 & -0.1204 & 2.6080 & 0.5470 & 0.9998 & [-2.15,  0.37] & 2.58  & $-$    & -5.059 & $-$    & -3.336 & -2.343 & -3.2484 & 2.2061 & 0.6539 & 1     & [-1.41,  0.37] \\
    1.83  & -3.237 & -2.794 & -1.851 & -0.270 & 0.819 & -0.1890 & 2.5750 & 0.5353 & 0.9998 & [-2.15,  0.37] & 2.59  & $-$    & -5.078 & $-$    & -3.361 & -2.368 & -3.2740 & 2.2060 & 0.6570 & 1     & [-1.41,  0.37] \\
    1.84  & -3.286 & -2.843 & -1.900 & -0.331 & 0.736 & -0.2560 & 2.5431 & 0.5238 & 0.9997 & [-2.15,  0.37] & 2.60  & $-$    & -5.097 & $-$    & -3.386 & -2.393 & -3.2990 & 2.2051 & 0.6594 & 1     & [-1.41,  0.37] \\
    1.85  & -3.336 & -2.890 & -1.948 & -0.391 & 0.655 & -0.3215 & 2.5122 & 0.5126 & 0.9996 & [-2.15,  0.37] & 2.61  & $-$    & -5.116 & $-$    & -3.411 & -2.418 & -3.3235 & 2.2032 & 0.6611 & 1     & [-1.41,  0.37] \\
    1.86  & -3.385 & -2.937 & -1.995 & -0.450 & 0.576 & -0.3856 & 2.4822 & 0.5017 & 0.9996 & [-2.15,  0.37] & 2.62  & $-$    & -5.134 & $-$    & -3.434 & -2.443 & -3.3473 & 2.2004 & 0.6620 & 1     & [-1.41,  0.37] \\
    1.87  & -3.433 & -2.982 & -2.041 & -0.509 & 0.498 & -0.4484 & 2.4532 & 0.4910 & 0.9995 & [-2.15,  0.37] & 2.63  & $-$    & -5.152 & $-$    & -3.457 & -2.467 & -3.3705 & 2.1967 & 0.6621 & 1     & [-1.41,  0.37] \\
    1.88  & -3.481 & -3.027 & -2.086 & -0.566 & 0.423 & -0.5097 & 2.4251 & 0.4806 & 0.9994 & [-2.15,  0.37] & 2.64  & $-$    & -5.169 & $-$    & -3.480 & -2.492 & -3.3931 & 2.1920 & 0.6613 & 1     & [-1.41,  0.37] \\
    1.89  & -3.529 & -3.071 & -2.131 & -0.622 & 0.350 & -0.5698 & 2.3978 & 0.4704 & 0.9994 & [-2.15,  0.37] & 2.65  & $-$    & -5.186 & $-$    & -3.501 & -2.516 & -3.4149 & 2.1862 & 0.6597 & 1     & [-1.41,  0.37] \\
    1.90  & -3.576 & -3.114 & -2.175 & -0.678 & 0.278 & -0.6286 & 2.3713 & 0.4604 & 0.9993 & [-2.15,  0.37] & 2.66  & $-$    & -5.203 & $-$    & -3.522 & -2.540 & -3.4361 & 2.1794 & 0.6572 & 1     & [-1.41,  0.37] \\
    1.91  & -3.623 & -3.157 & -2.218 & -0.733 & 0.209 & -0.6861 & 2.3457 & 0.4507 & 0.9992 & [-2.15,  0.37] & 2.67  & $-$    & -5.219 & $-$    & -3.542 & -2.564 & -3.4565 & 2.1716 & 0.6537 & 1     & [-1.41,  0.37] \\
    1.92  & -3.669 & -3.199 & -2.260 & -0.786 & 0.141 & -0.7425 & 2.3209 & 0.4412 & 0.9992 & [-2.15,  0.37] & 2.68  & $-$    & -5.234 & $-$    & -3.562 & -2.587 & -3.4761 & 2.1626 & 0.6494 & 1     & [-1.41,  0.37] \\
    1.93  & -3.715 & -3.240 & -2.302 & -0.840 & 0.075 & -0.7977 & 2.2968 & 0.4318 & 0.9991 & [-2.15,  0.37] & 2.69  & $-$    & $-$    & $-$    & -3.580 & -2.610 & 2.3650 & -3.4855 & $-$    & 1     & [-0.04,  0.37] \\
    1.94  & -3.761 & -3.280 & -2.343 & -0.892 & 0.010 & -0.8517 & 2.2735 & 0.4227 & 0.9991 & [-2.15,  0.37] & 2.70  & $-$    & $-$    & $-$    & -3.598 & -2.634 & 2.3515 & -3.5036 & $-$    & 1     & [-0.04,  0.37] \\
    1.95  & -3.806 & -3.320 & -2.383 & -0.944 & -0.053 & -0.9047 & 2.2509 & 0.4138 & 0.9991 & [-2.15,  0.37] & 2.71  & $-$    & $-$    & $-$    & -3.614 & -2.656 & 2.3366 & -3.5209 & $-$    & 1     & [-0.04,  0.37] \\
    1.96  & -3.851 & -3.359 & -2.422 & -0.994 & -0.114 & -0.9565 & 2.2289 & 0.4051 & 0.9990 & [-2.15,  0.37] & 2.72  & $-$    & $-$    & $-$    & -3.630 & -2.679 & 2.3200 & -3.5374 & $-$    & 1     & [-0.04,  0.37] \\
    1.97  & -3.895 & -3.397 & -2.461 & -1.045 & -0.174 & -1.0074 & 2.2077 & 0.3966 & 0.9990 & [-2.15,  0.37] & 2.73  & $-$    & $-$    & $-$    & -3.645 & -2.701 & 2.3018 & -3.5529 & $-$    & 1     & [-0.04,  0.37] \\
    1.98  & -3.939 & -3.435 & -2.499 & -1.094 & -0.232 & -1.0572 & 2.1871 & 0.3882 & 0.9990 & [-2.15,  0.37] & 2.74  & $-$    & $-$    & $-$    & -3.659 & -2.723 & 2.2820 & -3.5676 & $-$    & 1     & [-0.04,  0.37] \\
    1.99  & -3.983 & -3.472 & -2.537 & -1.143 & -0.289 & -1.1061 & 2.1671 & 0.3801 & 0.9990 & [-2.15,  0.37] & 2.75  & $-$    & $-$    & $-$    & -3.672 & -2.745 & 2.2604 & -3.5813 & $-$    & 1     & [-0.04,  0.37] \\
    2.00  & -4.026 & -3.509 & -2.574 & -1.192 & -0.345 & -1.1541 & 2.1478 & 0.3721 & 0.9990 & [-2.15,  0.37] & 2.76  & $-$    & $-$    & $-$    & -3.683 & -2.766 & 2.2371 & -3.5940 & $-$    & 1     & [-0.04,  0.37] \\
    2.01  & -4.069 & -3.545 & -2.610 & -1.239 & -0.399 & -1.2011 & 2.1290 & 0.3644 & 0.9990 & [-2.15,  0.37] & 2.77  & $-$    & $-$    & $-$    & -3.694 & -2.787 & 2.2120 & -3.6057 & $-$    & 1     & [-0.04,  0.37] \\
    2.02  & -4.111 & -3.580 & -2.646 & -1.287 & -0.452 & -1.2473 & 2.1109 & 0.3568 & 0.9990 & [-2.15,  0.37] & 2.78  & $-$    & $-$    & $-$    & -3.704 & -2.808 & 2.1851 & -3.6164 & $-$    & 1     & [-0.04,  0.37] \\
    2.03  & -4.153 & -3.615 & -2.681 & -1.333 & -0.504 & -1.2926 & 2.0933 & 0.3494 & 0.9991 & [-2.15,  0.37] & 2.79  & $-$    & $-$    & $-$    & -3.712 & -2.828 & 2.1564 & -3.6260 & $-$    & 1     & [-0.04,  0.37] \\
    2.04  & -4.194 & -3.650 & -2.716 & -1.379 & -0.554 & -1.3371 & 2.0762 & 0.3423 & 0.9991 & [-2.15,  0.37] & 2.80  & $-$    & $-$    & $-$    & -3.719 & -2.848 & 2.1257 & -3.6344 & $-$    & 1     & [-0.04,  0.37] \\
    2.05  & -4.235 & -3.684 & -2.750 & -1.425 & -0.604 & -1.3809 & 2.0598 & 0.3353 & 0.9991 & [-2.15,  0.37] & $-$    & $-$    & $-$    & $-$    & $-$    & $-$    & $-$    & $-$    & $-$    & \multicolumn{1}{c}{$-$} & \multicolumn{1}{c}{$-$} \\
\hline
    \end{tabular}
  \label{tab:addlabel}
}
\end{table*}

Now, we can fit the $M_{J}$ absolute magnitudes to the corresponding $[Fe/H]$ metallicity for a given $(V-J)_{0}$ colour index and obtain the required calibration. This is carried out for the colour indices $(V-J)_{0}=$ 1.30, 1.60, 1.90, 2.10, 2.25, 2.50, and 2.65 mag just for  exhibition of the procedure. The results are given in Table 9 and Fig. 4. The absolute magnitudes in all colour indices could be fitted to quadratic polynomials with high (squared) correlation coefficients, i.e. $R^{2}\geq0.9982$, including $(V-J)_{0}=1.90$, 2.10, 2.25 mag which cover the largest metallicity range, $-2.15\leq [Fe/H] \leq +0.37$ dex. High correlation coefficient plus (relatively) polynomial with small degree is a strong clue for accurate absolute magnitude estimation. 

This procedure can be applied to any $(V-J)_{0}$ colour interval for which the sample clusters are defined. The $(V-J)_{0}$ domain of the clusters are different. Hence, we adopted this interval as $1.3\leq(V-J)_{0}\leq2.8$ mag where at least two clusters are defined, and we evaluated $M_J$ absolute magnitude for each colour. $1.72\leq(V-J)_{0}\leq 2.28$ mag is the interval where all the clusters are defined, whereas the range of the interval where only two clusters (M67 and NGC 6791) are defined is rather small, i.e. $2.69\leq(V-J)_{0}\leq 2.80$ mag. However, this small interval can be used for estimation of the absolute magnitudes for red giants with metallicities $-0.04\leq [Fe/H] \leq +0.37$ dex. The general form of the equation for the calibration is as follows:

\begin{eqnarray}
M_J=c_{0} + c_{1}X + c_{2}X^{2}     
\end{eqnarray}
where $X=\lbrack Fe/H \rbrack$. $M_{J}$ could be fitted in terms of metallicity by a quadratic polynomial for all $(V-J)_{0}$ colour interval with high correlation coefficient, $R^{2}\geq 0.996$, except two intervals with small ranges, i.e. a linear fitting was sufficient for the colour intervals $1.34\leq(V-J)_{0}\leq 1.41$ and $2.69\leq(V-J)_{0} \leq 2.80$ mag where only two clusters are defined. The absolute magnitudes estimated via Eq. (7) for 151 $(V-J)_{0}$ colour indices and the corresponding $c_{i}~(i=0$, 1, 2) coefficients are given in Table 10. However, the diagrams for the calibrations are not given in the paper for avoiding space consuming.     

\begin{table}
\setlength{\tabcolsep}{3pt}
  \center
\scriptsize{
  \caption{$M_{K_{s}}$ absolute magnitudes estimated for a set of $(V-K_s)_0$ colours for five clusters used in the calibration.}
    \begin{tabular}{cccccc}
    \hline
    Cluster$\rightarrow$ & M92   & M13   & M71   & M67   & NGC 6791 \\
    \hline
    $(V-K_{s})_{0}$ & \multicolumn{5}{c}{$M_{K_s}$}             \\
    \hline
    1.75  &  0.957 &  1.456 &   $-$    & $-$    & \multicolumn{1}{c}{$-$} \\
    1.80  &  0.434 &  0.919 &  1.742 & $-$    & \multicolumn{1}{c}{$-$} \\
    1.85  & -0.045 &  0.424 &  1.250 & $-$    & \multicolumn{1}{c}{$-$} \\
    1.90  & -0.485 & -0.031 &  0.792 & $-$    & \multicolumn{1}{c}{$-$} \\
    1.95  & -0.890 & -0.449 &  0.367 & $-$    & \multicolumn{1}{c}{$-$} \\
    2.00  & -1.265 & -0.834 & -0.028 & $-$    & \multicolumn{1}{c}{$-$} \\
    2.05  & -1.613 & -1.189 & -0.394 & $-$    & \multicolumn{1}{c}{$-$} \\
    2.10  & -1.938 & -1.515 & -0.733 & $-$    & \multicolumn{1}{c}{$-$} \\
    2.15  & -2.242 & -1.816 & -1.048 & 0.960 & \multicolumn{1}{c}{$-$} \\
    2.20  & -2.529 & -2.094 & -1.339 & 0.635 & \multicolumn{1}{c}{$-$} \\
    2.25  & -2.800 & -2.352 & -1.608 & 0.320 & \multicolumn{1}{c}{$-$} \\
    2.30  & -3.059 & -2.591 & -1.857 & 0.016 & \multicolumn{1}{c}{$-$} \\
    2.35  & -3.306 & -2.814 & -2.088 & -0.274 & \multicolumn{1}{c}{$-$} \\
    2.40  & -3.543 & -3.023 & -2.302 & -0.550 & \multicolumn{1}{c}{$-$} \\
    2.45  & -3.771 & -3.219 & -2.500 & -0.811 & 0.619 \\
    2.50  & -3.992 & -3.403 & -2.683 & -1.057 & 0.256 \\
    2.55  & -4.204 & -3.578 & -2.854 & -1.288 & -0.072 \\
    2.60  & -4.408 & -3.745 & -3.012 & -1.505 & -0.369 \\
    2.65  & -4.605 & -3.905 & -3.160 & -1.709 & -0.637 \\
    2.70  & -4.793 & -4.058 & -3.299 & -1.901 & -0.880 \\
    2.75  & -4.971 & -4.206 & -3.428 & -2.083 & -1.101 \\
    2.80  & -5.138 & -4.350 & -3.550 & -2.255 & -1.302 \\
    2.85  & -5.292 & -4.490 & -3.666 & -2.421 & -1.486 \\
    2.90  & -5.432 & -4.626 & -3.776 & -2.581 & -1.656 \\
    2.95  & -5.556 & -4.759 & -3.881 & -2.736 & -1.813 \\
    3.00  & -5.659 & -4.890 & -3.981 & -2.890 & -1.960 \\
    3.05  & -5.740 & -5.017 & -4.078 & -3.042 & -2.098 \\
    3.10  & $-$    & -5.142 & -4.172 & -3.194 & -2.230 \\
    3.15  & $-$    & -5.264 &       & -3.347 & -2.357 \\
    3.20  & $-$    & -5.381 & $-$    & -3.501 & -2.479 \\
    3.30  & $-$    & -5.604 & $-$    & -3.815 & -2.717 \\
    3.35  & $-$    & -5.708 & $-$    & -3.973 & -2.834 \\
    3.40  & $-$    & -5.805 & $-$    & -4.130 & -2.951 \\
    3.45  & $-$    & -5.894 & $-$    & -4.285 & -3.067 \\
    3.50  & $-$    & -5.975 & $-$    & -4.435 & -3.184 \\
    3.55  & $-$    & -6.045 & $-$    & -4.576 & -3.301 \\
    3.60  & $-$    & -6.104 & $-$    & -4.705 & -3.418 \\
    3.65  & $-$    & -6.149 & $-$    & -4.817 & -3.535 \\
    3.70  & $-$    & -6.179 & $-$    & -4.905 & -3.651 \\
    3.75  & $-$    & $-$    & $-$    & -4.963 & -3.766 \\
    3.80  & $-$    & $-$    & $-$    & -4.984 & -3.878 \\
    3.85  & $-$    & $-$    & $-$    & $-$    & -3.988 \\
    3.90  & $-$    & $-$    & $-$    & $-$    & -4.092 \\
    3.95  & $-$    & $-$    & $-$    & $-$    & -4.191 \\
    4.00  & $-$    & $-$    & $-$    & $-$    & -4.283 \\
    4.05  & $-$    & $-$    & $-$    & $-$    & -4.365 \\
    4.10  & $-$    & $-$    & $-$    & $-$    & -4.436 \\
    \hline
    \end{tabular}
  \label{tab:addlabel}
}
\end{table}

\begin{table}
\setlength{\tabcolsep}{3pt}
  \center
\scriptsize{
  \caption{$M_{K_{s}}$ absolute magnitudes and $[Fe/H]$ metallicities for seven $(V-K_s)_0$ –intervals.}
    \begin{tabular}{ccc}
    \hline
    $(V-K_{s})_{0}$ & $[Fe/H]$ (dex)& $M_{K_s}$ \\
    \hline
    1.80  & -2.15 & 0.434 \\
          & -1.41 & 0.919 \\
          & -0.78 & 1.742 \\
    \hline
    2.00  & -2.15 & -1.265 \\
          & -1.41 & -0.834 \\
          & -0.78 & -0.028 \\
    \hline
    2.20  & -2.15 & -2.529 \\
          & -1.41 & -2.094 \\
          & -0.78 & -1.339 \\
          & -0.04 & 0.635 \\
    \hline
    2.40  & -2.15 & -3.543 \\
          & -1.41 & -3.023 \\
          & -0.78 & -2.302 \\
          & -0.04 & -0.550 \\
    \hline
    2.60  & -2.15 & -4.408 \\
          & -1.41 & -3.745 \\
          & -0.78 & -3.012 \\
          & -0.04 & -1.505 \\
          & 0.37  & -0.369 \\
    \hline
    3.00  & -2.15 & -5.659 \\
          & -1.41 & -4.890 \\
          & -0.78 & -3.981 \\
          & -0.04 & -2.890 \\
          & 0.37  & -1.960 \\
    \hline
    3.30  & -1.41 & -5.604 \\
          & -0.04 & -3.815 \\
          &  0.37 & -2.717 \\
    \hline
    \end{tabular}
  \label{tab:addlabel}
}
\end{table}

\begin{table*}
\setlength{\tabcolsep}{0.8pt}
\renewcommand{\baselinestretch}{0.5} 
  \center
\tiny{
  \caption{$M_{K_{s}}$ absolute magnitudes estimated for five Galactic clusters and the numerical values of $d_i$ ($i=$ 0, 1, 2) coefficients in Eq. (9). Eleventh and twenty second-th columns give the range of the metallicity $[Fe/H]$ (dex) for the star whose absolute magnitude would be estimated. $R^2$ is the square of the correlation coefficient.}
    \begin{tabular}{ccccccccccc|ccccccccccc}
    \hline
    Cluster$\rightarrow$& M92   & M13   & M71   & M67   & NGC 6791 &       &       &       &       &       &Cluster$\rightarrow$& M92   & M13   & M71   & M67   & NGC 6791 &       &       &       &       &  \\
    \hline
    $(V-K_{s})_{0}$ & \multicolumn{5}{c}{$M_{K_s}$}   & $d_{0}$    & $d_{1}$    & $d_{2}$    & $R{^2}$   & $[Fe/H]$-int. & $(V-K_{s})_{0}$ & \multicolumn{5}{c}{$M_{K_s}$}   & $d_{0}$    & $d_{1}$    & $d_{2}$    & $R{^2}$  & $[Fe/H]$-int.\\
    \hline
    1.75  & 0.957 & 1.456 & $-$    & $-$    & $-$    & 2.4069 & 0.6743 & $-$    & 1     & [-2.15, -1.41] & 2.78  & -5.072 & -4.293 & -3.502 & -2.187 & -1.224 & -2.0618 & 2.0624 & 0.3127 & 0.9993 & [-2.15, 0.37] \\
    1.76  & 0.849 & 1.345 & $-$    & $-$    & $-$    & 2.2905 & 0.6705 & $-$    & 1     & [-2.15, -1.41] & 2.79  & -5.105 & -4.322 & -3.527 & -2.222 & -1.263 & -2.0954 & 2.0480 & 0.3060 & 0.9993 & [-2.15, 0.37] \\
    1.77  & 0.742 & 1.236 & $-$    & $-$    & $-$    & 2.1757 & 0.6667 & $-$    & 1     & [-2.15, -1.41] & 2.80  & -5.138 & -4.350 & -3.550 & -2.255 & -1.302 & -2.1286 & 2.0342 & 0.2996 & 0.9993 & [-2.15, 0.37] \\
    1.78  & 0.638 & 1.128 & $-$    & $-$    & $-$    & 2.0626 & 0.6627 & $-$    & 1     & [-2.15, -1.41] & 2.81  & -5.170 & -4.378 & -3.574 & -2.289 & -1.340 & -2.1613 & 2.0209 & 0.2935 & 0.9993 & [-2.15, 0.37] \\
    1.79  & 0.535 & 1.023 & 1.845 & $-$    & $-$    & 3.3824 & 2.3393 & 0.4721 & 1     & [-2.15, -0.78] & 2.82  & -5.201 & -4.406 & -3.597 & -2.322 & -1.377 & -2.1937 & 2.0081 & 0.2876 & 0.9993 & [-2.15, 0.37] \\
    1.80  & 0.434 & 0.919 & 1.742 & $-$    & $-$    & 3.2861 & 2.3511 & 0.4766 & 1     & [-2.15, -0.78] & 2.83  & -5.232 & -4.434 & -3.621 & -2.355 & -1.414 & -2.2256 & 1.9957 & 0.2819 & 0.9993 & [-2.15, 0.37] \\
    1.81  & 0.335 & 0.816 & 1.641 & $-$    & $-$    & 3.1907 & 2.3618 & 0.4807 & 1     & [-2.15, -0.78] & 2.84  & -5.263 & -4.462 & -3.643 & -2.388 & -1.450 & -2.2572 & 1.9839 & 0.2766 & 0.9994 & [-2.15, 0.37] \\
    1.82  & 0.238 & 0.716 & 1.541 & $-$    & $-$    & 3.0961 & 2.3716 & 0.4847 & 1     & [-2.15, -0.78] & 2.85  & -5.292 & -4.490 & -3.666 & -2.421 & -1.486 & -2.2885 & 1.9725 & 0.2715 & 0.9994 & [-2.15, 0.37] \\
    1.83  & 0.142 & 0.617 & 1.443 & $-$    & $-$    & 3.0023 & 2.3804 & 0.4883 & 1     & [-2.15, -0.78] & 2.86  & -5.322 & -4.517 & -3.688 & -2.453 & -1.521 & -2.3194 & 1.9616 & 0.2666 & 0.9993 & [-2.15, 0.37] \\
    1.84  & 0.048 & 0.520 & 1.346 & $-$    & $-$    & 2.9094 & 2.3882 & 0.4917 & 1     & [-2.15, -0.78] & 2.87  & -5.350 & -4.545 & -3.711 & -2.485 & -1.555 & -2.3501 & 1.9512 & 0.2621 & 0.9993 & [-2.15, 0.37] \\
    1.85  & -0.045 & 0.424 & 1.250 & $-$    & $-$    & 2.8172 & 2.3951 & 0.4948 & 1     & [-2.15, -0.78] & 2.88  & -5.378 & -4.572 & -3.733 & -2.517 & -1.589 & -2.3804 & 1.9413 & 0.2578 & 0.9993 & [-2.15, 0.37] \\
    1.86  & -0.136 & 0.330 & 1.156 & $-$    & $-$    & 2.7259 & 2.4011 & 0.4976 & 1     & [-2.15, -0.78] & 2.89  & -5.406 & -4.599 & -3.754 & -2.549 & -1.623 & -2.4104 & 1.9318 & 0.2538 & 0.9993 & [-2.15, 0.37] \\
    1.87  & -0.225 & 0.237 & 1.063 & $-$    & $-$    & 2.6355 & 2.4061 & 0.5002 & 1     & [-2.15, -0.78] & 2.90  & -5.432 & -4.626 & -3.776 & -2.581 & -1.656 & -2.4402 & 1.9227 & 0.2501 & 0.9993 & [-2.15, 0.37] \\
    1.88  & -0.313 & 0.146 & 0.972 & $-$    & $-$    & 2.5458 & 2.4103 & 0.5025 & 1     & [-2.15, -0.78] & 2.91  & -5.459 & -4.653 & -3.797 & -2.612 & -1.688 & -2.4697 & 1.9142 & 0.2467 & 0.9992 & [-2.15, 0.37] \\
    1.89  & -0.400 & 0.057 & 0.881 & $-$    & $-$    & 2.4570 & 2.4136 & 0.5046 & 1     & [-2.15, -0.78] & 2.92  & -5.484 & -4.680 & -3.818 & -2.643 & -1.720 & -2.4989 & 1.9060 & 0.2436 & 0.9992 & [-2.15, 0.37] \\
    1.90  & -0.485 & -0.031 & 0.792 & $-$    & $-$    & 2.3689 & 2.4161 & 0.5064 & 1     & [-2.15, -0.78] & 2.93  & -5.509 & -4.706 & -3.839 & -2.675 & -1.751 & -2.5280 & 1.8983 & 0.2409 & 0.9991 & [-2.15, 0.37] \\
    1.91  & -0.569 & -0.118 & 0.705 & $-$    & $-$    & 2.2817 & 2.4177 & 0.5079 & 1     & [-2.15, -0.78] & 2.94  & -5.532 & -4.733 & -3.860 & -2.706 & -1.782 & -2.5568 & 1.8911 & 0.2384 & 0.9991 & [-2.15, 0.37] \\
    1.92  & -0.651 & -0.203 & 0.619 & $-$    & $-$    & 2.1952 & 2.4184 & 0.5091 & 1     & [-2.15, -0.78] & 2.95  & -5.556 & -4.759 & -3.881 & -2.736 & -1.813 & -2.5854 & 1.8842 & 0.2362 & 0.9990 & [-2.15, 0.37] \\
    1.93  & -0.732 & -0.286 & 0.534 & $-$    & $-$    & 2.1096 & 2.4184 & 0.5101 & 1     & [-2.15, -0.78] & 2.96  & -5.578 & -4.786 & -3.901 & -2.767 & -1.843 & -2.6138 & 1.8779 & 0.2344 & 0.9989 & [-2.15, 0.37] \\
    1.94  & -0.812 & -0.369 & 0.450 & $-$    & $-$    & 2.0247 & 2.4176 & 0.5108 & 1     & [-2.15, -0.78] & 2.97  & -5.600 & -4.812 & -3.921 & -2.798 & -1.873 & -2.6421 & 1.8719 & 0.2329 & 0.9989 & [-2.15, 0.37] \\
    1.95  & -0.890 & -0.449 & 0.367 & $-$    & $-$    & 1.9406 & 2.4160 & 0.5113 & 1     & [-2.15, -0.78] & 2.98  & -5.620 & -4.838 & -3.941 & -2.829 & -1.902 & -2.6707 & 1.8664 & 0.2317 & 0.9988 & [-2.15, 0.37] \\
    1.96  & -0.968 & -0.529 & 0.286 & $-$    & $-$    & 1.8573 & 2.4137 & 0.5115 & 1     & [-2.15, -0.78] & 2.99  & -5.640 & -4.864 & -3.961 & -2.859 & -1.931 & -2.6982 & 1.8613 & 0.2309 & 0.9986 & [-2.15, 0.37] \\
    1.97  & -1.044 & -0.607 & 0.206 & $-$    & $-$    & 1.7748 & 2.4106 & 0.5115 & 1     & [-2.15, -0.78] & 3.00  & -5.659 & -4.890 & -3.981 & -2.890 & -1.960 & -2.7260 & 1.8566 & 0.2304 & 0.9985 & [-2.15, 0.37] \\
    1.98  & -1.119 & -0.684 & 0.127 & $-$    & $-$    & 1.6931 & 2.4068 & 0.5112 & 1     & [-2.15, -0.78] & 3.01  & -5.677 & -4.916 & -4.001 & -2.920 & -1.988 & -2.7537 & 1.8524 & 0.2303 & 0.9984 & [-2.15, 0.37] \\
    1.99  & -1.192 & -0.760 & 0.049 & $-$    & $-$    & 1.6112 & 2.4023 & 0.5106 & 1     & [-2.15, -0.78] & 3.02  & -5.695 & -4.941 & -4.020 & -2.951 & -2.016 & -2.7813 & 1.8485 & 0.2303 & 0.9982 & [-2.15, 0.37] \\
    2.00  & -1.265 & -0.834 & -0.028 & $-$    & $-$    & 1.5320 & 2.3972 & 0.5098 & 1     & [-2.15, -0.78] & 3.03  & -5.711 & -4.967 & -4.040 & -2.981 & -2.044 & -2.8088 & 1.8451 & 0.2312 & 0.9980 & [-2.15, 0.37] \\
    2.01  & -1.337 & -0.908 & -0.103 & $-$    & $-$    & 1.4526 & 2.3913 & 0.5088 & 1     & [-2.15, -0.78] & 3.04  & -5.726 & -4.992 & -4.059 & -3.011 & -2.071 & -2.8362 & 1.8421 & 0.2322 & 0.9978 & [-2.15, 0.37] \\
    2.02  & -1.407 & -0.980 & -0.177 & $-$    & $-$    & 1.3740 & 2.3848 & 0.5075 & 1     & [-2.15, -0.78] & 3.05  & $-$    & -5.017 & -4.078 & -3.042 & -2.098 & -2.8651 & 1.8452 & 0.2401 & 0.9957 & [-1.41, 0.37] \\
    2.03  & -1.477 & -1.050 & -0.251 & $-$    & $-$    & 1.2961 & 2.3777 & 0.5060 & 1     & [-2.15, -0.78] & 3.06  & $-$    & -5.043 & -4.097 & -3.072 & -2.125 & -2.8910 & 1.8379 & 0.2359 & 0.9953 & [-1.41, 0.37] \\
    2.04  & -1.546 & -1.120 & -0.323 & $-$    & $-$    & 1.2191 & 2.3700 & 0.5042 & 1     & [-2.15, -0.78] & 3.07  & $-$    & -5.068 & -4.116 & -3.103 & -2.152 & -2.9168 & 1.8307 & 0.2319 & 0.9949 & [-1.41, 0.37] \\
    2.05  & -1.613 & -1.189 & -0.394 & $-$    & $-$    & 1.1427 & 2.3617 & 0.5022 & 1     & [-2.15, -0.78] & 3.08  & $-$    & -5.093 & -4.135 & -3.133 & -2.178 & -2.9426 & 1.8238 & 0.2281 & 0.9944 & [-1.41, 0.37] \\
    2.06  & -1.680 & -1.256 & -0.464 & $-$    & $-$    & 1.0672 & 2.3528 & 0.5000 & 1     & [-2.15, -0.78] & 3.09  & $-$    & -5.117 & -4.154 & -3.163 & -2.204 & -2.9682 & 1.8171 & 0.2244 & 0.9939 & [-1.41, 0.37] \\
    2.07  & -1.746 & -1.322 & -0.533 & $-$    & $-$    & 0.9924 & 2.3433 & 0.4976 & 1     & [-2.15, -0.78] & 3.10  & $-$    & -5.142 & -4.172 & -3.194 & -2.230 & -2.9938 & 1.8106 & 0.2210 & 0.9933 & [-1.41, 0.37] \\
    2.08  & -1.811 & -1.388 & -0.601 & $-$    & $-$    & 0.9184 & 2.3333 & 0.4949 & 1     & [-2.15, -0.78] & 3.11  & $-$    & -5.167 & $-$    & -3.224 & -2.256 & -3.1378 & 2.1872 & 0.5307 & 1     & [-1.41, 0.37] \\
    2.09  & -1.875 & -1.452 & -0.667 & $-$    & $-$    & 0.8451 & 2.3228 & 0.4920 & 1     & [-2.15, -0.78] & 3.12  & $-$    & -5.191 & $-$    & -3.255 & -2.281 & -3.1680 & 2.1964 & 0.5401 & 1     & [-1.41, 0.37] \\
    2.10  & -1.938 & -1.515 & -0.733 & $-$    & $-$    & 0.7726 & 2.3118 & 0.4889 & 1     & [-2.15, -0.78] & 3.13  & $-$    & -5.215 & $-$    & -3.286 & -2.307 & -3.1982 & 2.2061 & 0.5500 & 1     & [-1.41, 0.37] \\
    2.11  & -2.000 & -1.577 & -0.798 & $-$    & $-$    & 0.7008 & 2.3003 & 0.4856 & 1     & [-2.15, -0.78] & 3.14  & $-$    & -5.240 & $-$    & -3.316 & -2.332 & -3.2285 & 2.2162 & 0.5602 & 1     & [-1.41, 0.37] \\
    2.12  & -2.062 & -1.638 & -0.862 & $-$    & $-$    & 0.6298 & 2.2884 & 0.4821 & 1     & [-2.15, -0.78] & 3.15  & $-$    & -5.264 & $-$    & -3.347 & -2.357 & -3.2588 & 2.2266 & 0.5708 & 1     & [-1.41, 0.37] \\
    2.13  & -2.123 & -1.699 & -0.925 & $-$    & $-$    & 0.5595 & 2.2760 & 0.4784 & 1     & [-2.15, -0.78] & 3.16  & $-$    & -5.287 & $-$    & -3.378 & -2.382 & -3.2891 & 2.2375 & 0.5818 & 1     & [-1.41, 0.37] \\
    2.14  & -2.183 & -1.758 & -0.987 & $-$    & $-$    & 0.4900 & 2.2632 & 0.4744 & 1     & [-2.15, -0.78] & 3.17  & $-$    & -5.311 & $-$    & -3.408 & -2.406 & -3.3195 & 2.2488 & 0.5931 & 1     & [-1.41, 0.37] \\
    2.15  & -2.242 & -1.816 & -1.048 & 0.960 & $-$    & 1.0614 & 3.202 & 0.7802 & 0.9973 & [-2.15, -0.04] & 3.18  & $-$    & -5.335 & $-$    & -3.439 & -2.431 & -3.3499 & 2.2605 & 0.6048 & 1     & [-1.41, 0.37] \\
    2.16  & -2.301 & -1.874 & -1.108 & 0.895 & $-$    & 0.9953 & 3.1913 & 0.7768 & 0.9973 & [-2.15, -0.04] & 3.19  & $-$    & -5.358 & $-$    & -3.470 & -2.455 & -3.3804 & 2.2725 & 0.6169 & 1     & [-1.41, 0.37] \\
    2.17  & -2.359 & -1.930 & -1.167 & 0.829 & $-$    & 0.9295 & 3.1795 & 0.7731 & 0.9973 & [-2.15, -0.04] & 3.20  & $-$    & -5.381 & $-$    & -3.501 & -2.479 & -3.4110 & 2.2848 & 0.6293 & 1     & [-1.41, 0.37] \\
    2.18  & -2.416 & -1.986 & -1.225 & 0.764 & $-$    & 0.8640 & 3.1667 & 0.7689 & 0.9973 & [-2.15, -0.04] & 3.21  & $-$    & -5.405 & $-$    & -3.532 & -2.504 & -3.4415 & 2.2975 & 0.6421 & 1     & [-1.41, 0.37] \\
    2.19  & -2.473 & -2.041 & -1.282 & 0.700 & $-$    & 0.7987 & 3.1528 & 0.7643 & 0.9972 & [-2.15, -0.04] & 3.22  & $-$    & -5.427 & $-$    & -3.564 & -2.528 & -3.4722 & 2.3105 & 0.6551 & 1     & [-1.41, 0.37] \\
    2.20  & -2.529 & -2.094 & -1.339 & 0.635 & $-$    & 0.7338 & 3.1381 & 0.7594 & 0.9972 & [-2.15, -0.04] & 3.23  & $-$    & -5.450 & $-$    & -3.595 & -2.552 & -3.5029 & 2.3237 & 0.6685 & 1     & [-1.41, 0.37] \\
    2.21  & -2.584 & -2.148 & -1.394 & 0.572 & $-$    & 0.6693 & 3.1225 & 0.7541 & 0.9972 & [-2.15, -0.04] & 3.24  & $-$    & -5.473 & $-$    & -3.626 & -2.575 & -3.5337 & 2.3372 & 0.6822 & 1     & [-1.41, 0.37] \\
    2.22  & -2.639 & -2.200 & -1.449 & 0.508 & $-$    & 0.6052 & 3.1062 & 0.7485 & 0.9972 & [-2.15, -0.04] & 3.25  & $-$    & -5.495 & $-$    & -3.657 & -2.599 & -3.5645 & 2.3510 & 0.6962 & 1     & [-1.41, 0.37] \\
    2.23  & -2.694 & -2.251 & -1.503 & 0.445 & $-$    & 0.5414 & 3.0890 & 0.7425 & 0.9972 & [-2.15, -0.04] & 3.26  & $-$    & -5.517 & $-$    & -3.689 & -2.623 & -3.5953 & 2.3650 & 0.7105 & 1     & [-1.41, 0.37] \\
    2.24  & -2.747 & -2.302 & -1.556 & 0.382 & $-$    & 0.4781 & 3.0712 & 0.7363 & 0.9972 & [-2.15, -0.04] & 3.27  & $-$    & -5.539 & $-$    & -3.720 & -2.647 & -3.6262 & 2.3792 & 0.7250 & 1     & [-1.41, 0.37] \\
    2.25  & -2.800 & -2.352 & -1.608 & 0.320 & $-$    & 0.4152 & 3.0527 & 0.7298 & 0.9972 & [-2.15, -0.04] & 3.28  & $-$    & -5.561 & $-$    & -3.752 & -2.670 & -3.6572 & 2.3935 & 0.7398 & 1     & [-1.41, 0.37] \\
    2.26  & -2.853 & -2.401 & -1.660 & 0.258 & $-$    & 0.3527 & 3.0337 & 0.7230 & 0.9972 & [-2.15, -0.04] & 3.29  & $-$    & -5.583 & $-$    & -3.783 & -2.694 & -3.6881 & 2.4080 & 0.7548 & 1     & [-1.41, 0.37] \\
    2.27  & -2.905 & -2.450 & -1.710 & 0.197 & $-$    & 0.2908 & 3.0140 & 0.7160 & 0.9972 & [-2.15, -0.04] & 3.30  & $-$    & -5.604 & $-$    & -3.815 & -2.717 & -3.7192 & 2.4227 & 0.7700 & 1     & [-1.41, 0.37] \\
    2.28  & -2.957 & -2.498 & -1.760 & 0.136 & $-$    & 0.2293 & 2.9939 & 0.7087 & 0.9972 & [-2.15, -0.04] & 3.31  & $-$    & -5.625 & $-$    & -3.846 & -2.741 & -3.7502 & 2.4374 & 0.7854 & 1     & [-1.41, 0.37] \\
    2.29  & -3.008 & -2.545 & -1.809 & 0.076 & $-$    & 0.1683 & 2.9732 & 0.7012 & 0.9972 & [-2.15, -0.04] & 3.32  & $-$    & -5.646 & $-$    & -3.878 & -2.764 & -3.7812 & 2.4522 & 0.8010 & 1     & [-1.41, 0.37] \\
    2.30  & -3.059 & -2.591 & -1.857 & 0.016 & $-$    & 0.1079 & 2.9522 & 0.6935 & 0.9972 & [-2.15, -0.04] & 3.33  & $-$    & -5.667 & $-$    & -3.910 & -2.788 & -3.8123 & 2.4671 & 0.8168 & 1     & [-1.41, 0.37] \\
    2.31  & -3.109 & -2.637 & -1.905 & -0.043 & $-$    & 0.0480 & 2.9307 & 0.6856 & 0.9972 & [-2.15, -0.04] & 3.34  & $-$    & -5.688 & $-$    & -3.941 & -2.811 & -3.8433 & 2.4820 & 0.8327 & 1     & [-1.41, 0.37] \\
    2.32  & -3.159 & -2.683 & -1.952 & -0.102 & $-$    & -0.1140 & 2.9089 & 0.6774 & 0.9972 & [-2.15, -0.04] & 3.35  & $-$    & -5.708 & $-$    & -3.973 & -2.834 & -3.8744 & 2.4969 & 0.8487 & 1     & [-1.41, 0.37] \\
    2.33  & -3.209 & -2.727 & -1.998 & -0.160 & $-$    & -0.0702 & 2.8868 & 0.6692 & 0.9972 & [-2.15, -0.04] & 3.36  & $-$    & -5.728 & $-$    & -4.004 & -2.858 & -3.9054 & 2.5117 & 0.8647 & 1     & [-1.41, 0.37] \\
    2.34  & -3.257 & -2.771 & -2.043 & -0.217 & $-$    & -0.1285 & 2.8644 & 0.6607 & 0.9972 & [-2.15, -0.04] & 3.37  & $-$    & -5.747 & $-$    & -4.036 & -2.881 & -3.9364 & 2.5264 & 0.8809 & 1     & [-1.41, 0.37] \\
    2.35  & -3.306 & -2.814 & -2.088 & -0.274 & $-$    & -0.1862 & 2.8418 & 0.6521 & 0.9972 & [-2.15, -0.04] & 3.38  & $-$    & -5.767 & $-$    & -4.068 & -2.904 & -3.9673 & 2.5411 & 0.8971 & 1     & [-1.41, 0.37] \\
    2.36  & -3.354 & -2.857 & -2.132 & -0.331 & $-$    & -0.2433 & 2.8189 & 0.6434 & 0.9972 & [-2.15, -0.04] & 3.39  & $-$    & -5.786 & $-$    & -4.099 & -2.928 & -3.9982 & 2.5556 & 0.9133 & 1     & [-1.41, 0.37] \\
    2.37  & -3.402 & -2.900 & -2.176 & -0.386 & $-$    & -0.2998 & 2.7959 & 0.6345 & 0.9973 & [-2.15, -0.04] & 3.40  & $-$    & -5.805 & $-$    & -4.130 & -2.951 & -4.0290 & 2.5699 & 0.9294 & 1     & [-1.41, 0.37] \\
    2.38  & -3.450 & -2.941 & -2.218 & -0.442 & $-$    & -0.3558 & 2.7727 & 0.6255 & 0.9973 & [-2.15, -0.04] & 3.41  & $-$    & -5.823 & $-$    & -4.162 & -2.974 & -4.0597 & 2.5839 & 0.9455 & 1     & [-1.41, 0.37] \\
    2.39  & -3.497 & -2.982 & -2.260 & -0.496 & $-$    & -0.4111 & 2.7494 & 0.6164 & 0.9973 & [-2.15, -0.04] & 3.42  & $-$    & -5.841 & $-$    & -4.193 & -2.997 & -4.0903 & 2.5977 & 0.9616 & 1     & [-1.41, 0.37] \\
    2.40  & -3.543 & -3.023 & -2.302 & -0.550 & $-$    & -0.4658 & 2.7259 & 0.6072 & 0.9973 & [-2.15, -0.04] & 3.43  & $-$    & -5.859 & $-$    & -4.224 & -3.021 & -4.1208 & 2.6112 & 0.9775 & 1     & [-1.41, 0.37] \\
    2.41  & -3.590 & -3.063 & -2.343 & -0.604 & $-$    & -0.5200 & 2.7024 & 0.5980 & 0.9974 & [-2.15, -0.04] & 3.44  & $-$    & -5.877 & $-$    & -4.254 & -3.044 & -4.1511 & 2.6244 & 0.9932 & 1     & [-1.41, 0.37] \\
    2.42  & -3.636 & -3.103 & -2.383 & -0.656 & $-$    & -0.5735 & 2.6789 & 0.5886 & 0.9974 & [-2.15, -0.04] & 3.45  & $-$    & -5.894 & $-$    & -4.285 & -3.067 & -4.1813 & 2.6372 & 1.0088 & 1     & [-1.41, 0.37] \\
    2.43  & -3.681 & -3.142 & -2.422 & -0.709 & $-$    & -0.6264 & 2.6554 & 0.5792 & 0.9974 & [-2.15, -0.04] & 3.46  & $-$    & -5.911 & $-$    & -4.316 & -3.091 & -4.2112 & 2.6495 & 1.0241 & 1     & [-1.41, 0.37] \\
    2.44  & -3.727 & -3.181 & -2.461 & -0.760 & $-$    & -0.6787 & 2.6318 & 0.5697 & 0.9975 & [-2.15, -0.04] & 3.47  & $-$    & -5.927 & $-$    & -4.346 & -3.114 & -4.2410 & 2.6613 & 1.0392 & 1     & [-1.41, 0.37] \\
    2.45  & -3.771 & -3.219 & -2.500 & -0.811 & 0.619 & -0.6056 & 2.8777 & 0.6641 & 0.9974 & [-2.15, 0.37] & 3.48  & $-$    & -5.944 & $-$    & -4.376 & -3.137 & -4.2706 & 2.6726 & 1.0540 & 1     & [-1.41, 0.37] \\
    2.46  & -3.816 & -3.257 & -2.538 & -0.862 & 0.544 & -0.6626 & 2.8415 & 0.6495 & 0.9976 & [-2.15, 0.37] & 3.49  & $-$    & -5.959 & $-$    & -4.405 & -3.161 & -4.2998 & 2.6833 & 1.0684 & 1     & [-1.41, 0.37] \\
    2.47  & -3.860 & -3.294 & -2.575 & -0.911 & 0.470 & -0.7186 & 2.8060 & 0.6353 & 0.9977 & [-2.15, 0.37] & 3.50  & $-$    & -5.975 & $-$    & -4.435 & -3.184 & -4.3288 & 2.6934 & 1.0824 & 1     & [-1.41, 0.37] \\
    2.48  & -3.904 & -3.331 & -2.611 & -0.961 & 0.397 & -0.7736 & 2.7714 & 0.6213 & 0.9978 & [-2.15, 0.37] & 3.51  & $-$    & -5.990 & $-$    & -4.464 & -3.207 & -4.3575 & 2.7028 & 1.0960 & 1     & [-1.41, 0.37] \\
    2.49  & -3.948 & -3.367 & -2.648 & -1.009 & 0.326 & -0.8278 & 2.7377 & 0.6076 & 0.9979 & [-2.15, 0.37] & 3.52  & $-$    & -6.004 & $-$    & -4.493 & -3.231 & -4.3859 & 2.7114 & 1.1091 & 1     & [-1.41, 0.37] \\
    2.50  & -3.992 & -3.403 & -2.683 & -1.057 & 0.256 & -0.8810 & 2.7047 & 0.5941 & 0.9980 & [-2.15, 0.37] & 3.53  & $-$    & -6.018 & $-$    & -4.521 & -3.254 & -4.4139 & 2.7192 & 1.1216 & 1     & [-1.41, 0.37] \\
    2.51  & -4.035 & -3.439 & -2.718 & -1.104 & 0.188 & -0.9332 & 2.6725 & 0.5809 & 0.9981 & [-2.15, 0.37] & 3.54  & $-$    & -6.032 & $-$    & -4.549 & -3.278 & -4.4415 & 2.7262 & 1.1336 & 1     & [-1.41, 0.37] \\
    2.52  & -4.077 & -3.475 & -2.753 & -1.151 & 0.121 & -0.9846 & 2.6411 & 0.5679 & 0.9982 & [-2.15, 0.37] & 3.55  & $-$    & -6.045 & $-$    & -4.576 & -3.301 & -4.4687 & 2.7322 & 1.1449 & 1     & [-1.41, 0.37] \\
    2.53  & -4.120 & -3.510 & -2.787 & -1.198 & 0.055 & -1.0351 & 2.6105 & 0.5552 & 0.9982 & [-2.15, 0.37] & 3.56  & $-$    & -6.058 & $-$    & -4.603 & -3.324 & -4.4954 & 2.7373 & 1.1555 & 1     & [-1.41, 0.37] \\
    2.54  & -4.162 & -3.544 & -2.821 & -1.243 & -0.009 & -1.0848 & 2.5806 & 0.5427 & 0.9983 & [-2.15, 0.37] & 3.57  & $-$    & -6.070 & $-$    & -4.629 & -3.348 & -4.5217 & 2.7412 & 1.1654 & 1     & [-1.41, 0.37] \\
    2.55  & -4.204 & -3.578 & -2.854 & -1.288 & -0.072 & -1.1336 & 2.5515 & 0.5303 & 0.9984 & [-2.15, 0.37] & 3.58  & $-$    & -6.082 & $-$    & -4.655 & -3.371 & -4.5474 & 2.7441 & 1.1745 & 1     & [-1.41, 0.37] \\
    2.56  & -4.245 & -3.612 & -2.886 & -1.333 & -0.134 & -1.1816 & 2.5231 & 0.5184 & 0.9984 & [-2.15, 0.37] & 3.59  & $-$    & -6.093 & $-$    & -4.680 & -3.395 & -4.5725 & 2.7458 & 1.1826 & 1     & [-1.41, 0.37] \\
    2.57  & -4.287 & -3.646 & -2.919 & -1.377 & -0.195 & -1.2287 & 2.4954 & 0.5067 & 0.9985 & [-2.15, 0.37] & 3.60  & $-$    & -6.104 & $-$    & -4.705 & -3.418 & -4.5971 & 2.7462 & 1.1899 & 1     & [-1.41, 0.37] \\
    2.58  & -4.327 & -3.679 & -2.950 & -1.420 & -0.254 & -1.2751 & 2.4684 & 0.4951 & 0.9986 & [-2.15, 0.37] & 3.61  & $-$    & -6.114 & $-$    & -4.729 & -3.441 & -4.6210 & 2.7452 & 1.1961 & 1     & [-1.41, 0.37] \\
    2.59  & -4.368 & -3.712 & -2.981 & -1.463 & -0.312 & -1.3207 & 2.4422 & 0.4838 & 0.9986 & [-2.15, 0.37] & 3.62  & $-$    & -6.123 & $-$    & -4.752 & -3.465 & -4.6442 & 2.7429 & 1.2013 & 1     & [-1.41, 0.37] \\
    2.60  & -4.408 & -3.745 & -3.012 & -1.505 & -0.369 & -1.3656 & 2.4166 & 0.4728 & 0.9987 & [-2.15, 0.37] & 3.63  & $-$    & -6.132 & $-$    & -4.774 & -3.488 & -4.6667 & 2.7391 & 1.2053 & 1     & [-1.41, 0.37] \\
    2.61  & -4.448 & -3.778 & -3.043 & -1.547 & -0.425 & -1.4096 & 2.3917 & 0.4619 & 0.9987 & [-2.15, 0.37] & 3.64  & $-$    & -6.141 & $-$    & -4.796 & -3.512 & -4.6885 & 2.7337 & 1.2082 & 1     & [-1.41, 0.37] \\
    2.62  & -4.488 & -3.810 & -3.073 & -1.588 & -0.479 & -1.4530 & 2.3675 & 0.4513 & 0.9988 & [-2.15, 0.37] & 3.65  & $-$    & -6.149 & $-$    & -4.817 & -3.535 & -4.7094 & 2.7266 & 1.2097 & 1     & [-1.41, 0.37] \\
    2.63  & -4.527 & -3.842 & -3.102 & -1.629 & -0.533 & -1.4957 & 2.3439 & 0.4409 & 0.9988 & [-2.15, 0.37] & 3.66  & $-$    & -6.156 & $-$    & -4.836 & -3.558 & -4.7295 & 2.7178 & 1.2099 & 1     & [-1.41, 0.37] \\
    2.64  & -4.566 & -3.873 & -3.131 & -1.669 & -0.586 & -1.5376 & 2.3209 & 0.4307 & 0.9988 & [-2.15, 0.37] & 3.67  & $-$    & -6.163 & $-$    & -4.855 & -3.582 & -4.7486 & 2.7072 & 1.2087 & 1     & [-1.41, 0.37] \\
    2.65  & -4.605 & -3.905 & -3.160 & -1.709 & -0.637 & -1.5789 & 2.2986 & 0.4208 & 0.9989 & [-2.15, 0.37] & 3.68  & $-$    & -6.169 & $-$    & -4.873 & -3.605 & -4.7669 & 2.6947 & 1.2060 & 1     & [-1.41, 0.37] \\
    2.66  & -4.643 & -3.936 & -3.189 & -1.748 & -0.688 & -1.6195 & 2.2770 & 0.4111 & 0.9989 & [-2.15, 0.37] & 3.69  & $-$    & -6.174 & $-$    & -4.889 & -3.628 & -4.7841 & 2.6802 & 1.2016 & 1     & [-1.41, 0.37] \\
    2.67  & -4.681 & -3.967 & -3.217 & -1.787 & -0.737 & -1.6595 & 2.2559 & 0.4016 & 0.9990 & [-2.15, 0.37] & 3.70  & $-$    & $-$    & $-$    & -4.905 & -3.651 & -4.7826 & 3.0581 & $-$    & 1     & [-0.04, 0.37] \\
    2.68  & -4.719 & -3.997 & -3.244 & -1.826 & -0.786 & -1.6989 & 2.2354 & 0.3924 & 0.9990 & [-2.15, 0.37] & 3.71  & $-$    & $-$    & $-$    & -4.919 & -3.674 & -4.7978 & 3.0367 & $-$    & 1     & [-0.04, 0.37] \\
    2.69  & -4.756 & -4.028 & -3.272 & -1.864 & -0.833 & -1.7377 & 2.2156 & 0.3834 & 0.9990 & [-2.15, 0.37] & 3.72  & $-$    & $-$    & $-$    & -4.932 & -3.697 & -4.8118 & 3.0125 & $-$    & 1     & [-0.04, 0.37] \\
    2.70  & -4.793 & -4.058 & -3.299 & -1.901 & -0.880 & -1.7758 & 2.1963 & 0.3746 & 0.9991 & [-2.15, 0.37] & 3.73  & $-$    & $-$    & $-$    & -4.944 & -3.720 & -4.8247 & 2.9852 & $-$    & 1     & [-0.04, 0.37] \\
    2.71  & -4.829 & -4.088 & -3.325 & -1.938 & -0.926 & -1.8134 & 2.1776 & 0.3660 & 0.9991 & [-2.15, 0.37] & 3.74  & $-$    & $-$    & $-$    & -4.954 & -3.743 & -4.8363 & 2.9547 & $-$    & 1     & [-0.04, 0.37] \\
    2.72  & -4.865 & -4.118 & -3.351 & -1.975 & -0.971 & -1.8504 & 2.1595 & 0.3577 & 0.9991 & [-2.15, 0.37] & 3.75  & $-$    & $-$    & $-$    & -4.963 & -3.766 & -4.8466 & 2.9209 & $-$    & 1     & [-0.04, 0.37] \\
    2.73  & -4.901 & -4.148 & -3.377 & -2.011 & -1.015 & -1.8869 & 2.1420 & 0.3496 & 0.9992 & [-2.15, 0.37] & 3.76  & $-$    & $-$    & $-$    & -4.971 & -3.789 & -4.8555 & 2.8837 & $-$    & 1     & [-0.04, 0.37] \\
    2.74  & -4.936 & -4.177 & -3.403 & -2.047 & -1.058 & -1.9229 & 2.1250 & 0.3417 & 0.9992 & [-2.15, 0.37] & 3.77  & $-$    & $-$    & $-$    & -4.977 & -3.811 & -4.8631 & 2.8429 & $-$    & 1     & [-0.04, 0.37] \\
    2.75  & -4.971 & -4.206 & -3.428 & -2.083 & -1.101 & -1.9583 & 2.1085 & 0.3341 & 0.9992 & [-2.15, 0.37] & 3.78  & $-$    & $-$    & $-$    & -4.981 & -3.834 & -4.8691 & 2.7984 & $-$    & 1     & [-0.04, 0.37] \\
    2.76  & -5.005 & -4.235 & -3.453 & -2.118 & -1.142 & -1.9933 & 2.0926 & 0.3267 & 0.9992 & [-2.15, 0.37] & 3.79  & $-$    & $-$    & $-$    & -4.984 & -3.856 & -4.8736 & 2.7500 & $-$    & 1     & [-0.04, 0.37] \\
    2.77  & -5.039 & -4.264 & -3.478 & -2.153 & -1.183 & -2.0278 & 2.0772 & 0.3196 & 0.9993 & [-2.15, 0.37] & 3.80  & $-$    & $-$    & $-$    & -4.984 & -3.878 & -4.8765 & 2.6975 & $-$    & 1     & [-0.04, 0.37] \\
    \hline
    \end{tabular}
  \label{tab:addlabel}
}
\end{table*}

\begin{figure}
\begin{center}
\includegraphics[scale=0.35, angle=0]{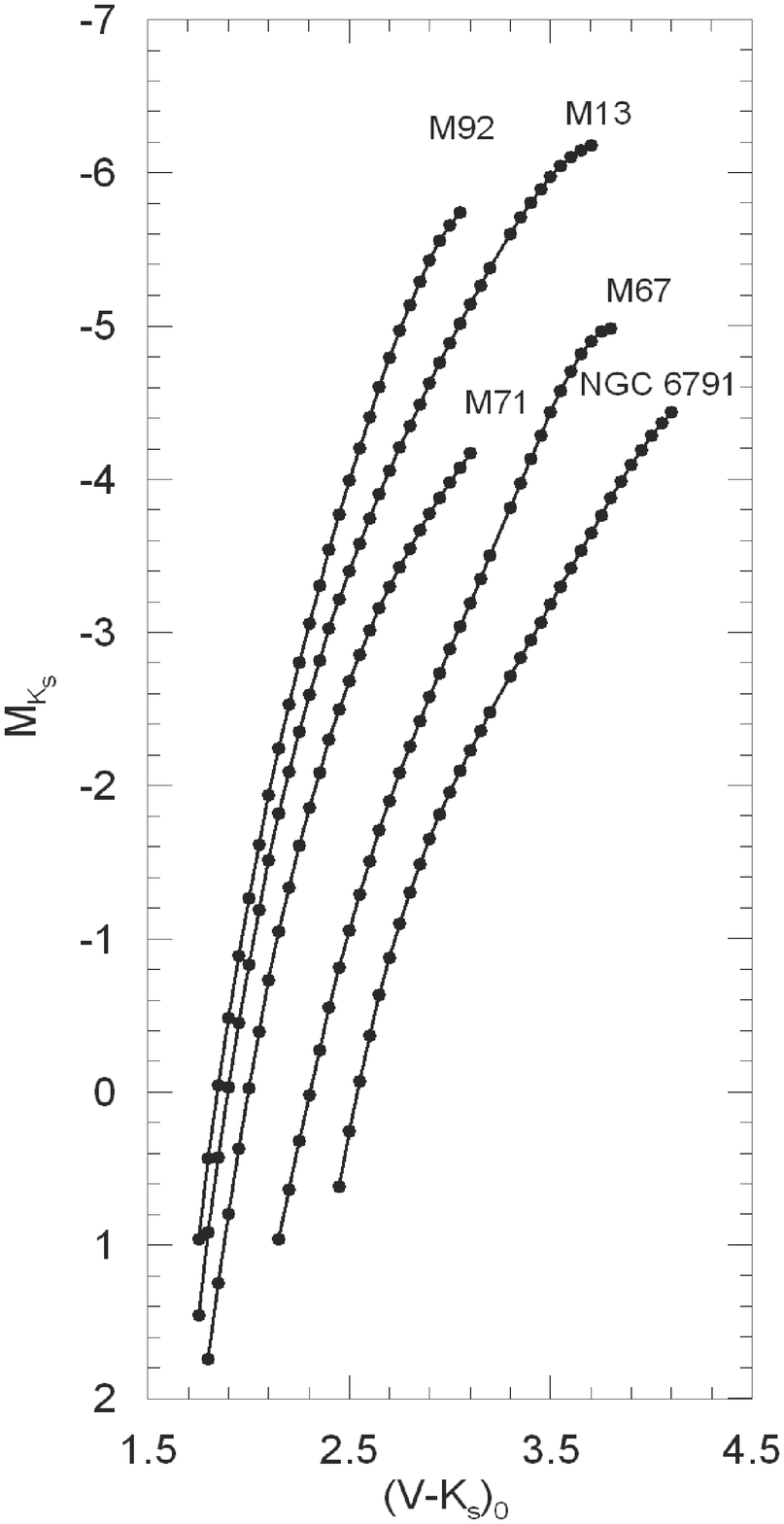} 
\caption[] {$M_{K_{s}}\times(V-K_{s})_{0}$ colour-absolute magnitude diagrams for five clusters used for the absolute magnitude calibration.} 
\label{his:col}
\end{center}
\end {figure}

\begin{figure}
\begin{center}
\includegraphics[scale=0.35, angle=0]{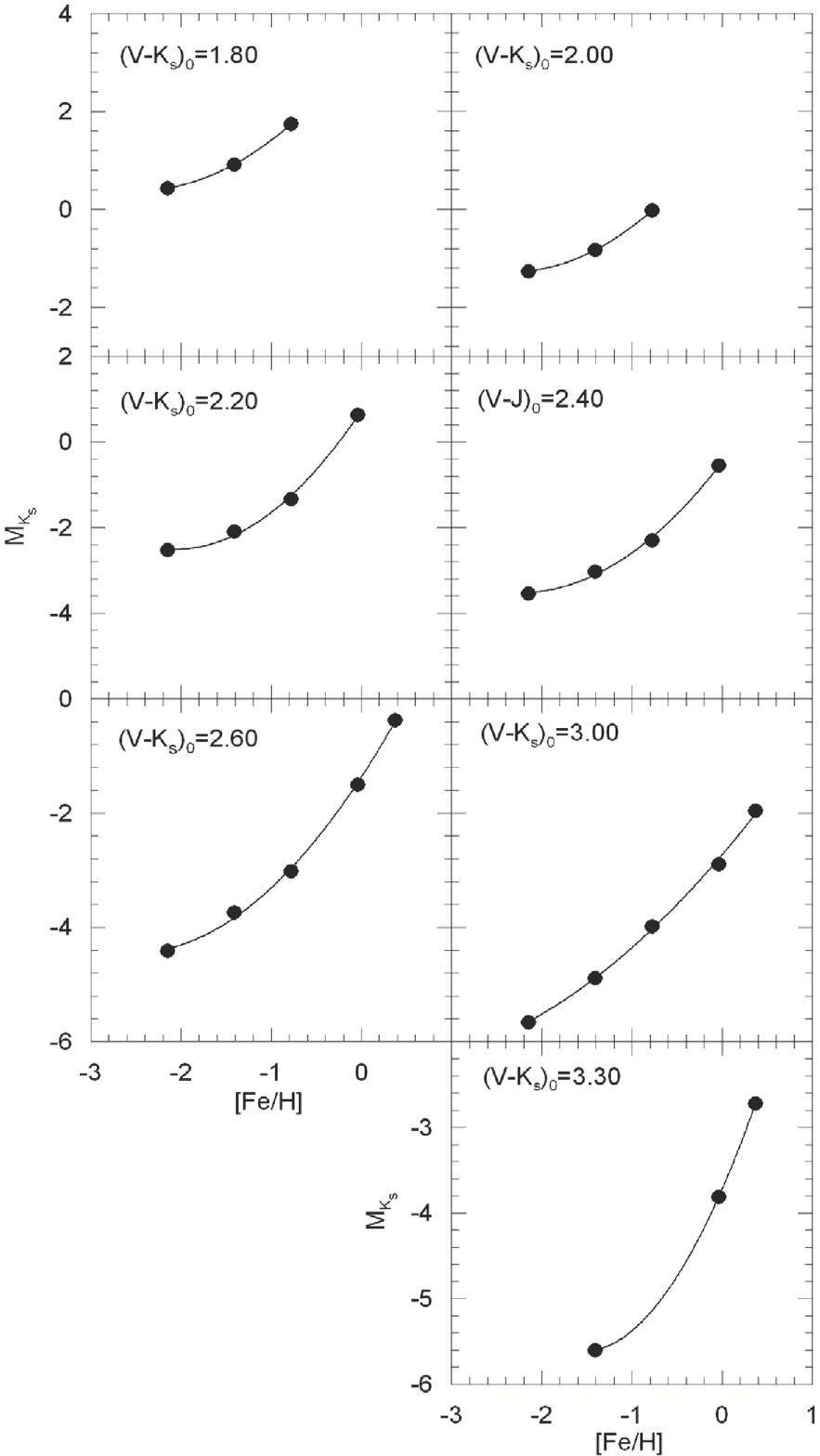} 
\caption[] {Calibration of the absolute magnitude $M_{K_{s}}$ as a function of metallicity $[Fe/H]$ for seven colour-indices.} 
\label{his:col}
\end{center}
\end {figure}

\subsubsection{Calibration of $M_{K_s}$ in terms of Metallicity}
We estimated the $M_{K_s}$ absolute magnitudes for the $(V-K_{s})_{0}$ colours given in Table 11 for the clusters M92, M13, M71, M67, and NGC 6791 by combining the $K_{{s}_0}$ apparent magnitudes evaluated by Eq. (5) and the true distance modulus ($\mu_{0}$) of the cluster in question, i.e.

\begin{eqnarray}
M_{{K}_s}=K_{{s}_0}-\mu_{0}.     
\end{eqnarray}
Then, we plotted the absolute magnitudes versus $(V-K_{s})_{0}$ colours. Fig. 5 shows a similar trend of Fig. 3, i.e. the absolute magnitude increases (algebraically) with increasing metallicity and decreasing colour.

We fitted the $M_{K_s}$ absolute magnitudes to the corresponding $[Fe/H]$ metallicity for the colour indices $(V-K_{s})_{0}=$ 1.80, 2.00, 2.20, 2.40, 2.60, 3.00, 3.30 mag just for the exhibition of the procedure. The results are given in Table 12 and Fig. 6. We can extend this fitting to a larger $(V-K_{s})_{0}$ interval for which the sample clusters are defined. The $(V-K_{s})_{0}$ domains of the clusters are different. Hence, we adopted this interval as 1.75 $\leq(V-K_{s})_{0} \leq$ 3.80 mag where at least two clusters are defined. However, the  range of the interval where only two clusters are defined is rather limited, i.e. $1.75 \leq(V-K_{s})_{0}\leq 1.78$ and $3.70 \leq(V-K_{s})_{0} \leq 3.80$ mag. The common domain of all the clusters is $2.45\leq(V-K_{s})_{0} \leq 3.04$ mag. The general form of the equation for the calibration is as follows:

\begin{eqnarray}
M_{K_s}=d_{0} + d_{1}X + d_{2}X^{2}     
\end{eqnarray}
where $X=\lbrack Fe/H \rbrack$. $M_{K_s}$ could be fitted in terms of metallicity by a quadratic polynomial for all $(V-K_{s})_{0}$ colour interval with high correlation coefficient, $R^{2} \geq$ 0.9972, except two intervals with small ranges, i.e. a linear fitting was sufficient for the colour intervals $1.75\leq(V-K_{s})_{0}\leq1.78$ and $3.70 \leq(V-K_{s})_{0} \leq 3.80$ mag where only two clusters are defined. The absolute magnitudes estimated via Eq. (9) for 206 $(V-K_{s})_{0}$ colour indices and the corresponding $d_{i}~(i=0$, 1, 2) coefficients are given in Table 13. However, the diagrams for the calibrations are not given in the paper for avoiding space consuming.     

The calibration of the absolute magnitude $M_{K_s}$ (and $M_J$) in terms of $[Fe/H]$ is carried out in steps of 0.01 mag. Small step is necessary to isolate an observational error on the colour $V-K_{s}$ (and $V-J$) plus an error due to reddening. The origin of the mentioned errors is the trend of the RGB. Although they are not as steep as in BV- and gr- photometries, a small error in $V-K_{s}$ (and $V-J$) implies a large change in the absolute magnitude. 

Iron abundance, $[Fe/H]$, is not the only parameter determining the chemistry of the star but also alpha enhancement, $[\alpha /Fe]$, is surely important. However, as it was stated in Paper I, there is a correlation between two sets of abundances. Hence, we do not expect any considerable change in the numerical values of the estimated absolute magnitudes in the case of additional of the alpha enhancement 
term in Eq. (7) and Eq. (9).

\subsection{Application of the Method}
\subsubsection{Application of the $M_J\times[Fe/H]$ Calibration}
We applied the $M_J\times [Fe/H]$ calibration to the clusters M5 and M68. The reason of choosing clusters instead of field giants is that clusters provide absolute magnitudes for comparison with the ones estimated by means of our method. The $J$ magnitudes and $V-J$ colours for the cluster M5 are taken from \cite{Brasseur10}. 2MASS photometric data are not available for the cluster M68. Hence, the $J_0$ magnitudes and $(V-J)_{0}$ colours are provided by transformation of the $V$, $B-V$, $V-I$ data of \cite{Walker94} to $J_{0}$ magnitudes and $(V-J)_{0}$ colours by means of the Eq. (1) and the procedure explained in Section 2.1. The data for the clusters are given in Table 14. Two references are given for the cluster M5. The colour excess $E(B-V)$ and the true distance modulus $\mu_{0}$  refer to the first author, whereas the metallicity which was tested in Paper I is taken from the second author. The $J_{0}$ magnitudes and $(V-J)_{0}$ colours as well as the original $V$, $B-V$, $V-I$ data are given in Table 15.  

\begin{table}[h]
  \center
  \caption{Data for the clusters used for the application of the procudure.}
    \begin{tabular}{ccccc}
    \hline
    Cluster & $E(B-V)$ & $\mu_{0}$ & $[Fe/H]$ & Ref.\\
            &  (mag)        & (mag)     & (dex) \\
    \hline
    M5      & 0.038 & 14.330 & -1.17 & (1), (2)  \\
    M68     & 0.060 & 14.994 & -2.01 & (3) \\
    NGC 188 & 0.087 & 11.130 & -0.01 & (4), (5) \\
    \hline
    \end{tabular}\\
(1) \cite{Brasseur10}, (2) \cite{Sandquist96}, (3) \cite{VC03}, (4) \cite{Stetson04}, (5) \cite{Meibom09}.
  \label{tab:addlabel}
\end{table}

\begin{table}[h]
\setlength{\tabcolsep}{3pt}
  \center
\scriptsize{
  \caption{$J_0\times(V-J)_0$ fiducial giant sequences for the clusters 
M5 and M68 used in the application of the procedure.}
    \begin{tabular}{cccccc}
    \hline
Cluster&   $V$   & $V-J$ & $J$& $(V-J)_{0}$ & $J_{0}$\\
    \hline
M5 & 12.31 & 2.59  & 9.72  & 2.505 &  9.687 \\
   & 12.50 & 2.47  & 10.03 & 2.385 &  9.997 \\
   & 12.68 & 2.36  & 10.32 & 2.275 & 10.287 \\
   & 12.86 & 2.27  & 10.59 & 2.185 & 10.557 \\
   & 13.09 & 2.18  & 10.91 & 2.095 & 10.877 \\
   & 13.29 & 2.12  & 11.17 & 2.035 & 11.137 \\
   & 13.49 & 2.06  & 11.43 & 1.975 & 11.397 \\
   & 13.71 & 2.00  & 11.71 & 1.915 & 11.677 \\
   & 13.91 & 1.95  & 11.96 & 1.865 & 11.927 \\
   & 14.14 & 1.89  & 12.25 & 1.805 & 12.217 \\
   & 14.35 & 1.84  & 12.51 & 1.755 & 12.477 \\
   & 14.57 & 1.79  & 12.78 & 1.705 & 12.747 \\
   & 14.80 & 1.74  & 13.06 & 1.655 & 13.027 \\
   & 15.04 & 1.70  & 13.34 & 1.615 & 13.307 \\
   & 15.27 & 1.67  & 13.60 & 1.585 & 13.567 \\
   & 15.55 & 1.63  & 13.92 & 1.545 & 13.887 \\
   & 15.81 & 1.60  & 14.21 & 1.515 & 14.177 \\
   & 16.05 & 1.57  & 14.48 & 1.485 & 14.447 \\
   & 16.33 & 1.54  & 14.79 & 1.455 & 14.757 \\
   & 16.57 & 1.51  & 15.06 & 1.425 & 15.027 \\
   & 16.83 & 1.50  & 15.33 & 1.415 & 15.297 \\
   & 17.08 & 1.48  & 15.60 & 1.395 & 15.567 \\
   & 17.32 & 1.47  & 15.85 & 1.385 & 15.817 \\
   & 17.54 & 1.46  & 16.08 & 1.375 & 16.047 \\
M68& 1.234 & 1.196 & 12.497 & 2.203 & 10.294 \\
   & 1.202 & 1.154 & 12.844 & 2.146 & 10.698 \\
   & 1.047 & 0.951 & 13.498 & 1.868 & 11.630 \\
   & 0.965 & 0.855 & 14.143 & 1.733 & 12.410 \\
   & 0.905 & 0.775 & 14.744 & 1.624 & 13.120 \\
   & 0.863 & 0.732 & 15.209 & 1.561 & 13.648 \\
   & 0.803 & 0.669 & 15.775 & 1.471 & 14.304 \\
   & 0.792 & 0.646 & 16.228 & 1.442 & 14.786 \\
   & 0.752 & 0.626 & 16.645 & 1.405 & 15.240 \\
   & 0.754 & 0.619 & 16.915 & 1.398 & 15.517 \\
   & 0.739 & 0.609 & 17.215 & 1.382 & 15.833 \\
    \hline
    \end{tabular}
  \label{tab:addlabel}
}
\end{table}

We evaluated the $M_{J}$ absolute magnitude by means of Eq. (7) for a 
set of $(V-J)_{0}$ colour indices where the clusters are defined. The 
results are given in Table 16. The columns give: (1) $(V-J)_{0}$ colour 
index, (2) $(M_J)_{cl}$, the absolute magnitude for a cluster estimated by 
its colour magnitude diagram, (3) $(M_{J})_{ev}$, the absolute magnitude 
estimated by the procedure, (4) $\Delta M_J$, absolute magnitude residuals. 
Also, the metallicity for each cluster is indicated near the name of the 
cluster. The differences between the absolute magnitudes estimated by the 
procedure presented in this study and the ones evaluated via the 
colour-magnitude diagrams of the clusters (the residuals) lie in a 
(relatively) short interval, i.e. -0.08 and +0.34 mag, and the range of 
94$\%$ of the absolute magnitude residuals is  only $0<M_J\leq 0.3$ mag. 
The mean and the standard deviation of (all) residuals are 
$<\Delta M_J>=0.137$ and $\sigma_{M_{J}}=0.080$ mag, respectively. The 
distribution of the residuals are given in Table 17 and Fig. 7. 

\begin{table}
\setlength{\tabcolsep}{2.5pt}
  \center
\scriptsize{
  \caption{$(M_J)_{ev}$ absolute magnitudes and $\Delta M$ residuals estimated 
by the procedure explained in our work. $(M_J)_{cl}$ denotes the absolute 
magnitude evaluated by means of the colour-magnitude diagram of the cluster.}
    \begin{tabular}{cccc|cccc}
    \hline
    $(V-J)_{0}$ & $(M_J)_{cl}$   & $(M_J)_{ev}$   & $\Delta M_J$ & $(V-J)_{0}$ & $(M_J)_{cl}$   & $(M_J)_{ev}$   & $\Delta M_J$\\
    \hline
    \multicolumn{4}{c|}{M5 ($[Fe/H]=$ -1.17 dex)} & \multicolumn{4}{c}{M5 (cont.) } \\
    \hline
    1.40  & 1.203 & 0.876 & 0.327 & 2.38  & -4.385 & -4.417 & 0.033 \\
    1.42  & 0.928 & 0.630 & 0.298 & 2.40  & -4.432 & -4.475 & 0.043 \\
    1.44  & 0.670 & 0.410 & 0.260 & 2.42  & -4.474 & -4.532 & 0.058 \\
    1.46  & 0.428 & 0.198 & 0.230 & 2.44  & -4.511 & -4.588 & 0.077 \\
    1.48  & 0.200 & -0.005 & 0.205 & 2.46  & -4.540 & -4.642 & 0.102 \\
    1.50  & -0.013 & -0.200 & 0.187 & 2.48  & -4.562 & -4.695 & 0.133 \\
    1.52  & -0.214 & -0.388 & 0.173 & 2.50  & -4.577 & -4.747 & 0.170 \\
\cline{5-8}1.54  & -0.403 & -0.568 & 0.164 & \multicolumn{4}{c}{M68 ($[Fe/H]=$ -2.01 dex)}  \\
\cline{5-8}1.56  & -0.581 & -0.741 & 0.159 & 1.40  & 0.437 & 0.095 & 0.342 \\
    1.58  & -0.749 & -0.947 & 0.198 & 1.42  & 0.140 & -0.146 & 0.286 \\
    1.60  & -0.907 & -1.099 & 0.192 & 1.44  & -0.136 & -0.373 & 0.236 \\
    1.62  & -1.056 & -1.245 & 0.188 & 1.46  & -0.394 & -0.588 & 0.195 \\
    1.64  & -1.198 & -1.385 & 0.188 & 1.48  & -0.633 & -0.792 & 0.159 \\
    1.66  & -1.332 & -1.520 & 0.189 & 1.50  & -0.855 & -0.986 & 0.131 \\
    1.68  & -1.459 & -1.650 & 0.191 & 1.52  & -1.063 & -1.170 & 0.108 \\
    1.70  & -1.580 & -1.775 & 0.195 & 1.54  & -1.256 & -1.346 & 0.090 \\
    1.72  & -1.696 & -1.920 & 0.224 & 1.56  & -1.436 & -1.513 & 0.077 \\
    1.74  & -1.807 & -2.027 & 0.220 & 1.58  & -1.604 & -1.690 & 0.086 \\
    1.76  & -1.913 & -2.131 & 0.217 & 1.60  & -1.761 & -1.839 & 0.078 \\
    1.78  & -2.016 & -2.231 & 0.215 & 1.62  & -1.909 & -1.982 & 0.073 \\
    1.80  & -2.115 & -2.329 & 0.214 & 1.64  & -2.047 & -2.119 & 0.071 \\
    1.82  & -2.211 & -2.423 & 0.212 & 1.66  & -2.178 & -2.250 & 0.072 \\
    1.84  & -2.304 & -2.514 & 0.210 & 1.68  & -2.301 & -2.376 & 0.075 \\
    1.86  & -2.396 & -2.603 & 0.207 & 1.70  & -2.417 & -2.498 & 0.081 \\
    1.88  & -2.485 & -2.689 & 0.204 & 1.72  & -2.529 & -2.614 & 0.086 \\
    1.90  & -2.573 & -2.773 & 0.200 & 1.74  & -2.635 & -2.728 & 0.094 \\
    1.92  & -2.659 & -2.854 & 0.195 & 1.76  & -2.736 & -2.839 & 0.103 \\
    1.94  & -2.744 & -2.933 & 0.189 & 1.78  & -2.835 & -2.946 & 0.112 \\
    1.96  & -2.828 & -3.010 & 0.181 & 1.80  & -2.929 & -3.051 & 0.121 \\
    1.98  & -2.912 & -3.085 & 0.173 & 1.82  & -3.022 & -3.153 & 0.131 \\
    2.00  & -2.995 & -3.158 & 0.163 & 1.84  & -3.112 & -3.251 & 0.140 \\
    2.02  & -3.077 & -3.229 & 0.152 & 1.86  & -3.200 & -3.348 & 0.148 \\
    2.04  & -3.159 & -3.298 & 0.139 & 1.88  & -3.287 & -3.442 & 0.156 \\
    2.06  & -3.240 & -3.365 & 0.125 & 1.90  & -3.373 & -3.535 & 0.162 \\
    2.08  & -3.321 & -3.431 & 0.110 & 1.92  & -3.458 & -3.625 & 0.167 \\
    2.10  & -3.402 & -3.496 & 0.094 & 1.94  & -3.542 & -3.714 & 0.172 \\
    2.12  & -3.482 & -3.559 & 0.077 & 1.96  & -3.626 & -3.800 & 0.174 \\
    2.14  & -3.561 & -3.620 & 0.059 & 1.98  & -3.709 & -3.885 & 0.176 \\
    2.16  & -3.640 & -3.680 & 0.040 & 2.00  & -3.792 & -3.968 & 0.176 \\
    2.18  & -3.717 & -3.739 & 0.021 & 2.02  & -3.875 & -4.049 & 0.174 \\
    2.20  & -3.794 & -3.796 & 0.002 & 2.04  & -3.957 & -4.127 & 0.170 \\
    2.22  & -3.869 & -3.887 & 0.018 & 2.06  & -4.039 & -4.205 & 0.165 \\
    2.24  & -3.943 & -3.907 & -0.036 & 2.08  & -4.121 & -4.280 & 0.159 \\
    2.26  & -4.014 & -3.960 & -0.054 & 2.10  & -4.202 & -4.352 & 0.150 \\
    2.28  & -4.084 & -4.012 & -0.072 & 2.12  & -4.281 & -4.422 & 0.141 \\
    2.30  & -4.151 & -4.075 & -0.076 & 2.14  & -4.360 & -4.489 & 0.129 \\
    2.32  & -4.215 & -4.237 & 0.022 & 2.16  & -4.437 & -4.554 & 0.117 \\
    2.34  & -4.275 & -4.298 & 0.023 & 2.18  & -4.512 & -4.616 & 0.103 \\
    2.36  & -4.332 & -4.358 & 0.026 & 2.20  & -4.585 & -4.673 & 0.088 \\
    \hline
    \end{tabular}
  \label{tab:addlabel}
}
\end{table}

\begin{table}[h]
\center
  \caption{Distribution of the residuals. $N$ denotes the number of stars.}
\begin{tabular}{ccc}
    \hline
    $\Delta M_J$ interval & $<\Delta M_J>$ & N \\
    \hline
    (-0.1, 0.0] & -0.059 & 4  \\
     (0.0, 0.1] &  0.060 & 26 \\
     (0.1, 0.2] &  0.158 & 50 \\
     (0.2, 0.3] &  0.229 & 15 \\
     (0.3, 0.4] &  0.335 & 2  \\
    \hline
    \end{tabular}
  \label{tab:addlabel}
\end{table}

\begin{figure}[h]
\begin{center}
\includegraphics[scale=0.4, angle=0]{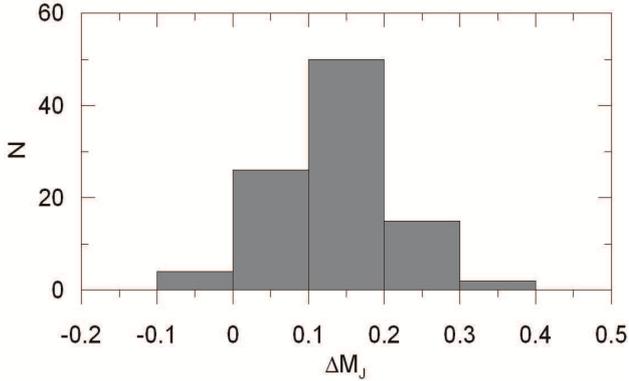}
\caption{Histogram of the residuals for $\Delta M_J$.}
\label{histogram}
\end{center}
\end{figure} 

\subsubsection{Application of the $M_{K_{s}}\times[Fe/H]$ Calibration}
NGC 1851 is the last cluster in \cite{Brasseur10} for which $K_{s}$ 
magnitudes and $V-K_{s}$ colours are available. However, the 
$M_{K_{s}}\times$ $(V-K_{s})_{0}$ colour-magnitude diagram of this 
cluster shows that the uncertainties in the data of this cluster are 
(relatively) large. Actually, the absolute magnitudes are fainter 
than the corresponding ones of cluster M71 for the colour interval 
$(V-K_{s})_{0}<2.5$ mag but brighter for $(V-K_{s})_{0}>2.5$ mag. 
This trend holds for distance modulus and colour excess in 
\cite{Brasseur10}, i.e. $\mu =$ 15.50 and $E(B-V)=0.034$ mag, as 
well as for alternative data such as $E(B-V)=0.02$ and 
$\mu_{0}=15.50\pm$0.20 mag of \cite{Saviane98}. If we regard the 
metallicity claimed by \cite{Brasseur10}, $[Fe/H]=-1.40$ dex, 
the absolute magnitude colour diagram of the cluster NGC 1851 
should coincide with the one of M13 ($[Fe/H]=-1.41$ dex). The 
alternative metallicity of \cite{Rosenberg99}, $[Fe/H]=-1.03\pm0.06$ 
dex, should replace the absolute magnitude colour diagram of NGC 1851 
between the diagrams of M13 and M71, which is not the case. 
$M_{K_{s}} \times (V-K_{s})_{0}$ diagrams for three clusters 
mentioned in this paragraph are not plotted in our paper for 
avoiding space consuming. 

\begin{table}
\setlength{\tabcolsep}{3pt}
  \center
\scriptsize{
\caption{$K_s\times (V-K_s)_0$ fiducial giant sequences for the 
clusters NGC 188 and M68 used in the application of the procedure.}
    \begin{tabular}{cccccc}
    \hline
 Cluster&    $V_{0}$ & $(B-V)_{0}$ & $(V-I)_{0}$ & $(V-K_{s})_{0}$ & $K_{{s}_0}$ \\
   \hline 
   NGC 188 &12.481 & 1.119 & 1.112 & 2.608 & 9.873 \\
            &13.031 & 1.052 & 1.063 & 2.474 & 10.557 \\
            &13.547 & 0.999 & 1.022 & 2.367 & 11.180 \\
            &14.069 & 0.954 & 0.991 & 2.277 & 11.792 \\
            &14.609 & 0.839 & 0.899 & 2.044 & 12.565 \\
            &14.786 & 0.819 & 0.885 & 2.004 & 12.782 \\

    M68     &12.497 & 1.196 & 1.234 & 3.040 & 9.457 \\
            &12.844 & 1.154 & 1.202 & 2.955 & 9.889 \\
            &13.498 & 0.951 & 1.047 & 2.546 & 10.952 \\
            &14.143 & 0.855 & 0.965 & 2.350 & 11.793 \\
            &14.744 & 0.775 & 0.905 & 2.189 & 12.555 \\
            &15.209 & 0.732 & 0.863 & 2.100 & 13.109 \\
            &15.775 & 0.669 & 0.803 & 1.970 & 13.805 \\
            &16.228 & 0.646 & 0.792 & 1.925 & 14.303 \\
            &16.645 & 0.626 & 0.752 & 1.878 & 14.767 \\
            &16.915 & 0.619 & 0.754 & 1.866 & 15.049 \\
            &17.215 & 0.609 & 0.739 & 1.843 & 15.372 \\
    \hline
    \end{tabular}
  \label{tab:addlabel}
}
\end{table}

Then, we transformed the $V$, $B-V$, $V-I$ data of NGC 188 \citep{Stetson04} 
and M68 \citep{Walker94} to the $K_{{s}_0}$ magnitudes and $(V-K_{s})_{0}$ 
colours by means of Eq. (3) and the procedure explained in Section 2.2. 
Colour excess $E(B-V)$, distance modulus $\mu_{0}$, and metallicity for the 
clusters are given in Table 14. The second reference for the cluster NGC 188 
is given for the metallicity only, whereas the first reference refers the 
colour excess and the distance modulus. The $K_{{s}_0}$ magnitudes and 
$(V-K_{s})_{0}$ colours as well as the original $V$, $B-V$, $V-I$ data are 
given in Table 18.   

\begin{table*}
  \center
\scriptsize{
  \caption{$(M_{K_{s}})_{ev}$ absolute magnitudes and $\Delta M_{K_s}$ residuals estimated by the procedure explained in our work. $(M_{K_{s}})_{cl}$ denotes the absolute magnitude evaluated by means of the colour-magnitude diagram of the cluster.}
    \begin{tabular}{cccc|cccc|cccc}
    \hline
    $(V-K_{s})_{0}$ & $(M_{K_s})_{cl}$ & $(M_{K_s})_{ev}$ & $\Delta M_{K_s}$ & $(V-K_{s})_{0}$ & $(M_{K_s})_{cl}$ & $(M_{K_s})_{ev}$ & $\Delta M_{K_s}$ & $(V-K_{s})_{0}$ & $(M_{K_s})_{cl}$ & $(M_{K_s})_{ev}$ & $\Delta M_{K_s}$\\
    \hline
    \multicolumn{4}{c|}{NGC 188 ($[Fe/H]=$ -0.01 dex)} & \multicolumn{4}{c|}{M68 (cont.)} & \multicolumn{4}{c}{M68 (cont.)} \\
    \hline
    2.15  & 1.092 & 1.029 & 0.063 & \multicolumn{1}{c}{1.93} & \multicolumn{1}{c}{-0.671} & \multicolumn{1}{c}{-0.691} & \multicolumn{1}{c|}{0.020} & \multicolumn{1}{c}{2.49} & \multicolumn{1}{c}{-3.735} & \multicolumn{1}{c}{-3.876} & \multicolumn{1}{c}{0.141} \\
    2.16  & 1.061 & 0.963 & 0.097 & \multicolumn{1}{c}{1.94} & \multicolumn{1}{c}{-0.765} & \multicolumn{1}{c}{-0.771} & \multicolumn{1}{c|}{0.006} & \multicolumn{1}{c}{2.50} & \multicolumn{1}{c}{-3.768} & \multicolumn{1}{c}{-3.917} & \multicolumn{1}{c}{0.149} \\
    2.17  & 1.029 & 0.898 & 0.131 & \multicolumn{1}{c}{1.95} & \multicolumn{1}{c}{-0.857} & \multicolumn{1}{c}{-0.850} & \multicolumn{1}{c|}{-0.007} & \multicolumn{1}{c}{2.51} & \multicolumn{1}{c}{-3.802} & \multicolumn{1}{c}{-3.958} & \multicolumn{1}{c}{0.156} \\
    2.18  & 0.996 & 0.832 & 0.163 & \multicolumn{1}{c}{1.96} & \multicolumn{1}{c}{-0.947} & \multicolumn{1}{c}{-0.928} & \multicolumn{1}{c|}{-0.020} & \multicolumn{1}{c}{2.52} & \multicolumn{1}{c}{-3.835} & \multicolumn{1}{c}{-3.999} & \multicolumn{1}{c}{0.164} \\
    2.19  & 0.962 & 0.767 & 0.194 & \multicolumn{1}{c}{1.97} & \multicolumn{1}{c}{-1.035} & \multicolumn{1}{c}{-1.004} & \multicolumn{1}{c|}{-0.031} & \multicolumn{1}{c}{2.53} & \multicolumn{1}{c}{-3.868} & \multicolumn{1}{c}{-4.039} & \multicolumn{1}{c}{0.171} \\
    2.20  & 0.926 & 0.702 & 0.224 & \multicolumn{1}{c}{1.98} & \multicolumn{1}{c}{-1.121} & \multicolumn{1}{c}{-1.079} & \multicolumn{1}{c|}{-0.041} & \multicolumn{1}{c}{2.54} & \multicolumn{1}{c}{-3.902} & \multicolumn{1}{c}{-4.079} & \multicolumn{1}{c}{0.178} \\
    2.21  & 0.890 & 0.638 & 0.252 & \multicolumn{1}{c}{1.99} & \multicolumn{1}{c}{-1.204} & \multicolumn{1}{c}{-1.155} & \multicolumn{1}{c|}{-0.050} & \multicolumn{1}{c}{2.55} & \multicolumn{1}{c}{-3.934} & \multicolumn{1}{c}{-4.120} & \multicolumn{1}{c}{0.185} \\
    2.22  & 0.852 & 0.574 & 0.278 & \multicolumn{1}{c}{2.00} & \multicolumn{1}{c}{-1.286} & \multicolumn{1}{c}{-1.227} & \multicolumn{1}{c|}{-0.059} & \multicolumn{1}{c}{2.56} & \multicolumn{1}{c}{-3.967} & \multicolumn{1}{c}{-4.159} & \multicolumn{1}{c}{0.191} \\
    2.23  & 0.812 & 0.511 & 0.302 & \multicolumn{1}{c}{2.01} & \multicolumn{1}{c}{-1.365} & \multicolumn{1}{c}{-1.298} & \multicolumn{1}{c|}{-0.067} & \multicolumn{1}{c}{2.57} & \multicolumn{1}{c}{-4.000} & \multicolumn{1}{c}{-4.197} & \multicolumn{1}{c}{0.198} \\
    2.24  & 0.771 & 0.447 & 0.324 & \multicolumn{1}{c}{2.02} & \multicolumn{1}{c}{-1.443} & \multicolumn{1}{c}{-1.369} & \multicolumn{1}{c|}{-0.074} & \multicolumn{1}{c}{2.58} & \multicolumn{1}{c}{-4.032} & \multicolumn{1}{c}{-4.236} & \multicolumn{1}{c}{0.204} \\
    2.25  & 0.728 & 0.385 & 0.344 & \multicolumn{1}{c}{2.03} & \multicolumn{1}{c}{-1.518} & \multicolumn{1}{c}{-1.439} & \multicolumn{1}{c|}{-0.080} & \multicolumn{1}{c}{2.59} & \multicolumn{1}{c}{-4.065} & \multicolumn{1}{c}{-4.275} & \multicolumn{1}{c}{0.210} \\
    2.26  & 0.684 & 0.322 & 0.361 & \multicolumn{1}{c}{2.04} & \multicolumn{1}{c}{-1.592} & \multicolumn{1}{c}{-1.508} & \multicolumn{1}{c|}{-0.085} & \multicolumn{1}{c}{2.60} & \multicolumn{1}{c}{-4.097} & \multicolumn{1}{c}{-4.313} & \multicolumn{1}{c}{0.216} \\
    2.27  & 0.637 & 0.261 & 0.377 & \multicolumn{1}{c}{2.05} & \multicolumn{1}{c}{-1.664} & \multicolumn{1}{c}{-1.575} & \multicolumn{1}{c|}{-0.089} & \multicolumn{1}{c}{2.61} & \multicolumn{1}{c}{-4.129} & \multicolumn{1}{c}{-4.351} & \multicolumn{1}{c}{0.221} \\
    2.28  & 0.589 & 0.199 & 0.390 & \multicolumn{1}{c}{2.06} & \multicolumn{1}{c}{-1.735} & \multicolumn{1}{c}{-1.642} & \multicolumn{1}{c|}{-0.093} & \multicolumn{1}{c}{2.62} & \multicolumn{1}{c}{-4.162} & \multicolumn{1}{c}{-4.388} & \multicolumn{1}{c}{0.227} \\
    2.29  & 0.539 & 0.139 & 0.401 & \multicolumn{1}{c}{2.07} & \multicolumn{1}{c}{-1.803} & \multicolumn{1}{c}{-1.707} & \multicolumn{1}{c|}{-0.096} & \multicolumn{1}{c}{2.63} & \multicolumn{1}{c}{-4.194} & \multicolumn{1}{c}{-4.426} & \multicolumn{1}{c}{0.232} \\
    2.30  & 0.488 & 0.078 & 0.409 & \multicolumn{1}{c}{2.08} & \multicolumn{1}{c}{-1.870} & \multicolumn{1}{c}{-1.772} & \multicolumn{1}{c|}{-0.098} & \multicolumn{1}{c}{2.64} & \multicolumn{1}{c}{-4.226} & \multicolumn{1}{c}{-4.463} & \multicolumn{1}{c}{0.237} \\
    2.31  & 0.434 & 0.019 & 0.415 & \multicolumn{1}{c}{2.09} & \multicolumn{1}{c}{-1.936} & \multicolumn{1}{c}{-1.836} & \multicolumn{1}{c|}{-0.100} & \multicolumn{1}{c}{2.65} & \multicolumn{1}{c}{-4.258} & \multicolumn{1}{c}{-4.499} & \multicolumn{1}{c}{0.241} \\
    2.32  & 0.379 & -0.143 & 0.522 & \multicolumn{1}{c}{2.10} & \multicolumn{1}{c}{-2.000} & \multicolumn{1}{c}{-1.899} & \multicolumn{1}{c|}{-0.101} & \multicolumn{1}{c}{2.66} & \multicolumn{1}{c}{-4.290} & \multicolumn{1}{c}{-4.535} & \multicolumn{1}{c}{0.246} \\
    2.33  & 0.322 & -0.099 & 0.421 & \multicolumn{1}{c}{2.11} & \multicolumn{1}{c}{-2.062} & \multicolumn{1}{c}{-1.961} & \multicolumn{1}{c|}{-0.102} & \multicolumn{1}{c}{2.67} & \multicolumn{1}{c}{-4.322} & \multicolumn{1}{c}{-4.571} & \multicolumn{1}{c}{0.250} \\
    2.34  & 0.264 & -0.157 & 0.421 & \multicolumn{1}{c}{2.12} & \multicolumn{1}{c}{-2.124} & \multicolumn{1}{c}{-2.022} & \multicolumn{1}{c|}{-0.101} & \multicolumn{1}{c}{2.68} & \multicolumn{1}{c}{-4.354} & \multicolumn{1}{c}{-4.607} & \multicolumn{1}{c}{0.253} \\
    2.35  & 0.204 & -0.215 & 0.418 & \multicolumn{1}{c}{2.13} & \multicolumn{1}{c}{-2.183} & \multicolumn{1}{c}{-2.082} & \multicolumn{1}{c|}{-0.101} & \multicolumn{1}{c}{2.69} & \multicolumn{1}{c}{-4.386} & \multicolumn{1}{c}{-4.642} & \multicolumn{1}{c}{0.256} \\
    2.36  & 0.142 & -0.271 & 0.414 & \multicolumn{1}{c}{2.14} & \multicolumn{1}{c}{-2.242} & \multicolumn{1}{c}{-2.142} & \multicolumn{1}{c|}{-0.099} & \multicolumn{1}{c}{2.70} & \multicolumn{1}{c}{-4.418} & \multicolumn{1}{c}{-4.677} & \multicolumn{1}{c}{0.259} \\
    2.37  & 0.080 & -0.328 & 0.407 & \multicolumn{1}{c}{2.15} & \multicolumn{1}{c}{-2.299} & \multicolumn{1}{c}{-2.223} & \multicolumn{1}{c|}{-0.076} & \multicolumn{1}{c}{2.71} & \multicolumn{1}{c}{-4.449} & \multicolumn{1}{c}{-4.712} & \multicolumn{1}{c}{0.262} \\
    2.38  & 0.016 & -0.383 & 0.399 & \multicolumn{1}{c}{2.16} & \multicolumn{1}{c}{-2.355} & \multicolumn{1}{c}{-2.281} & \multicolumn{1}{c|}{-0.074} & \multicolumn{1}{c}{2.72} & \multicolumn{1}{c}{-4.481} & \multicolumn{1}{c}{-4.746} & \multicolumn{1}{c}{0.265} \\
    2.39  & -0.049 & -0.439 & 0.389 & \multicolumn{1}{c}{2.17} & \multicolumn{1}{c}{-2.409} & \multicolumn{1}{c}{-2.338} & \multicolumn{1}{c|}{-0.071} & \multicolumn{1}{c}{2.73} & \multicolumn{1}{c}{-4.513} & \multicolumn{1}{c}{-4.780} & \multicolumn{1}{c}{0.267} \\
    2.40  & -0.115 & -0.493 & 0.378 & \multicolumn{1}{c}{2.18} & \multicolumn{1}{c}{-2.463} & \multicolumn{1}{c}{-2.395} & \multicolumn{1}{c|}{-0.068} & \multicolumn{1}{c}{2.74} & \multicolumn{1}{c}{-4.545} & \multicolumn{1}{c}{-4.814} & \multicolumn{1}{c}{0.269} \\
    2.41  & -0.181 & -0.547 & 0.366 & \multicolumn{1}{c}{2.19} & \multicolumn{1}{c}{-2.515} & \multicolumn{1}{c}{-2.451} & \multicolumn{1}{c|}{-0.064} & \multicolumn{1}{c}{2.75} & \multicolumn{1}{c}{-4.577} & \multicolumn{1}{c}{-4.847} & \multicolumn{1}{c}{0.270} \\
    2.42  & -0.248 & -0.600 & 0.352 & \multicolumn{1}{c}{2.20} & \multicolumn{1}{c}{-2.566} & \multicolumn{1}{c}{-2.506} & \multicolumn{1}{c|}{-0.060} & \multicolumn{1}{c}{2.76} & \multicolumn{1}{c}{-4.608} & \multicolumn{1}{c}{-4.880} & \multicolumn{1}{c}{0.271} \\
    2.43  & -0.316 & -0.653 & 0.337 & \multicolumn{1}{c}{2.21} & \multicolumn{1}{c}{-2.616} & \multicolumn{1}{c}{-2.560} & \multicolumn{1}{c|}{-0.056} & \multicolumn{1}{c}{2.77} & \multicolumn{1}{c}{-4.640} & \multicolumn{1}{c}{-4.912} & \multicolumn{1}{c}{0.272} \\
    2.44  & -0.383 & -0.705 & 0.322 & \multicolumn{1}{c}{2.22} & \multicolumn{1}{c}{-2.665} & \multicolumn{1}{c}{-2.614} & \multicolumn{1}{c|}{-0.051} & \multicolumn{1}{c}{2.78} & \multicolumn{1}{c}{-4.672} & \multicolumn{1}{c}{-4.944} & \multicolumn{1}{c}{0.272} \\
    2.45  & -0.450 & -0.634 & 0.184 & \multicolumn{1}{c}{2.23} & \multicolumn{1}{c}{-2.714} & \multicolumn{1}{c}{-2.668} & \multicolumn{1}{c|}{-0.046} & \multicolumn{1}{c}{2.79} & \multicolumn{1}{c}{-4.704} & \multicolumn{1}{c}{-4.976} & \multicolumn{1}{c}{0.272} \\
    2.46  & -0.517 & -0.691 & 0.174 & \multicolumn{1}{c}{2.24} & \multicolumn{1}{c}{-2.761} & \multicolumn{1}{c}{-2.720} & \multicolumn{1}{c|}{-0.041} & \multicolumn{1}{c}{2.80} & \multicolumn{1}{c}{-4.735} & \multicolumn{1}{c}{-5.007} & \multicolumn{1}{c}{0.272} \\
    2.47  & -0.583 & -0.747 & 0.163 & \multicolumn{1}{c}{2.25} & \multicolumn{1}{c}{-2.807} & \multicolumn{1}{c}{-2.772} & \multicolumn{1}{c|}{-0.035} & \multicolumn{1}{c}{2.81} & \multicolumn{1}{c}{-4.767} & \multicolumn{1}{c}{-5.038} & \multicolumn{1}{c}{0.271} \\
    2.48  & -0.649 & -0.801 & 0.152 & \multicolumn{1}{c}{2.26} & \multicolumn{1}{c}{-2.853} & \multicolumn{1}{c}{-2.824} & \multicolumn{1}{c|}{-0.029} & \multicolumn{1}{c}{2.82} & \multicolumn{1}{c}{-4.798} & \multicolumn{1}{c}{-5.068} & \multicolumn{1}{c}{0.270} \\
    2.49  & -0.713 & -0.855 & 0.142 & \multicolumn{1}{c}{2.27} & \multicolumn{1}{c}{-2.898} & \multicolumn{1}{c}{-2.875} & \multicolumn{1}{c|}{-0.023} & \multicolumn{1}{c}{2.83} & \multicolumn{1}{c}{-4.830} & \multicolumn{1}{c}{-5.098} & \multicolumn{1}{c}{0.268} \\
    2.50  & -0.775 & -0.908 & 0.133 & \multicolumn{1}{c}{2.28} & \multicolumn{1}{c}{-2.941} & \multicolumn{1}{c}{-2.925} & \multicolumn{1}{c|}{-0.016} & \multicolumn{1}{c}{2.84} & \multicolumn{1}{c}{-4.861} & \multicolumn{1}{c}{-5.127} & \multicolumn{1}{c}{0.266} \\
    2.51  & -0.836 & -0.960 & 0.124 & \multicolumn{1}{c}{2.29} & \multicolumn{1}{c}{-2.985} & \multicolumn{1}{c}{-2.975} & \multicolumn{1}{c|}{-0.010} & \multicolumn{1}{c}{2.85} & \multicolumn{1}{c}{-4.893} & \multicolumn{1}{c}{-5.156} & \multicolumn{1}{c}{0.264} \\
    2.52  & -0.894 & -1.011 & 0.117 & \multicolumn{1}{c}{2.30} & \multicolumn{1}{c}{-3.027} & \multicolumn{1}{c}{-3.024} & \multicolumn{1}{c|}{-0.003} & \multicolumn{1}{c}{2.86} & \multicolumn{1}{c}{-4.924} & \multicolumn{1}{c}{-5.185} & \multicolumn{1}{c}{0.261} \\
    2.53  & -0.950 & -1.061 & 0.111 & \multicolumn{1}{c}{2.31} & \multicolumn{1}{c}{-3.069} & \multicolumn{1}{c}{-3.073} & \multicolumn{1}{c|}{0.004} & \multicolumn{1}{c}{2.87} & \multicolumn{1}{c}{-4.955} & \multicolumn{1}{c}{-5.213} & \multicolumn{1}{c}{0.258} \\
    2.54  & -1.003 & -1.111 & 0.107 & \multicolumn{1}{c}{2.32} & \multicolumn{1}{c}{-3.110} & \multicolumn{1}{c}{-3.224} & \multicolumn{1}{c|}{0.114} & \multicolumn{1}{c}{2.88} & \multicolumn{1}{c}{-4.986} & \multicolumn{1}{c}{-5.241} & \multicolumn{1}{c}{0.255} \\
    2.55  & -1.053 & -1.159 & 0.106 & \multicolumn{1}{c}{2.33} & \multicolumn{1}{c}{-3.150} & \multicolumn{1}{c}{-3.169} & \multicolumn{1}{c|}{0.019} & \multicolumn{1}{c}{2.89} & \multicolumn{1}{c}{-5.017} & \multicolumn{1}{c}{-5.268} & \multicolumn{1}{c}{0.251} \\
    2.56  & -1.099 & -1.207 & 0.108 & \multicolumn{1}{c}{2.34} & \multicolumn{1}{c}{-3.190} & \multicolumn{1}{c}{-3.217} & \multicolumn{1}{c|}{0.026} & \multicolumn{1}{c}{2.90} & \multicolumn{1}{c}{-5.048} & \multicolumn{1}{c}{-5.294} & \multicolumn{1}{c}{0.247} \\
    2.57  & -1.141 & -1.254 & 0.113 & \multicolumn{1}{c}{2.35} & \multicolumn{1}{c}{-3.230} & \multicolumn{1}{c}{-3.264} & \multicolumn{1}{c|}{0.034} & \multicolumn{1}{c}{2.91} & \multicolumn{1}{c}{-5.078} & \multicolumn{1}{c}{-5.321} & \multicolumn{1}{c}{0.242} \\
    2.58  & -1.178 & -1.300 & 0.122 & \multicolumn{1}{c}{2.36} & \multicolumn{1}{c}{-3.268} & \multicolumn{1}{c}{-3.310} & \multicolumn{1}{c|}{0.042} & \multicolumn{1}{c}{2.92} & \multicolumn{1}{c}{-5.109} & \multicolumn{1}{c}{-5.346} & \multicolumn{1}{c}{0.237} \\
    2.59  & -1.210 & -1.345 & 0.135 & \multicolumn{1}{c}{2.37} & \multicolumn{1}{c}{-3.307} & \multicolumn{1}{c}{-3.356} & \multicolumn{1}{c|}{0.050} & \multicolumn{1}{c}{2.93} & \multicolumn{1}{c}{-5.139} & \multicolumn{1}{c}{-5.370} & \multicolumn{1}{c}{0.231} \\
    2.60  & -1.237 & -1.390 & 0.153 & \multicolumn{1}{c}{2.38} & \multicolumn{1}{c}{-3.344} & \multicolumn{1}{c}{-3.402} & \multicolumn{1}{c|}{0.057} & \multicolumn{1}{c}{2.94} & \multicolumn{1}{c}{-5.169} & \multicolumn{1}{c}{-5.395} & \multicolumn{1}{c}{0.225} \\
\cline{1-4}    \multicolumn{4}{c|}{M68 ([Fe/H]=-2.01 dex)} & \multicolumn{1}{c}{2.39} & \multicolumn{1}{c}{-3.382} & \multicolumn{1}{c}{-3.447} & \multicolumn{1}{c|}{0.065} & \multicolumn{1}{c}{2.95} & \multicolumn{1}{c}{-5.199} & \multicolumn{1}{c}{-5.418} & \multicolumn{1}{c}{0.219}\\
\cline{1-4}    1.84  & 0.295 & 0.096 & 0.199 & \multicolumn{1}{c}{2.40} & \multicolumn{1}{c}{-3.419} & \multicolumn{1}{c}{-3.492} & \multicolumn{1}{c|}{0.073} & \multicolumn{1}{c}{2.96} & \multicolumn{1}{c}{-5.229} & \multicolumn{1}{c}{-5.441} & \multicolumn{1}{c}{0.212}\\
    1.85  & 0.177 & 0.002 & 0.175 & \multicolumn{1}{c}{2.41} & \multicolumn{1}{c}{-3.455} & \multicolumn{1}{c}{-3.536} & \multicolumn{1}{c|}{0.081} & \multicolumn{1}{c}{2.97} & \multicolumn{1}{c}{-5.259} & \multicolumn{1}{c}{-5.464} & \multicolumn{1}{c}{0.205} \\
    1.86  & 0.061 & -0.090 & 0.151 & \multicolumn{1}{c}{2.42} & \multicolumn{1}{c}{-3.491} & \multicolumn{1}{c}{-3.580} & \multicolumn{1}{c|}{0.089} & \multicolumn{1}{c}{2.98} & \multicolumn{1}{c}{-5.288} & \multicolumn{1}{c}{-5.486} & \multicolumn{1}{c}{0.198} \\
    1.87  & -0.051 & -0.180 & 0.129 & \multicolumn{1}{c}{2.43} & \multicolumn{1}{c}{-3.527} & \multicolumn{1}{c}{-3.624} & \multicolumn{1}{c|}{0.097} & \multicolumn{1}{c}{2.99} & \multicolumn{1}{c}{-5.317} & \multicolumn{1}{c}{-5.507} & \multicolumn{1}{c}{0.190} \\
    1.88  & -0.161 & -0.269 & 0.108 & \multicolumn{1}{c}{2.44} & \multicolumn{1}{c}{-3.562} & \multicolumn{1}{c}{-3.667} & \multicolumn{1}{c|}{0.105} & \multicolumn{1}{c}{3.00} & \multicolumn{1}{c}{-5.346} & \multicolumn{1}{c}{-5.527} & \multicolumn{1}{c}{0.181} \\
    1.89  & -0.268 & -0.356 & 0.088 & \multicolumn{1}{c}{2.45} & \multicolumn{1}{c}{-3.597} & \multicolumn{1}{c}{-3.707} & \multicolumn{1}{c|}{0.110} & \multicolumn{1}{c}{3.01} & \multicolumn{1}{c}{-5.374} & \multicolumn{1}{c}{-5.547} & \multicolumn{1}{c}{0.173} \\
    1.90  & -0.372 & -0.442 & 0.069 & \multicolumn{1}{c}{2.46} & \multicolumn{1}{c}{-3.632} & \multicolumn{1}{c}{-3.750} & \multicolumn{1}{c|}{0.118} & \multicolumn{1}{c}{3.02} & \multicolumn{1}{c}{-5.402} & \multicolumn{1}{c}{-5.566} & \multicolumn{1}{c}{0.164} \\
    1.91  & -0.474 & -0.526 & 0.052 & \multicolumn{1}{c}{2.47} & \multicolumn{1}{c}{-3.666} & \multicolumn{1}{c}{-3.792} & \multicolumn{1}{c|}{0.126} & \multicolumn{1}{c}{3.03} & \multicolumn{1}{c}{-5.430} & \multicolumn{1}{c}{-5.583} & \multicolumn{1}{c}{0.154} \\
    1.92  & -0.574 & -0.609 & 0.035 & \multicolumn{1}{c}{2.48} & \multicolumn{1}{c}{-3.701} & \multicolumn{1}{c}{-3.834} & \multicolumn{1}{c|}{0.133} & \multicolumn{1}{c}{3.04} & \multicolumn{1}{c}{-5.457} & \multicolumn{1}{c}{-5.601} & \multicolumn{1}{c}{0.144} \\
    \hline

    \end{tabular}
  \label{tab:addlabel}
}
\end{table*}

We evaluated the $M_{K_{s}}$ absolute magnitude by means of the Eq. (9) for a 
set of $(V-K_{s})_{0}$ colour indices where the clusters are defined. The 
results are given in Table 19. The columns give: (1) $(V-K_{s})_{0}$ colour 
index, (2) $(M_{K_{s}})_{cl}$, the absolute magnitude for a cluster estimated 
by its colour magnitude diagram, (3) $(M_{K_{s}})_{ev}$, the absolute magnitude 
estimated by the procedure, (4) $\Delta M$, absolute magnitude residuals. 
Also, the metallicity for each cluster is indicated near the name of the 
cluster. The differences between the absolute magnitudes estimated by the 
procedure presented in this study and the ones evaluated via the colour-magnitude 
diagrams of the clusters (the residuals) lie in a short interval, i.e. -0.10 
and +0.27 mag. The mean and the standard deviation of the residuals are 
$<\Delta M_{K_{s}}>=0.109$ and $\sigma_{M_{K_{s}}}=0.123$ mag, respectively. 
The distribution of the residuals are given in Table 20 and Fig. 8. The 
residuals for the cluster NGC 188 for the colour interval 
$2.15\leq(V-K_{s})_{0}\leq 2.44$ where the metallicity of NGC 188 
($[Fe/H]=-0.01$ dex) fall off the metallicity interval corresponding to the 
cited colour interval, $-2.15\leq[Fe/H]\leq-0.04$ dex, are a bit larger than 
 the ones for $2.45\leq(V-K_{s})_{0}\leq2.60$ where the metallicity interval, 
$-2.15\leq[Fe/H]\leq +0.37$ dex, covers the metallicity of NGC 188. Although 
the larger residuals are given in the Table 19, they are not considered 
in the statistics. 

\begin{figure}[h]
\begin{center}
\includegraphics[scale=0.4, angle=0]{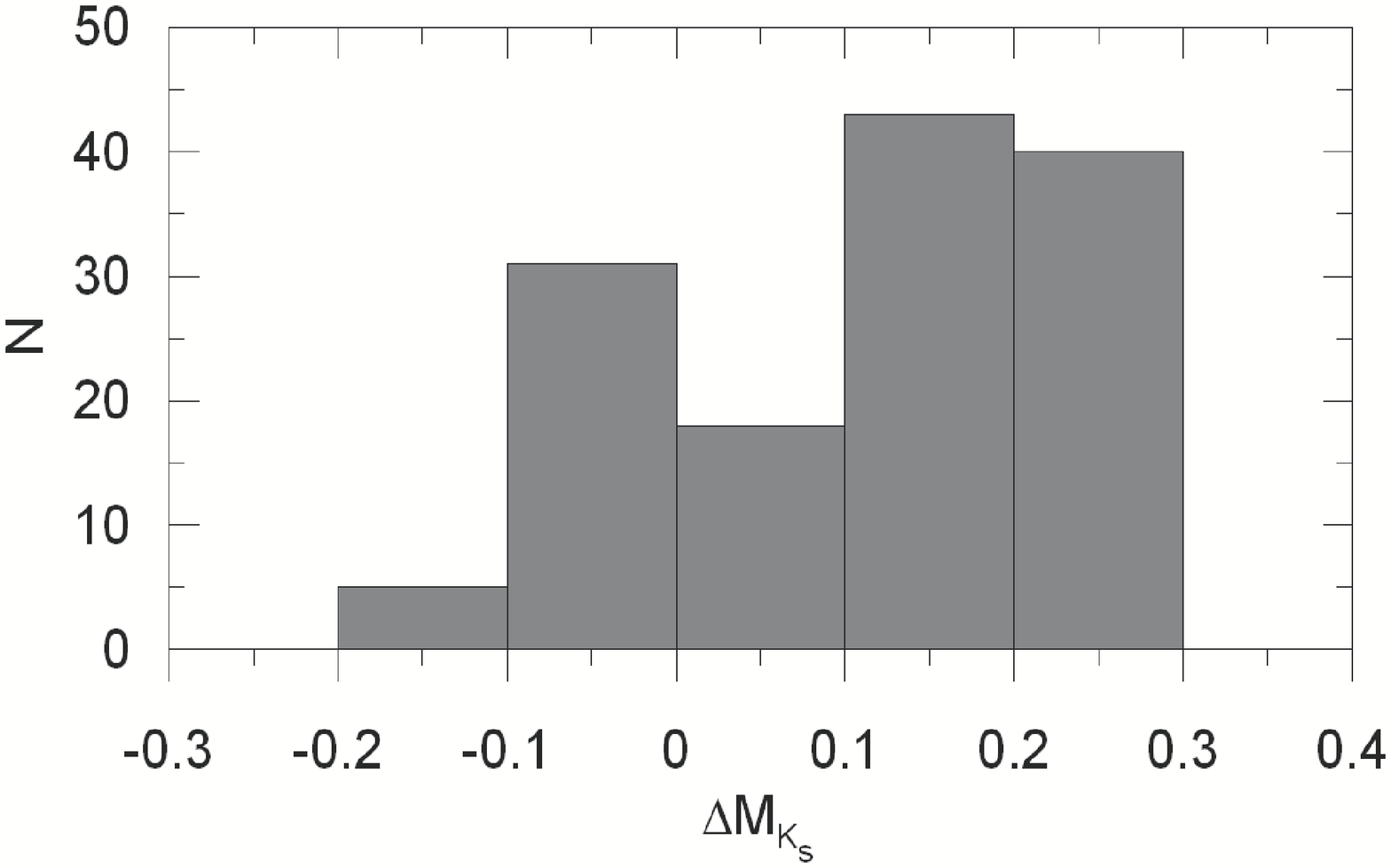}
\caption{Histogram of the residuals for $\Delta M_{K_{s}}$.}
\label{histogram}
\end{center}
\end{figure}

\begin{table}[h]
  \centering
  \caption{Distribution of the residuals. $N$ denotes the number of stars.}
    \begin{tabular}{ccc}
    \hline
  $\Delta M_{K_{s}}$ interval & $<\Delta M_{K_{s}}>$ & N \\
    \hline
    (-0.2, -0.1] & -0.101 &  5\\
    (-0.1,  0.0] & -0.055 & 31\\
    ( 0.0,  0.1] &  0.050 & 18\\
    ( 0.1,  0.2] &  0.148 & 43\\
    ( 0.2,  0.3] &  0.248 & 40\\
    \hline
    \end{tabular}%
  \label{tab:addlabel}%
\end{table}%

\section{Summary and Discussion}
We calibrated the absolute magnitudes $M_{J}$ and $M_{K_{s}}$ for red giants 
in terms of metallicity by means of the colour magnitude diagrams of the 
clusters M92, M13, M71, M67, and NGC 6791 with different metallicities. 
The $J \times (V-J)$ and $K_{s} \times (V-K_{s})$ sequences used for the 
calibration of $M_J$ and $M_{K_{s}}$ are provided from different sources 
and by different procedures, as explained in the following. The main source 
is the paper of \cite{Brasseur10}. The $J \times (V-J)$ and $K_{s} \times 
(V-K_{s})$ sequences for the clusters M92, M13 and  M71 are taken from the 
tables in \cite{Brasseur10}, whereas the $J_{0} \times (V-J)_{0}$ sequence 
for M67 and  NGC 6791 are obtained by transformation of $V$, $B-V$, $V-I$ 
data in \cite{Montgomery93} and by means of the $M_{V} \times (V-J)_{0}$ 
diagram in \cite{Brasseur10}, respectively. Also, the $K_{s} \times (V-K_{s})$ 
sequence for M67 are transformed from $V$, $B-V$, $V-I$ data in 
\cite{Montgomery93}. The fiducial sequence for NGC 6791 is given in $K_{s} 
\times (J-K_{s})$ in \cite{Brasseur10}. We transformed the $J_{0}$ magnitudes 
to the $V_{0}$ ones obtained from the $M_V \times(V-J)_{0}$ diagram and 
altered the $K_{s} \times (J-K_{s})$ data to $K_{s} \times (V-K_{s}$) ones. 
Thus, we obtained two sets of data for two absolute magnitude calibration, 
i.e. $J_{0} \times (V-J)_{0}$  and $K_{{s}_{0}} \times (V-K_{s})_{0}$ for 
$M_{J}$ and $M_{K_s}$, respectively. We combined each set of data for each 
 cluster with their true distance modulus and evaluated two sets of 
absolute magnitudes for the $(V-J)_{0}$ and $(V-K_{s})_{0}$ ranges of each 
cluster. Then, we fitted $M_{J}$ and $M_{K_s}$ absolute magnitudes in terms 
of iron metallicity, $[Fe/H]$, by quadratic polynomials, for a given 
$(V-J)_{0}$ and $(V-K_{s})_{0}$ colour index, respectively. The calibrations 
cover large ranges, i.e. $1.30 \leq (V-J)_{0}\leq 2.80$ and $1.75 
\leq (V-K_{s})_{0}\leq 3.80$ mag for $M_{J}$ and $M_{K_s}$, respectively.   

We evaluated the $M_J$ absolute magnitudes of the clusters M5 
($[Fe/H]=-1.17$ dex) and M68 ($[Fe/H]=-2.01$ dex) by the procedure 
presented in our study for a set of $(V-J)_{0}$ colour index and compared 
them with the ones estimated via combination of the fiducial 
$J_{0}\times(V-J)_{0}$ sequence and the true distance modulus for each 
cluster. The total of the residuals lie between -0.08 and +0.34 mag, 
and the range of 94$\%$ of them is $0<\Delta M_J \leq 0.3$ mag. The mean 
and the standard deviation of (all) the residuals are $<\Delta M_J>=0.137$ 
and $\sigma_{M_{J}}= 0.080$ mag. For the evaluation of the $M_{K_s}$ 
absolute magnitudes, we applied the corresponding procedure to the clusters 
NGC 188 ($[Fe/H]=-0.01$ dex) and M68 ($[Fe/H]=-2.01$ dex). Here again, 
the range of the residuals, their mean and standard deviation are small, 
i.e. $-0.10<\Delta M_{K_s} \leq+0.27$ mag, $<\Delta M_{K_s}>=0.109$ and 
$\sigma_{M_{K_s}}=0.123$ mag.

We compared the statistical results obtained in this study with the ones 
in Paper I and Paper II. Table 21 shows that $M_V$, $M_g$, $M_{J}$, and 
$M_{K_s}$ absolute magnitudes can be estimated with an error less than 
0.3 mag. However, one can notice an improvement on $M_{J}$ and $M_{K_s}$ 
with respect to $M_V$ and $M_g$. The main difference between the data of 
the clusters in three studies is the large domain of the clusters in 
$(V-J)_{0}$ and $(V-K_{s})_{0}$ which probably contributed to more accurate 
calibrations of the apparent $J_{0}$ and $K_{s}$ magnitudes in terms of the 
corresponding colours with respect to $(B-V)_{0}$ and $(g-r)_{0}$ ones. 
Accurate calibration in apparent magnitude provided accurate absolute 
magnitudes. The magnitudes and colours for the cluster M67 used in 
calibration of $M_{J}$ and $M_{K_s}$  absolute magnitudes are not original, 
but they are transformed from the $V$, $B-V$, and $V-I$ data by means of 
the equations of \cite{Yaz10}. The same case holds for the clusters NGC 188 
and M68 which are used in the application of the procedure. Calibrations 
with high correlation coefficients and small residuals confirm also the 
equations of \cite{Yaz10}. As claimed in Paper I and Paper II, there was 
an improvement on the results therein respect to the ones of \cite{Hog98}. 
Hence, the same improvement holds for this study. We quote also the work 
of \cite{Ljunggren66}.

\begin{table}
\setlength{\tabcolsep}{3pt} 
  \center
\scriptsize{  
\caption{Comparison of the results in three studies. The word ``all'' indicates 
to all the residuals. A subset of the residuals is denoted by a percentage, 
such as 91\% or 94\%.}
    \begin{tabular}{cccc}
    \hline
    $\Delta M$ – range  & $< \Delta M >$ & $\sigma$    & Study \\
    \hline
    $(-0.61, +0.66)$ (all) & 0.05 (91$\%$) & 0.190 (91$\%$) & Paper I\\
    $[-0.40, 0.40]$ (91$\%$)  &       &       &\\
    \hline
    $(-0.28, +0.43]$ (all) & 0.169 (all) & 0.140 (all) & Paper II\\
    $( 0.10, 0.40]$, (94$\%$) &       &       &  \\
    \hline
  $(-0.08, +0.34]$ (all) & 0.137 (all) & 0.080 (all) & This study, $\Delta M_J$\\
  $( 0.00, 0.30]$ 94$\%$ &       &       &  \\
 \hline
    $(-0.10, +0.27]$  & 0.109 & 0.123 & This study, $\Delta M_{K_{s}}$\\
     \hline
    \end{tabular}
  \label{tab:addlabel}
}
\end{table}

Although age plays an important role in the trend of the fiducial sequence of the
 RGB, we have not used it as a parameter in the calibration of absolute magnitude. 
A quadratic calibration in terms of (only) metallicity provides absolute magnitudes 
with high accuracy. Another problem may originate from the red clump (RC) stars. 
These stars lie very close to the RGB but they present a completely different 
group of stars. Tables 16 and 19, and Figures 7 and 8 summarize how reliable are 
our absolute magnitudes. If age and possibly the mix with RC stars would affect 
our results this should up. Additionally, we should add that the fiducial 
sequences used in our study were properly selected as RGB. However, the 
researchers should identify and exclude the RC stars when they apply our 
calibrations to the field stars.  

The accuracy of the estimated absolute magnitudes depends mainly on the accuracy 
of metallicity. We altered the metallicity by $[Fe/H]+\Delta[Fe/H]$ in evaluation 
of the absolute magnitudes by the procedure presented in our study and we checked 
its effect on the absolute magnitude. We adopted $[Fe/H]=$-2.01, -1.117, -0.01 dex 
and $\Delta [Fe/H]=$0.05, 0.10, 0.15, 0.20 dex and re-evaluated the absolute 
magnitudes for ten $(V-J)_{0}$ and nine $(V-K_{s})$ colour indices for this 
purpose. The differences between the absolute magnitudes evaluated in this way 
and the corresponding ones evaluated without $\Delta [Fe/H]$ increments are 
given in Table 22. The maximum absolute magnitude differences corresponding to 
$\Delta[Fe/H]=0.20$ dex in $M_{J}$ and $M_{K_s}$ lie in the intervals 
0.09 $\leq \Delta M_{J}\leq$ 0.31 and 0.05 $\leq \Delta M_{K_s}\leq$ 0.14, 
respectively, for the metallicities $[Fe/H]=-1.117$ and $[Fe/H]=-2.01$ dex. 
Whereas, they are about 0.5 mag for the metallicity $[Fe/H]=-0.01$ dex. The 
mean error in metallicity for 42 globular and 33 open clusters in the catalogue 
of \cite{Santos04} is $\sigma=0.19$ dex. If we assume the same error for the 
field stars, the probable error in $M_{J}$ and $M_{K_s}$ would be less than 
0.3 mag for relatively metal poor stars. Whereas, for the solar metallicity 
stars, the metallicity error should be $\sigma_{[Fe/H]}<0.15$ dex in order to 
estimate accurate absolute magnitudes. That is, the solar metallicities should 
be determined more preciously.

\begin{table*}
\centering
\scriptsize{
  \caption{Absolute magnitudes estimated by altering the metallicity as $[Fe/H]+\Delta [Fe/H]$. The numerical values of $[Fe/H]$ are indicated in the last column. The absolute magnitudes in column (1) are the original ones taken from Table 16 and Table 19, whereas those in the columns (2)-(5) correspond to the increments 0.05, 0.10, 0.15, and 0.20 dex. The differences between the original absolute magnitudes and those evaluated by means of the increments are given in columns (6)-(9).}
    \begin{tabular}{c|ccccc|cccc|r}
    \hline
    \multicolumn{1}{c}{} & \multicolumn{5}{c}{$M_{J}$}              & \multicolumn{4}{c}{$\Delta M_J$}   &  \\
    \hline
    \multicolumn{1}{c}{$(V-J)_{0}$} & (1)  & (2)   & (3)  & (4)     & \multicolumn{1}{c}{(5)} & (6)  & (7)  & (8)  & \multicolumn{1}{c}{(9)} & \\
    \hline
    1.50  & -0.200 & -0.139 & -0.076 & -0.012 & 0.055 & 0.061 & 0.124 & 0.189 & 0.255 &  \\
    1.70  & -1.775 & -1.703 & -1.629 & -1.550 & -1.469 & 0.072 & 0.147 & 0.225 & 0.306 &  \\
    1.90  & -2.773 & -2.707 & -2.639 & -2.568 & -2.496 & 0.066 & 0.134 & 0.204 & 0.277 & $[Fe/H]=-1.117+\Delta [Fe/H]$\\
    2.10  & -3.496 & -3.431 & -3.365 & -3.298 & -3.229 & 0.064 & 0.130 & 0.198 & 0.267 &  \\
    2.30  & -4.075 & -4.007 & -3.938 & -3.868 & -3.798 & 0.068 & 0.137 & 0.207 & 0.277 &  \\
    2.50  & -4.747 & -4.707 & -4.664 & -4.619 & -4.570 & 0.040 & 0.082 & 0.128 & 0.177 &  \\
    \hline
    1.50  & -0.986 & -0.952 & -0.917 & -0.879 & -0.841 & 0.034 & 0.070 & 0.107 & 0.146 &  \\
    1.70  & -2.498 & -2.481 & -2.460 & -2.436 & -2.408 & 0.018 & 0.038 & 0.062 & 0.090 & $[Fe/H]= -2.01+\Delta [Fe/H]$\\
    1.90  & -3.535 & -3.508 & -3.478 & -3.446 & -3.412 & 0.027 & 0.057 & 0.088 & 0.123 &  \\
    2.10  & -4.352 & -4.313 & -4.272 & -4.230 & -4.187 & 0.039 & 0.079 & 0.121 & 0.165 &  \\
    \hline
    \multicolumn{1}{c}{} & \multicolumn{5}{c}{$M_{K_s}$}      & \multicolumn{4}{c}{$\Delta M_{K_{s}}$}        & \multicolumn{1}{c}{} \\
    \hline
\multicolumn{1}{c}{$(V-K_{s})_{0}$} & (1)   & (2) & (3) & (4) & \multicolumn{1}{c}{(5)} & (6)     & (7)     & (8)     & \multicolumn{1}{c}{(9)} & \multicolumn{1}{c}{}\\
    \hline
    2.45  & -0.634 & -0.489 & -0.341 & -0.190 & -0.035 & 0.145 & 0.293 & 0.445 & 0.599 &  \\
    2.50  & -0.908 & -0.772 & -0.633 & -0.491 & -0.346 & 0.136 & 0.275 & 0.417 & 0.562 & \multicolumn{1}{c}{$[Fe/H] = -0.01+\Delta [Fe/H]$} \\
    2.55  & -1.159 & -1.031 & -0.900 & -0.766 & -0.630 & 0.128 & 0.259 & 0.393 & 0.529 &  \\
    2.60  & -1.390 & -1.268 & -1.144 & -1.018 & -0.889 & 0.122 & 0.245 & 0.372 & 0.500 &  \\
    \hline
    1.85  & 0.002 & 0.024 & 0.048 & 0.074 & 0.103 & 0.022 & 0.046 & 0.072 & 0.101 &  \\
    2.05  & -1.575 & -1.557 & -1.536 & -1.513 & -1.487 & 0.018 & 0.039 & 0.063 & 0.089 &  \\
    2.25  & -2.772 & -2.764 & -2.753 & -2.738 & -2.719 & 0.008 & 0.019 & 0.034 & 0.053 & \multicolumn{1}{c}{$[Fe/H] = -2.01+\Delta [Fe/H]$} \\
    2.45  & -3.707 & -3.695 & -3.679 & -3.661 & -3.639 & 0.012 & 0.027 & 0.046 & 0.068 &  \\
    2.65  & -4.499 & -4.468 & -4.434 & -4.398 & -4.361 & 0.031 & 0.065 & 0.101 & 0.138 & \multicolumn{1}{c}{} \\
    \hline
    \end{tabular}
}
\end{table*}

The absolute magnitudes can be calibrated as a function of ultraviolet excess 
instead of metallicity, in general. However, an ultraviolet band is not defined 
in 2MASS photometry. Hence, we calibrated the $M_{J}$ and $M_{K_s}$ absolute 
magnitudes in term of metallicity which can be determined by means of atmospheric 
model parameters. Age is a secondary parameter for the old clusters and does not 
influence much the position of their RGB. The youngest cluster in our study is 
M67 with age 4 Gyr (Paper I). However, the field stars may be younger. We should 
remind that the derived relations are applicable to stars older than 4 Gyr.  

We conclude that the two absolute magnitudes, $M_{J}$ and $M_{K_s}$, in 2MASS 
photometry can be estimated for the red giants in terms of metallicity with 
accuracy of $\Delta M \leq $ 0.3 mag. Our target in near future would be to 
adopt this procedure to RC stars.          
 
\section*{Acknowledgments}
This research has made use of NASA's Astrophysics Data System and the SIMBAD 
database, operated at CDS, Strasbourg, France


\begin{thebibliography}{}

\bibitem[Anthony-Twarog, Twarog \& Mayer(2007)]{Anthony-Twarog07}
Anthony-Twarog, B. J., Twarog, B. A., Mayer, L. 2007, AJ, 133, 1585

\bibitem[Bilir et al.(2008)]{Bilir08}
Bilir, S., Karaali, S., Ak, S., Yaz, E., Cabrera-Lavers, A., Co\c skuno\u glu, K. B. 2008, MNRAS, 390, 1569

\bibitem[Bilir et al.(2009)]{Bilir09}
Bilir S., Karaali S., Ak S., Co\c skuno\u glu K. B., Yaz E., Cabrera-Lavers A. 2009, MNRAS, 396, 1589

\bibitem[Bilir et al.(2012)]{Bilir12} 
Bilir, S., Karaali, S., Da{\v g}tekin, N.~D., {\"O}nal, {\"O}., Ak, S., Ak, T., Cabrera-Lavers, A., 2012, PASA, 29, 121 

\bibitem[Brasseur et al.(2010)]{Brasseur10}
Brasseur, C. M., Stetson, P. B., VandenBerg, D. A., Casagrande, L., Bono, G., Dall'Ora, M. 2010, AJ, 140, 1672

\bibitem[Breddels et al.(2010)]{Breddels10}
Breddels, M. A., et al. 2010, A\&A, 511A, 90 

\bibitem[Chen et al.(2001)]{Chen01}
Chen, B., et al. 2001 ApJ, 553, 184

\bibitem[ESA(1997)]{ESA97}
ESA, 1997, The Hipparcos and Tycho Catalogues, ESA SP-1200.ESA Noordwijk

\bibitem[Fiorucci \& Munari(2003)]{FM03}
Fiorucci, M., Munari, U. 2003, A\&A, 401, 781

\bibitem[Gratton et al.(1997)]{Gratton97}
Gratton, R. G., Fusi Pecci, F., Carretta, E., Clementini, G., Corsi, C. E., Lattanzi, M. 1997, ApJ, 491, 749

\bibitem[Harris(2010)]{Harris10}
Harris, W. E. 2010, arXiv1012.3224

\bibitem[Hodder et al.(1992)]{Hodder92}
Hodder, P. J. C., Nemec, J. M., Richer, H. B., Fahlman, G. G. 1992, AJ, 103, 460

\bibitem[Hog \& Flynn(1998)]{Hog98} 
Hog, E., Flynn, C. 1998, MNRAS, 294, 28

\bibitem[Karaali et al.(2003)]{Karaali03}
Karaali, S., Karata\c s, Y., Bilir, S., Ak, S. G., Hamzao\u glu, E. 2003, PASA, 20, 270

\bibitem[Karaali et al.(2012a)]{Karaali12a}
Karaali, S., Bilir, S., Yaz G\"ok\c ce, E., 2012a, PASA, (Paper I), arXiv1204.4291K, http://dx.doi.org/10.1071/AS12003 
 
\bibitem[Karaali et al.(2012b)]{Karaali12b}
Karaali, S., Bilir, S., Yaz G\"ok\c ce, E., 2012b, PASA, (accepted, Paper II), arXiv1206.2752K 

\bibitem[Laird, Carney \& Latham(1988)]{Laird88}
Laird, J. B., Carney, B. W., Latham, D. W. 1988, AJ, 96, 1908

\bibitem[Ljunggren \& Oja(1966)]{Ljunggren66}
Ljunggren, B., Oja, T. 1966, IAUS, 24, 317

\bibitem[McCall (2004)]{McCall04}
McCall, M. L. 2004, AJ, 128, 2144

\bibitem[Meibom et al.(2009)]{Meibom09}
Meibom, S., Grundahl, F., Clausen, J. V., Mathieu, R., Frandsen, S., Pigulski, A., Narwid, A., Steslicki, M., Lefever, K. 2009, AJ, 137, 5086

\bibitem[Montgomery, Marschall \& Janes(1993)]{Montgomery93}
Montgomery, K. A., Marschall, L. A., Janes, K. A. 1993, AJ, 106, 181

\bibitem[Nissen \& Schuster(1991)Nissen \& Schuster]{NS91}
Nissen, P. E., Schuster, W. J. 1991, A\&A, 251, 457

\bibitem[Phleps et al.(2000)]{Phleps00}
Phleps, S., Meisenheimer, K., Fuchs, B., Wolf, C. 2000, A\&A, 356, 108

\bibitem[Rosenberg et al.(1999)]{Rosenberg99}
Rosenberg, A., Saviane, I., Piotto, G., Aparicio, A. 1999, AJ, 118, 2306

\bibitem[Sandquist et al.(1996)]{Sandquist96}
Sandquist, E. L., Bolte, M., Stetson, P. B., Hesser, J. E. 1996, ApJ, 470, 910

\bibitem[Santos \& Piatti(2004)]{Santos04}
Santos, J. F. C. Jr., Piatti, A. E. 2004, A\&A, 428, 79

\bibitem[Sarajedini, Dotter \& Kirkpatrick(2009)]{Sarajedini09}
Sarajedini, A., Dotter, A., Kirkpatrick, A. 2009, ApJ, 698, 1872

\bibitem[Sandage, Lubin, \&  VandenBerg(2003)]{Sandage03}
Sandage, A., Lubin, L.M., VandenBerg, D.A. 2003, PASP, 115, 1187

\bibitem[Savage \& Mathis(1979)]{SM79}
Savage, B. D., Mathis, J. S. 1979, ARA\&A, 17, 73

\bibitem[Saviane et al.(1998)]{Saviane98}
Saviane, I., Piotto, G., Fagotto, F., Zaggia, S., Capaccioli, M., Aparicio, A. 1998, A\&A, 333, 479

\bibitem[Siebert et al.(2011)]{Siebert11}
Siebert A., et al., 2011, AJ, 141, 187

\bibitem[Siegel et al.(2002)]{Siegel02}
Siegel, M. H., Majewski, S. R., Reid, I. N., Thompson, I. B. 2002, ApJ, 578, 151
 
\bibitem[Skrutskie et al.(2006)]{Skrutskie06}
Skrutskie, M.~F., et al. 2006, AJ, 131, 1163

\bibitem[Smith(1987)]{Smith87}
Smith, R. G. 1987, MNRAS, 227, 943

\bibitem[Stetson, McClure \& VandenBerg(2004)]{Stetson04}
Stetson, P. B., McClure, R. D., VandenBerg, D. 2004, PASP, 116, 1012

\bibitem[Stetson, McClure \& VandenBerg(2004)]{Stetson04}
Stetson, P. B., McClure, R. D., VandenBerg, D. 2004, PASP, 116, 1012

\bibitem[Turner(2011)]{Turner11}
Turner, D.G. 2011, RMxAA, 47, 127

\bibitem[Turner(2012)]{Turner12}
Turner, D.G. 2012, Ap\&SS, 337, 303

\bibitem[VandenBerg \& Clem(2003)]{VC03}
VandenBerg, D. A., Clem, J. L. 2003, AJ, 126, 778

\bibitem[van Leeuwen(2007)]{Leeuwen07}
van Leeuwen, F. 2007, A\&A, 474, 653

\bibitem[Walker(1994)]{Walker94}
Walker, A. R. 1994, AJ, 108, 555

\bibitem[Yaz et al.(2010)]{Yaz10}
Yaz, E., Bilir, S., Karaali, S., Ak, S., Co\c skuno\u glu, B., Cabrera-Lavers, A. 2010, AN, 331, 807

\bibitem[Zwitter et al.(2010)]{Zwitter10}
Zwitter, T., et al. 2010, A\&A, 522A, 54

\end{thebibliography}
\end{document}